%% file: main.tex
\documentclass[11pt,letterpaper]{article}
\input{preamble}

\title{Decentralized Dynamic Stable Matching}
\author{
  Jing Chen \qquad Mengtong Gao \qquad Zhenhe Zhang \\
  jchencs@tsinghua.edu.cn\\
  \{gmt25, zhenhe-z21\}@mails.tsinghua.edu.cn\\
  Department of Computer Science and Technology\\
  Tsinghua University, Beijing, China
}
\date{}

\begin{document}
\maketitle
\thispagestyle{empty}


\begin{abstract}
\normalsize
\input{sections/00_abstract}
\end{abstract}

\clearpage
\input{sections/01_introduction}
\input{sections/02_model}
\input{sections/03_known-preferences}

\input{sections/04_tradeoffs}

\input{sections/05_unknown-preferences}

\clearpage
\appendix
\input{appendices/A_algorithm1}
\clearpage
\input{appendices/B_tradeoffs}
\clearpage
\input{appendices/C_algorithm2}
\clearpage
\input{appendices/D_Claim1}

\input{sections/ai-methodology}
\clearpage

\begingroup
\raggedright
\bibliographystyle{alpha}
\bibliography{references}
\endgroup
\end{document}

%% file: preamble.tex
\usepackage[margin=1in]{geometry}
\usepackage[T1]{fontenc}
\usepackage{lmodern}
\usepackage{microtype}
\usepackage{amsmath,amssymb,amsthm}
\usepackage{booktabs,array}
\usepackage{graphicx}
\usepackage{tikz}
\usepackage{enumitem}
\usepackage{algorithm}
\usepackage[noend]{algpseudocode}
\usepackage[hidelinks]{hyperref}
\usepackage[nameinlink,noabbrev]{cleveref}
\makeatletter
\providecommand*{\theHALG@line}{}
\renewcommand*{\theHALG@line}{\thealgorithm.\arabic{ALG@line}}
\makeatother
\newtheorem{claim}{Claim}
\newtheorem{theorem}{Theorem}
\newtheorem{lemma}{Lemma}

\newtheorem{corollary}{Corollary}

\newtheorem*{restatedlemma}{\restatedlemmaname}

\theoremstyle{definition}
\newtheorem{definition}{Definition}
\crefname{algorithm}{Algorithm}{Algorithms}
\Crefname{algorithm}{Algorithm}{Algorithms}
\crefname{sublemma}{Lemma}{Lemmas}
\Crefname{sublemma}{Lemma}{Lemmas}
\input{macros}
\setlist[itemize]{topsep=4pt,itemsep=2pt,parsep=0pt}
\setlist[enumerate]{topsep=4pt,itemsep=2pt,parsep=0pt}

%% file: macros.tex
\newcommand{\BUS}{\mathsf{BUS}}
\newcommand{\PARK}{\mathsf{PARK}}

\newcommand{\phase}[1]{\textnormal{\textsc{#1}}}
\newcommand{\SEEK}{\phase{Seek}}
\newcommand{\SWEEP}{\phase{Sweep}}
\newcommand{\WORK}{\phase{Work}}
\newcommand{\REPLAY}{\phase{Replay}}
\newcommand{\A}{\mathsf{A}}
\newcommand{\R}{\mathsf{R}}
\newcommand{\none}{\mathsf{none}}
\newcommand{\true}{\mathsf{true}}
\newcommand{\false}{\mathsf{false}}
\newcommand{\protocol}{\textup{\textsc{SC-DA}}}
\newcommand{\indicator}{\mathbf{1}}

%% file: sections/00_abstract.tex
In this work we study decentralized computing of the player-optimal stable matching (POSM) under unannounced player arrivals and departures. Without a common clock, arrival/departure events information, or direct communication, players can only make individual proposals to the arms and observe individual accept/reject outcomes. We measure the mismatch cost --- the total number of rounds in which the realized matching differs from the POSM of that round, as the time horizon $T$ grows. For known preferences, we develop a deterministic polynomial-time self-calibrating protocol that achieves a mismatch cost of $O(t_*+K^2\log T)$, greatly improving the existing upper-bound $O(T^{1-1/K}(\log T)^{2/K})$. Here $K$ and $t_*$ are respectively the number of arms and the time of the last event in the problem instance. Each player individually runs the protocol based on his/her own local time and proposal-outcome history, without knowing~$t_*$ or $T$. More generally, our protocol can be configured with a non-decreasing and polynomial-time-computable function $g(t)=\Omega(t)$, so as to achieve a mismatch cost of $O(g(t_*) + K^2 N_g(T))$, where $N_g(T)$ counts iterations of $t\mapsto t+g(t)$ below $T$. Here,~$g(t)$ reflects the protocol's convergence speed. The relationship between $g(t)$ and~$N_g(T)$ reflects the tradeoff between convergence speed and asymptotic mismatch cost.

We next show that any randomized protocol with convergence speed $g(t)\geq 1$ has a worst-case expected cost of $\Omega(t_* + N_g(T))$. This further implies that, up to a $K^2$ factor, our protocol with $g(t)=t$ is optimal among all protocols with convergence speed $O(t_*)$. Finally, for unknown preferences on both sides, each player and each arm individually observe their own sampled utilities after being matched to the other side. We develop a decentralized learning protocol that achieves expected mismatch cost $O(t_*+K^4\Delta_{\mathrm{min}}^{-2}\log T)$, where $\Delta_{\mathrm{min}}$ measures the preference gaps in the learning problem. A complementary static-setting lower-bound adapted from the literature establishes that the dependencies on $T$ and $\Delta_{\mathrm{min}}$ are tight.

%% file: sections/01_introduction.tex
\section{Introduction}
\label{sec:introduction}

Classical studies of two-sided matching markets focus on computing a stable matching for two fixed sets of participants (``players'' and ``arms''). When preferences are known, player-proposing deferred acceptance (DA) computes the player-optimal stable matching (POSM), which assigns to each player its most preferred arm among all stable matchings~\cite{GS62}. In dynamic markets, player arrivals and departures may change the POSM.
Without any notification of these changes, remaining players may continue being matched to 
their previously assigned arms, even when the resulting matching is no longer the current POSM. 
In this work, we study decentralized protocols for maintaining the POSM under such changes, where each player acts on its own local clock and proposal-outcome history. We measure the cumulative mismatch cost by the number of rounds in which the realized matching differs from the POSM of that round, as the time horizon $T$ grows.

As a toy example, consider two players \(p,q\) and two arms \(a,b\). Both players prefer \(a\) to \(b\), arm~\(a\) prefers \(p\) to \(q\), and arm \(b\) prefers \(q\) to \(p\). With both players present, the POSM assigns \(p\) to \(a\) and \(q\) to \(b\). 
If \(p\) departs, the new POSM should have assigned \(q\) to \(a\), but player \(q\) continues to be accepted when proposing to \(b\). In fact, as long as \(q\) continues proposing to \(b\), it receives the same acceptance outcome whether \(p\) stays or departs. To distinguish the two cases, therefore, \(q\) must propose to \(a\), which disrupts the correct matching if \(p\) is still present. This creates a trade-off between recovery from hidden departures and maintenance of a correct matching in a static scenario where the population doesn't change.
%

Besides detecting population changes, players must also coordinate in a decentralized manner when computing the POSM. They may have different arrival times and local states, without a common clock, population roster, or direct communication. Consequently, coordination must be achieved through individual proposals and accept/reject outcomes. 
To tackle these difficulties, our approach alternates between deferred acceptance and 
replay of the resulting matching, 
using only a player's local information to schedule his/her subsequent computation.

\subsection{Our Contributions}
\label{sec:contributions}

\paragraph{Dynamic matching with known preferences.}
We develop Self-Calibrating Deferred Acceptance (\protocol{}), a deterministic decentralized protocol for maintaining the POSM when preferences are known. With \(K\) arms and finitely many player arrivals and departures, \protocol{} achieves a cumulative mismatch cost \(O(t_*+K^2\log T)\), where 
\(t_*\) is the last arrival/departure event time. Using only each player's local clock and proposal-outcome history, the protocol coordinates deferred acceptance computing and calibrates replay lengths without requiring any player to know \(T\) or \(t_*\).

\paragraph{Recovery-maintenance trade-offs.}
More generally, \protocol{} can be configured with a non-decreasing and polynomial-time-computable function $g(t)=\Omega(t)$, so as to achieve a mismatch cost of $O(g(t_*) + K^2 N_g(T))$, where $N_g(T)$ counts iterations of 
\(t\mapsto t+g(t)\) below~\(T\).
Here $g(t)$ reflects the protocol's convergence speed to recover the POSM after a population event, and the relationship between $g(t)$ and $N_g(T)$ highlights the tradeoff between recovery and maintaining POSM.
We further establish an information-theoretic lower-bound on the cost of maintenance: 
any protocol with convergence speed $g(t)\geq 1$ and a sublinear mismatch cost
has a worst-case expected cost of 
$\Omega(t_* + N_g(T))$, or \(\Omega(N_g(T))\) for static scenarios.
The lower-bound establishes the asymptotic optimality of \protocol{} (where \(g(t)=t\)) up to a factor \(K^2\).

\paragraph{Dynamic matching with two-sided learning.}
For multi-armed bandits where both sides' preferences are unknown, we develop Self-Calibrating Preference Certification (\textsc{SC-PC}), a decentralized learning protocol for computing and maintaining the POSM.
\textsc{SC-PC} achieves expected cumulative mismatch \(O(t_*+K^4\Delta_{\min}^{-2}\log T)\), where \(\Delta_{\mathrm{min}}\) denotes the minimum relevant preference gap for the preference distributions
after $t_*$.
A lower-bound adapted from prior work shows that the dependence on \(T\) and \(\Delta_{\min}\) is tight even in static markets.

\subsection{Related Work}
\label{sec:related-work}

\paragraph{Dynamic matching markets.}
Dynamic matching has been broadly studied under evolving preferences and rewards, with objectives including the number of blocking pairs, changes of stable matching, and nonstationary regret~\cite{kanade2016stable,bredereck2020adapting,ghosh2024nonstationary,muthirayan2023competing}. In particular, \cite{ghosh2024nonstationary} studied asynchronous matching bandits 
with a global clock.
\cite{boehmer2022theory} and \cite{faenza2024twostage} studied centralized matching with announced agent arrivals and departures. 
\cite{blum1997vacancy} studied decentralized matching with firm entries and worker departures, 
but 
the recovery period doesn't allow new arrivals or departures, and 
the matching at the beginning of recovery is publicly known.
Closest to our setting, \cite{gao2026decentralized} studied decentralized matching bandits with unannounced player arrivals and departures and no global clock. 
It achieves an expected per-player POSM regret with an asymptotic dependence $O(T^{1-1/K}(\log T)^{2/K})$ under two-sided learning, whereas the same dependence also holds for their cumulative mismatch cost.
Our work greatly improves this asymptotic dependence, for known preferences as well as for two-sided learning.

\paragraph{Recovery and maintenance in distributed protocols.}
A central question in distributed computing is how quickly correctness can be restored after changes and how long it can be maintained. 
Dynamic size-counting protocols study convergence and holding times under adversarial population changes~\cite{doty2022dynamic,kaaser2024dynamic}, while loosely-stabilizing leader election establishes a tight trade-off between these times~\cite{sudo2021timeoptimal}. For matching, self-stabilizing stable-marriage algorithms recover stable matchings from arbitrary configurations~\cite{laveau2017self,beauquier2018timeefficient}, and dynamic maximal-matching algorithms bound cumulative incorrect rounds with topology-change notifications and message passing~\cite{censorhillel2021highlydynamic}. Silent-failure monitoring separately studies the trade-off between probing overhead and detection delay~\cite{cohen2014probe}. 
In our setting, detecting potential hidden departures may require proposals that disrupt an otherwise correct POSM, coupling recovery requirements with
maintenance costs.

\paragraph{Learning in matching markets.}
Bandit matching markets study preference learning through repeated interactions~\cite{liu2020competing}. For player-side learning in static markets, decentralized approaches differ in their feedback and coordination mechanisms, including publicly observed matching outcomes~\cite{liu2021bandit,kong2023player,kong2024improved}, phase-based exploration~\cite{basu2021beyond}, and deliberate collision-based communication~\cite{zhang2022matching}. Coordination-free learning has been studied under structural assumptions on preferences~\cite{maheshwari2022decentralized} and through game-theoretic dynamics~\cite{etesami2025uncoordinated}. For two-sided preference learning in static markets, existing methods employ player indexing and coordinated exploration~\cite{pagare2024explore,zhang2024twosided}, or use super-stability to certify stable matchings from partial preference rankings~\cite{basu2025super}. Completely decentralized trial-and-error learning has also been studied with noiseless utility feedback~\cite{shah2025trial}. 
In comparison, our study on two-sided learning is in a dynamic setting with unannounced population changes, using local clocks and noisy utility feedback.


%% file: sections/02_model.tex
\section{Preliminaries}
\label{sec:model}

\subsection{Dynamic Market and Stable Matching}
\label{sec:market}

We consider a dynamic two-sided matching market with \(K\ge 2\) fixed arms, indexed by \(\mathcal K=\{0, 1, \ldots,K-1\}\), where \(K\) is publicly known. Time proceeds in discrete rounds \(t=1,2,\ldots\). Each player \(i\) enters at round \(b_i\ge1\) and departs at round \(d_i>b_i\), with \(d_i=\infty\) if the player never leaves. The set of active players at round \(t\) is $\mathcal P_t=\{i:b_i\le t<d_i\}$, and \(|\mathcal P_t|\le K\) for all \(t\). We consider markets with finitely many player arrivals and departures. 
Simultaneous arrivals and departures are allowed.
Round~$t$ has a population event if some players arrive or depart at $t$. 
The initial market is considered as a population event in round~\(1\), and we denote the last population-event time by \(t_*\). 

Both sides have complete, strict preference rankings over the other side, represented by fixed true utilities. For each player \(i\), the utility of arm \(k\) is \(\mu_{i,k}\in[0,1]\); for each arm \(k\), the utility of player \(i\) is \(\pi_{k,i}\in[0,1]\). Their preferences satisfy 
$k\succ_i\ell\iff\mu_{i,k}>\mu_{i,\ell}$, and $i\succ_k j\iff\pi_{k,i}>\pi_{k,j}$.
All agents prefer having a partner to being unmatched. Let \(p_i(r)\) denote player \(i\)'s \(r\)-th most preferred arm ($r = 1, 2, \dots, K$). Moreover, an arm's ranking is over all players who are ever active at any time in the instance.

At each round \(t\), a matching \(m\) consists of disjoint player-arm pairs from \(\mathcal P_t\times\mathcal K\). We write \(m(i)\) for player \(i\)'s assigned arm, with \(m(i)=\bot\) when \(i\) is unmatched. 
A pair \((i,k)\) is a blocking pair of \(m\) if player \(i\) and arm \(k\) prefer each other to their respective partners under \(m\). A matching is stable if all active players are matched and no blocking pair exists.
Among all stable matchings at round \(t\), the {\em player-optimal stable matching (POSM)}, denoted by~\(m_t^*\), assigns each active player its most preferred arm attainable in any stable matching:
\[
 \mu_{i,m_t^*(i)}\ge\mu_{i,m(i)} \quad \text{for every } i\in\mathcal P_t \text{ and stable matching } m.
\]
It is well known that the POSM is unique and can be computed by player-proposing deferred acceptance~\cite{GS62}.

\subsection{Decentralized Matching Protocol}
\label{sec:information}

A decentralized matching protocol specifies how players make proposals to arms, and how arms decide which to accept, using only a player's or an arm's locally available information.
In each round \(t\), every active player proposes to exactly one arm. Each arm observes its proposers and accepts exactly one if the proposal set is nonempty; all other proposers are rejected. Each player observes the outcome on whether its proposal is accepted or rejected (A/R) and, when accepted, its realized utility. The accepted player-arm pairs form the realized matching \(M_t\).

Upon entry at round $b_i$, player \(i\) initializes its local state according to the protocol,
and sets its local clock to \(0\). Its local clock in round~\(t\geq b_i\) is \(a_i=t-b_i\). 
Active players' local states are not affected by population changes.
Neither players nor arms have access to a global clock for $t$. 
Players cannot observe the active player set \(\mathcal P_t\) or other players' actions or outcomes. No direct communication, population-event notification, or public coordination signal is available. 

In the \emph{known-preference setting}, each player \(i\) knows its true utilities \(\{\mu_{i,k}\}_{k\in\mathcal K}\), and each arm \(k\) knows its true utilities \(\{\pi_{k,i}\}_i\). Each arm accepts its most preferred current proposer.

In the \emph{two-sided learning setting}, both players and arms initially lack knowledge of their own true utilities and preference rankings. Each arm recognizes returning proposers and maintains its own local clock and interaction history. Its acceptance decisions may depend on this local information.
When player \(i\) is matched with arm \(k\), both sides observe their respective stochastic utilities in \([0,1]\), with 
expectations \(\mu_{i,k}\) and \(\pi_{k, i}\), respectively. Rejected proposals yield no utility observations.


\subsection{Performance Measurement}
\label{sec:objective}

A dynamic matching instance \(I\) specifies the true utilities of both sides and
finitely many population changes, 
fixed independently of the protocol's randomness and utility observations. 
%
Wlog, $I$ has a {\em non-empty final market}: \(\mathcal P_{t_*}\neq \emptyset\).
With known preferences and a deterministic protocol~$\Pi$,
for an instance \(I\) and a time horizon \(T\ge1\), we define the cumulative POSM mismatch cost as
\begin{equation}
  D_{\Pi}(T;I)=\sum_{t=1}^{T}\indicator\{M_t\ne m_t^*\}.
  \label{eq:cost}
\end{equation}
We write \(D(T; I)\) when the protocol is clear. This cost counts the number of rounds in which the realized matching differs from the POSM of that round, whereas the binary stable regret~\cite{basu2025super} counts the number of rounds in which the resulting matching is unstable.

When protocol decisions or utility observations are randomized, we consider the expected mismatch cost \(\mathbb E[D_{\Pi}(T; I)]\). Finally, the protocol is specified first,
before the instance and the time horizon are chosen; and agents are not given \(T\), \(t_*\), or the population changes. 

%% file: sections/03_known-preferences.tex
\section{Dynamic Matching with Known Preferences}
\label{sec:known}

We now introduce Self-Calibrating Deferred Acceptance (\protocol{}), a decentralized matching protocol that combines 
player-proposing deferred acceptance 
with ``replay'' of the resulting matching. 
Each player calibrates its own replay lengths,
allowing the protocol to recover after population changes while limiting cumulative mismatch.

\subsection{The Self-Calibrating Deferred Acceptance Protocol}
\label{sec:known-algorithm}

\begin{figure}[htbp]
    \centering
    \includegraphics[width=0.65\linewidth]{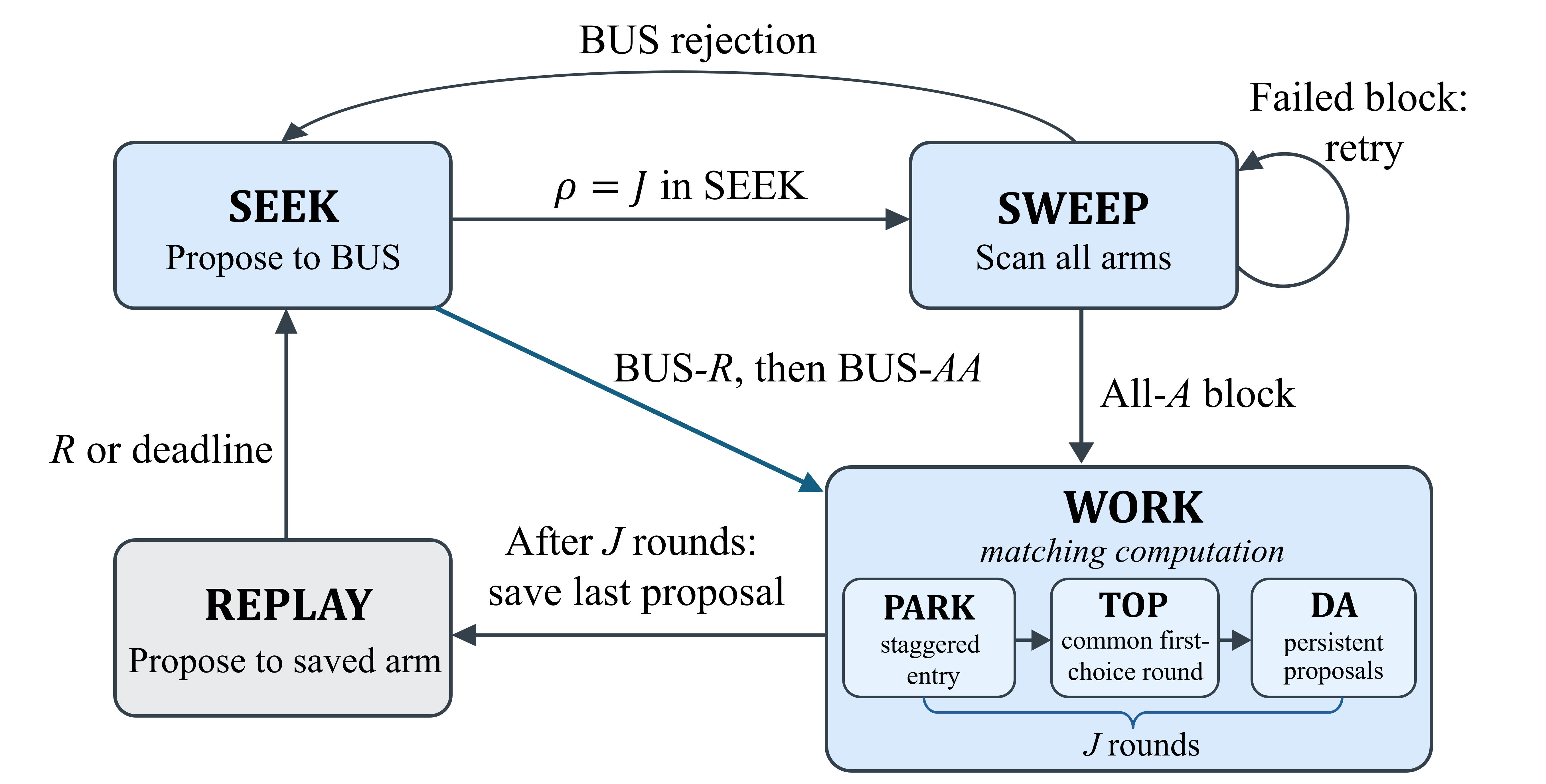}
    \caption{State transitions in \protocol{}.
    The three stages in \WORK{} are also shown. 
    Most transitions (except, e.g., \REPLAY{}$\rightarrow$\SEEK{}) 
    are triggered by outcome feedback and take effect in the next round.}
\label{fig:scda-modes}
\end{figure}

\protocol{} 
has four local modes for each player: \SEEK{}, \SWEEP{}, \WORK{}, and \REPLAY{}. 
Each newly arriving player $i$ starts in \SEEK{} and individually runs the protocol. \Cref{fig:scda-modes} illustrates the transitions between these modes at a high level, and \Cref{alg:scda-core} specifies the player-local operations. Below we explain the notions and operations used in the protocol.

\input{sections/known/local-rules}

In particular, we designate arm \(0\) as a special BUS role for the purpose of coordinating players, and arm \(1\) as a special PARK role for the initial stage of \WORK{}. 
The protocol defines
\begin{equation}
  J=K^2+5K+1,\qquad H=4J+4.
  \label{eq:parameters}
\end{equation}
Here \(J\) sets both the duration of \WORK{} and the number of consecutive BUS acceptances required to enter \SWEEP{}, and \(H\) provides a lower bound on the replay length.

\paragraph{\SEEK.}
A player in \SEEK{} repeatedly proposes to BUS. 
%
Each player maintains a counter \(\rho\), initially zero and capped at \(J\), recording consecutive BUS acceptances. An accepted BUS proposal increments~\(\rho\), 
while a BUS rejection or a proposal to another arm resets it to zero (Line~\ref{line:scda-bus-counter} of \Cref{alg:scda-core}). The counter is updated in every mode and preserved across mode transitions.  
At the start of a round, a player in \SEEK{} with \(\rho=J\) enters \SWEEP{} (Lines~\ref{line:scda-scan-qualification}--\ref{line:scda-scan-qualification-end} of \Cref{alg:scda-core}).
In a given round,
call the active player most preferred by BUS the \emph{root}. 
If the root remains the same for a while, then competitions at the BUS 
enable it to accumulate $J$ consecutive acceptances and eventually enter \SWEEP{}.

On entering \SEEK{}, a player clears its rejection flag, which is set upon receiving a BUS rejection. Once the flag is set, two consecutive BUS acceptances trigger an \emph{AA handoff}, and the player enters \WORK{} in the following round (Lines~\ref{line:scda-aa-handoff}--\ref{line:scda-aa-handoff-end} of \Cref{alg:scda-core}).
Roughly speaking, the transition path of a long-lasting root is \SEEK{}$\rightarrow$\SWEEP{}$\rightarrow$\WORK{}, and that of the other players is \SEEK{}$\rightarrow$\WORK{} via AA handoff.

\paragraph{\SWEEP.}
After entering \SWEEP{}, a player repeatedly executes \(2K\)-round blocks, each with the proposal sequence
\begin{equation}
  \sigma=(0,1,0,2,\ldots,0,K-1,0,0),
  \label{eq:sweep}
\end{equation}
indexed by $1, 2, \dots, 2K$.
A rejection at BUS causes the player to directly return to \SEEK{} in the next round. A rejection at a non-BUS arm marks the current block as unsuccessful; 
but the player still completes the block and starts a new one unless rejected at BUS. If every proposal in a block is accepted, 
the block is successful and the player enters \WORK{} in the following round (see \hyperlink{scda:sweep}{\textbf{SweepFeedback}}, called in Line~\ref{line:scda-sweep-advance} of \Cref{alg:scda-core}).

The sweep visits every arm while leaving BUS for only one round at a time. Intuitively, a successful block in \SWEEP{} by some player (likely, the root)
brings all other players in \REPLAY{} back to \SEEK{}, preparing them for 
subsequent entry into \WORK{}; see also the discussion below on \REPLAY{}.
The alternating visits between BUS and other arms prevent other players from entering \WORK{} via AA handoff too early ---that is, earlier than the root does.

\paragraph{\WORK.}
A player enters \WORK{} following either an AA handoff or a successful sweep. Each \WORK{} mode lasts exactly \(J\) rounds and consists of three consecutive stages: \phase{Park}, \phase{Top}, and \phase{DA} (see \hyperlink{scda:select}{\textbf{SelectArm}} in \Cref{alg:scda-core}).

In the \phase{Park} stage, the player proposes to PARK for \(2K\) rounds, leaving BUS available for other players to enter \WORK{}. In \phase{Top}, it proposes to its most preferred arm 
$p_i(1)$
for the next \(2K+1\) rounds. Intuitively, when players enter \WORK{} at staggered two-round offsets because of AA handoff, their \phase{Top} stages share a round in which all players propose to their most preferred arms.

In \phase{DA}, the player uses an index \(k\), initially \(1\), to propose to arm \(p_i(k)\) for the remaining \(J-4K-1\) rounds. Upon acceptance, it continues proposing to the same arm. After a rejection in a \phase{DA} round that is not the final one, it increases \(k\) by one, capped at \(K\) (Lines~\ref{line:scda-da-advance}--\ref{line:scda-da-advance-end} of \Cref{alg:scda-core}). These repeated proposals implement the acceptance and rejection steps of deferred acceptance.

In the final \WORK{} round, regardless of the feedback $z$, the player records its proposed arm as the output \(y\) 
, and updates
\begin{equation}
  L=H+2(a-\ell),\qquad q=a+L,\qquad \ell\gets a,
  \label{eq:output}
\end{equation}
where \(a\) is its current \WORK{} completion time and \(\ell\) is 
its previous \WORK{} completion time,
initially~$0$. 
The protocol thus recursively uses the completion times of \WORK{} to set the maximum length of \REPLAY{} (Lines~\ref{line:scda-output}--\ref{line:scda-enter-replay} of \Cref{alg:scda-core}).
The player enters \REPLAY{} in the following round. 

\paragraph{\REPLAY.}
In this mode, a player repeatedly proposes to its saved arm \(y\). A rejection causes it to enter \SEEK{} in the next round, whereas reaching its replay deadline \(q\) causes it to enter \SEEK{} at the start of that round (Lines~\ref{line:scda-replay}--\ref{line:scda-replay-end} and~\ref{line:scda-replay-expiry}--\ref{line:scda-replay-expiry-end}, respectively, in \Cref{alg:scda-core}). 
The deadline prevents indefinite replay even when proposals to \(y\) continue to be accepted, enabling recovery from hidden population changes.

\subsection{Performance Upper-Bounds}
\label{sec:known-analysis}

We now establish 
the performance guarantees of the \protocol{} protocol.

\begin{theorem}[Cumulative Mismatch of \protocol{}]
\label{thm:known}
In the known-preference setting, for every \(K\ge2\), dynamic matching instance~\(I\),
and time horizon \(T>H\), \protocol{} satisfies
\[
 D(T;I)=O(t_*+K^2\log T).
\]
\end{theorem}

To prove Theorem~\ref{thm:known}, we first intuitively explain a property that we will prove.
We then establish an upper-bound for more general replay lengths
and derive \Cref{thm:known} as a special case.
Consider a static market or a dynamic market after the unknown $t_*$.
Suppose that after a period of time, every active player has finished two \WORK{} modes and obtained two outputs.
%
%
Further suppose that each player's two outputs 
have the same time offset from the root's corresponding outputs.
Let \(o_1<o_2\) be the root's output rounds, and~\(\delta_i\) be player \(i\)'s time offset for its two outputs. Let \(a_i'\) and~\(a_i\) denote~\(i\)'s local clocks at its first and second outputs, respectively. Using its entry round \(b_i\), one easily has
\[
  a_i'=o_1+\delta_i-b_i,\quad a_i=o_2+\delta_i-b_i, \quad \mbox{ and thus } \quad
  a_i-a_i'=
  o_2-o_1.
\]
That is, the elapsed time between the two outputs is the same for every player. Since the two outputs are consecutive in each player's computation history, at the second output, the value of $\ell$ before updating is equal to~$a_i'$.
%
%
The definition of $L$
in Equation~\eqref{eq:output} therefore gives all players {\em the same replay length} after their second outputs, using only their local computation histories. We say that the players are {\em calibrated} when this property holds.

Notice that the equality property of replay lengths holds as long as all players apply a common function to their own \(a-\ell\) in the definition of $L$. 
More generally, let function \(g:\mathbb N_{\ge1}\to\mathbb N_{\ge1}\) be non-decreasing and polynomial-time computable, with \(g(t)=\Omega(t)\). 
We call such a function {\em feasible}.
Consider the following recursion.
\begin{equation}
 \begin{aligned}
  s_0&=H,\qquad s_{j+1}=s_j+g(s_j)\quad \forall j\ge0,\\
  F_g(x)&=H+2x+\max\Bigl(\{g(s_j)-s_j:j\ge0,\ 64s_j\le x\}\cup\{0\}\Bigr)
  \quad \forall x\ge1.
 \end{aligned}
 \label{eq:budget-waiting-rule}
\end{equation}
By construction, $F_g$ is non-decreasing and polynomial-time computable for a feasible $g$, and $F_g(x)\geq H+2x$. 
For each \(s_j\), the term \(g(s_j)-s_j\) can extend the replay beyond \(H+2x\) only when \(64s_j\leq x\). 
This requirement controls how long replay can be at most.

Given $g$, we define a deterministic decentralized matching protocol \(\Pi_g\) by replacing $L$ in Equation~\eqref{eq:output} and Algorithm~\ref{alg:scda-core} with \(L\leftarrow F_g(a-\ell)\). 
When \(g(t)=t\), we have \(F_g(x)=H+2x\), 
and~\(\Pi_g\) coincides with protocol \protocol{}.
Let \(\Psi_{F_g}(x)=x+F_g(x+H)+3H\) and $f^{\circ3}(x) = f(f(f(x)))$ be a function $f$ composed three times.
Lemma~\ref{lem:known-recovery} below guarantees the existence of the two \WORK{} outputs per player for calibration. 
Lemma~\ref{lem:known-maintenance} then shows that subsequent maintenance preserves calibration. 
The proofs are in Appendices~\ref{sec:proof-recovery} and~\ref{sec:proof-maintenance}, respectively.

\begin{lemma}[Recovery and Calibration]
\label{lem:known-recovery}
$\forall$ feasible function $g$ and instance \(I\), in protocol \(\Pi_g\),
each active player $i$ in the final market finishes two consecutive \WORK{} modes, initiated after \(t_*\) and outputting the arm $m_{t_*}^*(i)$.
Player $i$'s outputs
have the same time offset from the root's outputs. 
Let \(t_{\mathrm{cal}}\) be the round when all active players have entered \REPLAY{} after the second \WORK{} outputs. At this round, the players' replay lengths are all equal, and the realized matching equals $m_{t_*}^*$, with
$t_*<t_{\mathrm{cal}}\le\Psi_{F_g}^{\circ3}(t_*+1)$.
\end{lemma}

By Lemma~\ref{lem:known-recovery}, not only the players will be calibrated, but the realized matching at round $t_{\mathrm{cal}}$ is actually the final POSM $m_{t_*}^*$.
After calibration, there will be repeated periods in which all players are in \REPLAY{}. 
We refer to the time windows in between such periods as {\em maintenance windows}. Intuitively, 
a maintenance window starts at the root's replay deadline and ends when all active players have returned to \REPLAY{} after their next \WORK{} outputs.

\begin{lemma}[Maintenance after Calibration]
\label{lem:known-maintenance}
After \(t_{\mathrm{cal}}\) and outside maintenance windows, the realized matching under \(\Pi_g\) is always~\(m_{t_*}^*\), and all players have the same replay lengths.
Each maintenance window lasts \(O(K^2)\) rounds.
Let \(L_0\) be the common replay length at \(t_{\mathrm{cal}}\), and for \(j\ge1\), let \(L_j\) be the common replay length between the \(j\)-th and the $j+1$-st maintenance windows.
\[
 L_0\ge H,\qquad L_{j+1}\ge F_g(L_j)\qquad \forall j\ge0.
\]
\end{lemma}

To bound the total mismatch cost incurred by the maintenance windows by a time horizon $T$, we also need to bound the number of such windows. By \Cref{lem:known-maintenance} and the monotonicity of~\(F_g\), \(L_j\ge F_g^{\circ j}(H)\). 
If a maintenance window starts on or before round \(T\), then the common replay length proceeding it satisfies \(L_j<T\), 
and hence \(F_g^{\circ j}(H)<T\). 
Define
\begin{equation}
N_g(T)=\#\{j\ge0:s_j<T\}.
\label{eq:budget-iteration-count}
\end{equation}
Without loss of generality the time horizon is long enough, with $T> H$. Then $N_g(T)\geq 1$.

\begin{lemma}[Properties of the Replay Length \(F_g\)]
\label{lem:known-scale-rule}
 $\forall K\ge2, T\geq 1$, and $1\leq t\leq T$, we have
\[
 \Psi_{F_g}^{\circ3}(t+1)=O\!\left(g(t)+K^2\right), \quad \text{ and }
 \quad
 \#\{j\ge0:F_g^{\circ j}(H)<T\}=O\!\left(N_g(T)\right).
\]
\end{lemma}

The proof is in Appendix~\ref{sec:tradeoff-upper-proof}.
Lemma~\ref{lem:known-scale-rule} implies that there will be at most $O(N_g(T))$ maintenance windows before $T$.
Together, these lemmas bound the mismatch cost before calibration and during subsequent maintenance, yielding the following theorem.

\begin{theorem}[Cumulative Mismatch of \(\Pi_g\)]
\label{thm:general-budget-upper}
$\forall K\ge2$, feasible function \(g\),
dynamic matching instance~\(I\) and time horizon $T>H$, the decentralized matching protocol \(\Pi_g\) satisfies
\begin{equation}
 D_{\Pi_g}(T;I)=O(g(t_*)+K^2N_g(T)).
 \label{eq:tradeoff-dynamic}
\end{equation}
\end{theorem}

\begin{proof}[Proof of \Cref{thm:known,thm:general-budget-upper}]
We first prove \Cref{thm:general-budget-upper} for the protocol \(\Pi_g\) defined above.
%
%
\Cref{lem:known-recovery,lem:known-scale-rule} give \(t_{\mathrm{cal}}=O(g(t_*)+K^2)\). If \(T<t_{\mathrm{cal}}\), then \(D_{\Pi_g}(T;I)\le T<t_{\mathrm{cal}} = O(g(t_*)+K^2 N_g(T))\). 

For $T\geq t_{\mathrm{cal}}$, 
charge at most~\(t_{\mathrm{cal}}\) for all rounds before calibration. After calibration, \Cref{lem:known-maintenance} guarantees that mismatch occurs only during maintenance windows, and the length of each such window is $O(K^2)$. 
It further implies that the $j+1$-st maintenance window that occurs on or before~$T$ satisfies $F_g^{\circ j}(H) \leq L_j<T$.
%
%
By \Cref{lem:known-scale-rule},
the number of such windows is \(O(N_g(T))\), thus the total mismatch cost is \(O(K^2 N_g(T))\) in them. 


Combining everything together, the cumulative mismatch cost for $T\geq t_{\mathrm{cal}}$ is 
\[
 D_{\Pi_g}(T;I)=O\left(t_{\mathrm{cal}} + K^2N_g(T)\right) = O\left(g(t_*)+K^2N_g(T)\right).
\]
This proves \Cref{thm:general-budget-upper}.
Finally, taking \(g(t)=t\) gives the \protocol{} protocol. In this case, \(s_{j+1}=2s_j\), $F_g(x) = H+ 2x$, and we have \(N_g(T)=O(\log T)\), establishing \Cref{thm:known}.
\end{proof}

Given the instance $I$, the value $g(t_*)$ does not asymptotically grow in the time horizon~$T$. We keep it in the upper-bound to reflect the tradeoff between the protocol's ``convergence speed'' and how much asymptotic mismatch cost it incurs as time grows.

%% file: sections/known/local-rules.tex
\makeatletter
\begin{algorithm}[htbp]
\caption{The Self-Calibrating Deferred Acceptance Protocol (\protocol{})}
\label{alg:scda-core}
\label{alg:outer}
\label{alg:work}
\normalsize
\algrenewcommand{\algorithmicindent}{0.75em}
\algrenewcommand{\algorithmiccomment}[1]{\hfill\(\triangleright\)~#1}
\algrenewtext{EndIf}{\algorithmicend\ \algorithmicif}
\algrenewtext{EndWhile}{\algorithmicend\ \algorithmicwhile}
\setlength{\fboxsep}{4pt}
\setlength{\fboxrule}{0.3pt}
\newcommand{\scdaSubroutineBox}[3]{%
  \begingroup
  \renewcommand*{\theHALG@line}{\thealgorithm.#2.\arabic{ALG@line}}%
  \fbox{\begin{tabular}{@{}p{\dimexpr\linewidth-2\fboxsep-2\fboxrule\relax}@{}}
    \raggedright\hypertarget{scda:#2}{}\textbf{#1}\par\smallskip
    #3
  \end{tabular}}%
  \endgroup
}
\begin{algorithmic}[1]
\Statex Upon player $i$'s arrival:
\State \textbf{Input:} Arms \(\mathcal{K}=\{0,\ldots,K-1\}\), \(K\ge2\); player \(i\)'s strict ranking \(p_i(1),\ldots,p_i(K)\).
\State {\thinmuskip=1.5mu \medmuskip=2mu \thickmuskip=2mu \mbox{\textbf{Parameters:} \(\BUS\gets0;\, \PARK\gets1;\, J\gets K^2+5K+1;\, H\gets4J+4;\, \sigma=(0,1,0,2,\ldots,0,K-1,0,0)\).}}
\State \textbf{Initialization:} \(a,\ell,\rho,h,y\gets0;\quad s,k\gets1;\quad m\gets\SEEK;\quad \nu\gets\none;\quad \mathit{seenR}\gets\false\).
\algstore{scda-main}
\end{algorithmic}
\vspace{-0.6em}
\noindent\begin{tabular}[t]{@{}p{0.59\linewidth}@{\hspace{0.04\linewidth}}p{0.37\linewidth}@{}}
\raggedright\vspace{0pt}
\begin{algorithmic}[1]
\algrestore{scda-main}
\Statex Upon a new round:
\While{player \(i\) is active}
  \If{\(\nu\ne\none\)}\label{line:scda-round-start} \Comment{State transition.}
    \State \(m\gets\nu;\quad\nu\gets\none\)
    \If{\(m=\SEEK\)}
      \State \(\mathit{seenR}\gets\false\)
    \ElsIf{\(m=\WORK\)}
      \State \(h\gets0;\quad k\gets1\)
    \EndIf
  \EndIf
  \If{\(m=\REPLAY\) and \(a\ge q\)}\label{line:scda-replay-expiry}
    \State \(m\gets\SEEK;\quad\mathit{seenR}\gets\false\)\label{line:scda-replay-expiry-end}
  \EndIf
  \If{\(m=\SEEK\) and \(\rho=J\)}\label{line:scda-scan-qualification}
    \State \(m\gets\SWEEP;\quad s\gets1;\quad\mathit{blockOK}\gets\true\)\label{line:scda-scan-qualification-end}
  \EndIf
  \State \(x\gets\)\,\hyperlink{scda:select}{\Call{\textbf{SelectArm}()}{}} 
  \State \textbf{Propose} to \(x\); observe \(z\in\{\A,\R\}\).
  \State \(\rho\gets\begin{cases}
    \min\{J,\rho+1\}, & \text{if } x=\BUS\text{ and }z=\A,\\
    0, & \text{otherwise}.
  \end{cases}\) \label{line:scda-bus-counter}
  
  \If{\(m=\SEEK\) and \(z=\R\)}\label{line:scda-seek} \label{line:scda-aa-handoff}
  \State \(\mathit{seenR}\gets\true\)
  \ElsIf{\(m=\SEEK \land z=\A \land \mathit{seenR} \land \rho\ge2\)}
    \State \(\nu\gets\WORK\)\Comment{AA handoff.}\label{line:scda-aa-handoff-end}
  \ElsIf{\(m=\SWEEP\)}\label{line:scda-sweep}
    \State \((s,\mathit{blockOK},\nu)\gets\) \hyperlink{scda:sweep}{\Call{\textbf{SweepFeedback}}{$x,z,\mathit{blockOK}$}}\label{line:scda-sweep-advance}
  \ElsIf{\(m=\WORK\)}\label{line:scda-work}
    \If{\(h=J-1\)}
      \State \(y\gets x\)\label{line:scda-output}
      \State \(L\gets H+2(a-\ell)\) \Comment{Maximum length of \REPLAY.}\label{line:Lupdate}
      \State \(q\gets a+L;\quad\ell\gets a;\quad\nu\gets\REPLAY\)\label{line:scda-enter-replay}
    \Else
      \If{\(h>4K\) and \(z=\R\)}\label{line:scda-da-advance}
        \State \(k\gets\min\{k+1,K\}\)\label{line:scda-da-advance-end}
      \EndIf
      \State \(h\gets h+1\)\label{line:scda-work-advance}
    \EndIf
    \ElsIf{\(m=\REPLAY\) and \(z=\R\)}\label{line:scda-replay}
    \State \(\nu\gets\SEEK\)\label{line:scda-replay-end}
  \EndIf
  \State \(a\gets a+1\) \Comment{Player $i$'s local clock.}
\EndWhile
\end{algorithmic}
&
\raggedright\vspace{0pt}
\scdaSubroutineBox{\textsc{SelectArm}\(()\)}{select}{%
  \textbf{Return:} 
  The arm specified below.
  \begin{tabular}{@{}l@{\quad}l@{}}
    \(\BUS\) & if \(m=\SEEK\);\\
    \(\sigma_s\) & if \(m=\SWEEP\);\\
    \(y\) & if \(m=\REPLAY\);\\[2pt]
    \multicolumn{2}{@{}l@{}}{If \(m=\WORK\):}\\
    \(\PARK\) & if \(0\le h<2K\); \Comment{\phase{Park}}\\
    \(p_i(1)\) & if \(2K\le h\le4K\); \Comment{\phase{Top}}\\
    \(p_i(k)\) & if \(4K<h<J\) \Comment{\phase{DA}}\\

  \end{tabular}
}
\par\medskip
\scdaSubroutineBox{\textsc{SweepFeedback}\((x,z,\mathit{blockOK})\)}{sweep}{%
  \begin{algorithmic}
  \If{$x=\BUS \land z=\R$}
      \State \Return $(s,\mathit{blockOK},\SEEK)$
    \EndIf

    \State $b\gets\mathit{blockOK}\land z=\A$

    \If{$s<2K$}
      \State \Return $(s+1,b,\none)$
    \ElsIf{$b$}
      \State \Return $(s,b,\WORK)$
    \Else
      \State \Return $(1,\true,\none)$
    \EndIf
  \end{algorithmic}
}
\end{tabular}
\end{algorithm}
\makeatother

%% file: sections/04_tradeoffs.tex
\section{Lower-Bounds and Optimality}
\label{sec:lower-bounds}
\label{sec:tradeoffs}
We continue in the known-preference setting. 
As the example in \Cref{sec:introduction} illustrates, a player may have to propose to another arm to track the POSM after an unannounced departure.
If no one has left, the same proposal disrupts the current POSM.
We therefore study how much cumulative mismatch cost a protocol must incur to meet a given recovery requirement.


\subsection{Recovery and Necessary Mismatches}
\label{sec:tradeoff-problem}
\label{sec:tradeoff-bounds}

For a protocol \(\Pi\) and an instance \(I\), let \(L_\Pi(h; I)\) denote the expected number of mismatched rounds in the post-event window \(t_*+1,\ldots,t_*+h\).
\begin{equation}
 L_\Pi(h;I)=\mathbb E\!\left[D_\Pi(t_*+h;I)-D_\Pi(t_*;I)\right],
 \qquad \forall h\ge1.
 \label{eq:recovery-window}
\end{equation}

Section~\ref{sec:known} constructs \(\Pi_g\) for feasible functions \(g\). 
We now define convergence speed for a broader class of functions, to which the lower bounds below also apply.

\begin{definition}[Convergence speed]
\label{def:sublinear-mismatch}
Let \(g:\mathbb N_{\ge1}\to\mathbb N_{\ge1}\) be a non-decreasing function.
A decentralized matching protocol \(\Pi\) has \emph{convergence speed \(g\)} if, for every \(\varepsilon\in(0,1)\), there is a constant \(c_\varepsilon>0\) such that, for every instance $I$, we have
\[
 \frac{L_\Pi(h;I)}{h}\le\varepsilon,
 \qquad \forall  h\ge c_\varepsilon g(t_*).
\]
\end{definition}

This condition ensures sublinear mismatch cost after 
the last population event: 
\(g(t_*)\) controls the post-event window length needed
to make the average expected mismatch cost per round
arbitrarily small. 
\begin{corollary}[Convergence speed of \(\Pi_g\)]
\label{cor:flexible-sublinear}
For every \(K\ge2\) and every feasible \(g\) as defined in Section~\ref{sec:known-analysis}, the protocol \(\Pi_g\) has convergence speed \(g\).
\end{corollary}

The proof is in Appendix~\ref{sec:known-recovery-guarantee}. In particular, \protocol{} has convergence speed \(g(t)=t\). 
For the lower-bounds, we use the same definition of \(N_g(T)\) as in
Equation~\eqref{eq:budget-iteration-count}. 

\begin{theorem}[Necessary mismatch cost]
\label{thm:recovery-maintenance}
For every \(K\ge2\), every non-decreasing function \(g:\mathbb N_{\ge1}\to\mathbb N_{\ge1}\), and every randomized decentralized matching protocol \(\Pi\) with convergence speed \(g\), the following lower-bounds hold.

There exists a two-player static matching instance \(I_{0}\) such that
\label{claim:static-maintenance}
\[
 \mathbb E [ D_\Pi(T;I_{0})] =\Omega(N_g(T)).
\]

For every \(t_*\ge2\), there exists a dynamic matching instance \(I'\) with last population-event time \(t_*\) such that, for every \(T\ge t_*\),
\label{claim:dynamic-mismatch}
\label{cor:cumulative-mismatch-lower}
\begin{equation}
 \mathbb E [D_\Pi(T;I')] =\Omega\left(t_*+N_g(T)\right).
 \label{eq:cumulative-mismatch-lower}
\end{equation}
\end{theorem}

\paragraph{Proof ideas.}
First, to show the lower-bound for static matching instances, we construct a two-player market where detecting one player's departure requires the other to propose to the departing player's arm in the POSM with both players' presence.
Until this proposal, the remaining player's local feedback is identical whether the other player stays or leaves.
Such proposals cause mismatches when both players remain, and the convergence speed requirement forces sufficiently frequent proposals to yield the lower-bound even though no player leaves in the static case.

For dynamic matching instances, we arrange population changes that cause \(\Omega(t_*)\) expected mismatch cost by round \(t_*\), after which two players stay forever.
As the protocol does not know that population changes have ended,
proposals needed to detect possible departures continue to cause mismatches, so mismatches arise as in the static instance.

The full proof appears in Appendix~\ref{sec:tradeoff-lower-proof}.
%
For fixed \(K\ge2\) and feasible \(g\), 
Corollary~\ref{cor:flexible-sublinear} and \Cref{thm:general-budget-upper,thm:recovery-maintenance} imply that \(\Pi_g\) has worst-case mismatch cost \(\Theta(N_g(T))\) for static matching instances.
For the linear function \(g(t)=t\), \(N_{g}(T)=\Theta(\log T)\), and \Cref{thm:recovery-maintenance} gives the dynamic lower-bound \(\Omega(t_*+\log T)\). This matches \protocol{}’s upper-bound in Theorem~\ref{thm:known}.
Indeed, among all decentralized protocols with convergence speed $O(t)$, \protocol{} is optimal for both the asymptotic part (up to a $K^2$ factor)
and the additive part in terms of $t_*$.
And for any fixed feasible function $g(t) = \Omega(t)$, 
the protocol $\Pi_g$ has the optimal asymptotic mismatch cost among all protocols with convergence speed $g$, also up to a $K^2$ factor. Its additive term may be larger than $t_*$.



\subsection{Representative Convergence Speed Functions}
\label{sec:tradeoff-insights}
For fixed \(K\) and feasible \(g\), \Cref{tab:representative-budgets} gives the optimal asymptotic worst-case expected cumulative mismatch cost
for protocols with convergence speed $g$. 
Larger \(g\) allows 
slower convergence
and can reduce the asymptotic cost
by permitting less frequent maintenance.

\begin{table}[ht]
\centering
\begin{tabular}{@{}ll@{}}
\toprule
Convergence speed function \(g(t)\) & \(N_g(T)\), as \(T\to\infty\)\\
\midrule
\(t\) & \(\Theta(\log T)\)\\
\(t(1+\log t)\) & \(\Theta(\log T/\log\log T)\)\\
\(t\,2^{(\log t)^\delta},\quad0<\delta<1\)
 & \(\Theta((\log T)^{1-\delta})\)\\
\(t^\alpha,\quad\alpha>1\) & \(\Theta(\log\log T)\)\\
\(t\,2^{(\log t)^\delta},\quad\delta>1\)
 & \(\Theta(\log\log\log T)\)\\
\bottomrule
\end{tabular}
\caption{Iteration counts $N_g(T)$ for representative convergence speed functions $g$. All logarithms in the table are base two.
The parameters $\alpha$ and $\delta$ are constant rational numbers.}
\label{tab:representative-budgets}
\label{cor:representative-budgets}
\end{table}


The five functions in Table~\ref{tab:representative-budgets} are non-decreasing, polynomial-time computable,
and satisfy \(g(t)\ge t\).
Thus Theorem~\ref{thm:general-budget-upper}, Corollary~\ref{cor:flexible-sublinear} and Theorem~\ref{thm:recovery-maintenance} apply to these functions.%
\footnote{The exponential function \(g(t)=2^t\) is not polynomial-time computable (given~\(t\) in binary), and thus doesn't satisfy the conditions in Theorem~\ref{thm:general-budget-upper}.
Still, a variant of the protocol $\Pi_g$ can use $g(t)$ and proceed in polynomial time, so that the conclusions of the theorem continue to hold; 
see Appendix~\ref{sec:tradeoff-exponential}.}
Moreover, for every feasible function \(g\), the following relations hold as $T\to\infty$:

\begin{equation}
\begin{aligned}
 g(t)=t^{1+o(1)}
 &\quad\Longrightarrow\quad N_g(T)=\omega(\log\log T),\\
 g(t)=t^{\alpha+o(1)},\quad\alpha>1
 &\quad\Longrightarrow\quad N_g(T)=\Theta(\log\log T),\\
 g(t)=t^{\omega(1)}
 &\quad\Longrightarrow\quad N_g(T)=o(\log\log T).
\end{aligned}
\label{eq:budget-growth-regimes}
\end{equation}
The proofs of these relations and the table entries appear in Appendix~\ref{sec:tradeoff-growth-proof}.

%% file: sections/05_unknown-preferences.tex
\section{Dynamic Matching with Two-Sided Learning}
\label{sec:unknown}

Finally, we introduce Self-Calibrating Preference Certification (\textsc{SC-PC}), a decentralized matching protocol in the dynamic two-sided learning setting introduced in Section~\ref{sec:information}.

\subsection{The Setting and The Performance Guarantee}
\label{sec:learning-model}
\label{sec:learning-analysis}

As before, each agent only uses \(K\), its local clock, its own action history and observations in the computation.
In the learning setting,
conditional on all past actions and observations and on the current proposals and acceptance decisions, the utilities observed by a matched player \(i\) and the corresponding arm \(k\), have means \(\mu_{i,k}\) and \(\pi_{k,i}\), respectively.

In the final market, compare each player's mean utilities for distinct arms and each arm's mean utilities for distinct players. The smallest absolute difference is
\begin{equation}
\begin{aligned}
\Delta_{\min}=\min\Bigl(&\{|\mu_{i,k}-\mu_{i,\ell}|:i\in\mathcal P_{t_*},\
                                  k,\ell\in\mathcal K,\ k\ne\ell\}\\
&{}\cup\{|\pi_{k,i}-\pi_{k,j}|:k\in\mathcal K,\
                                  i,j\in\mathcal P_{t_*},\ i\ne j\}\Bigr).
\end{aligned}
\label{eq:learning-gap}
\end{equation}
Strict preferences give \(\Delta_{\min}>0\). A decentralized protocol does not require \(T\), \(t_*\), or \(\Delta_{\min}\) as input.
We first state the performance guarantee of \textup{\textsc{SC-PC}}.

\begin{theorem}[Two-sided learning and dynamic POSM]
\label{thm:learning}
In the two-sided learning setting,
for every \(K\ge2\), every dynamic matching instance \(I\), and time horizon \(T\ge1\), \textup{\textsc{SC-PC}} satisfies
\[
\mathbb E[D(T;I)]=O\!\left(t_*+K^4\Delta_{\min}^{-2}\log T\right).
\]
\end{theorem}

We also have the following lower-bound even for {\em static Bernoulli instances}, where there are no population changes and the utilities all follow independent Bernoulli distributions.
In particular, 
call a protocol \(\Pi\) \emph{uniformly efficient} if (1) its rules are fixed independently of the unknown means,
and (2) \(\mathbb E[D_\Pi(T;I)]=o(T^\alpha)\) for every constant \(\alpha>0\) 
on every static Bernoulli instance \(I\).
It will become clear that the \textsc{SC-PC} protocol is uniformly efficient.

\begin{claim}[Static learning lower bound]
\label{cor:learning-static-lower}
For every \(K\ge2\) and \(0<\Delta\le\min\{1/8,1/[4(K-1)]\}\), there exists a static \(K\)-player Bernoulli instance \(I\) with a unique stable matching and \(\Delta_{\min}=\Delta\),
on which every uniformly efficient protocol \(\Pi\) satisfies
\[
 \liminf_{T\to\infty}\frac{\mathbb E[D_\Pi(T;I)]}{\log_2 T}
 \ge\frac{K-1}{4\Delta^2}.
\]
\end{claim}

The claim shows that the \(\Delta_{\min}^{-2}\log T\) dependence in Theorem~\ref{thm:learning} is tight,
and that preference learning alone can incur this cost. See Appendix~\ref{app:learning-lower} for the proof.
The same dependence on $T$ and~$\Delta_{\min}$ 
appeared in a lower-bound on individual reward regret for static markets with player-side learning~\cite{SBA21}.
The lower-bound in the claim above is adapted from there.

\subsection{The \textsc{SC-PC} Protocol}
\label{sec:learning-overview}

We now introduce the high-level ideas of \textsc{SC-PC}.
Recall that in \protocol{}, a rejection means that the arm preferred another proposer. In \textsc{SC-PC}, an arm may accept a proposer to learn its utility, so a rejected player cannot draw the same conclusion.
Both sides instead compare their options using confidence bounds based on their own utility observations from past matches.
Appendix~\ref{sec:rcpc:rules} gives the complete \textsc{SC-PC} protocol, starting with the main protocol for the player side, Algorithm~\ref{alg:rcpc-player}, then the protocol for the arm side, Algorithm~\ref{alg:rcpc-arm}. The subroutines used by Algorithm~\ref{alg:rcpc-player} are then introduced in Algorithms~\ref{alg:rcpc-state-update},~\ref{alg:rcpc-collection},~\ref{alg:rcpc-work-action}, and~\ref{alg:rcpc-work-feedback}.

\paragraph{Arm service.}
Each arm 
switches between two service states, \phase{Control} and \phase{Business}. 
An empty round has no acceptance.
In \phase{Control}, only the first proposer in the arm's private order is eligible, except on periodic rounds when all current proposers are eligible.
In \phase{Business}, the confidence bounds determine which current proposers would be the best. Among eligible proposers, the arm first serves those it has never accepted, then those accepted least recently.

\paragraph{Player modes.}
A player starts in \SEEK{} and repeatedly proposes to BUS (arm~\(0\)). Its BUS acceptances and rejections determine whether it enters \WORK{} or \phase{Root} (Algorithm~\ref{alg:rcpc-collection}).
Here \phase{Root} is a new player mode, similar to but different from the previous \SWEEP{} mode. BUS's private order may differ from the preference ranking that defines the root in \protocol{}.
A player entering \phase{Root} waits if its local clock is still below its clearing deadline. During the scan operations it alternates between BUS and the other arms to bring players in \REPLAY{} back to \SEEK{} (Algorithms~\ref{alg:rcpc-state-update} and~\ref{alg:rcpc-collection}).

On entering \WORK{}, the player resets its remaining-arm set to \(\mathcal K\) (Algorithm~\ref{alg:rcpc-state-update}).
It explores arms whose ranking is still uncertain. When its confidence bounds identify a single best remaining arm, the player can \emph{hold} it by repeatedly proposing to it (Algorithm~\ref{alg:rcpc-work-action}).
Removing an arm from its remaining-arm set requires \(K\) consecutive rejections of proposals to it during the part of \WORK{} that permits 
arm deletion (Algorithm~\ref{alg:rcpc-work-feedback}).
The player saves the arm \(y\) that it proposed to in its final \WORK{} round, 
and repeatedly proposes to \(y\) in \REPLAY{} (Algorithms~\ref{alg:rcpc-work-feedback} and~\ref{alg:rcpc-player}).
It updates a local 
\emph{replay scale} 
using the time between consecutive \WORK{} outputs (Algorithm~\ref{alg:rcpc-work-feedback} and Equation \eqref{eq:rcpc:output}). If it finishes \WORK{} holding an arm that accepted its final proposal, this scale sets the waiting length; otherwise the player uses a fixed minimum length (Algorithm~\ref{alg:rcpc-work-feedback}). A rejection can shorten the wait. When the deadline expires, the player returns to \SEEK{} (Algorithm~\ref{alg:rcpc-player}).

\paragraph{Correctness and the learning progress.}
After the unknown \(t_*\), consider a mature episode (in Appendix~\ref{sec:rcpc:control}, similar to a maintenance window in previous sections). One player enters \WORK{} from \phase{Root} and the others follow from \SEEK{}. The episode ends at the first round when all players are in \REPLAY{}.
Suppose the confidence bounds cover both sides' true utility differences throughout the episode. Arm service and the spacing of exploratory proposals guarantee a {\em witness} to each deletion. In at least one rejection round, another player whom the arm prefers to the rejected player held it and proposed to it. Thus every pair of every stable matching survives.
If all players finish holding distinct arms, they replay the POSM. By the next mature episode's first common replay round, their smallest replay scale has grown by a constant factor. An episode that ends without distinct held arms for all players adds samples on player--arm pairs with too few observations.
The proof of \Cref{thm:learning} combines 
the rough ideas above
with an upper-bound on the cost of recovery from statistical errors (Appendix~\ref{sec:rcpc:theorem}).

%% file: appendices/A_algorithm1.tex
\section{Analysis of \protocol{}}
\label{app:known-proofs}
\label{sec:proof}

This appendix proves \Cref{lem:known-recovery,lem:known-maintenance,lem:known-scale-rule} and \Cref{cor:flexible-sublinear}. We first establish matching correctness and calibration for mature executions, then prove recovery and bounded maintenance under a general replay rule. We finally verify the properties of the function \(F_g\) defined in Equation~\eqref{eq:budget-waiting-rule} and prove its post-event guarantee.

Fix \(K\ge2\), with \(J=K^2+5K+1\) and \(H=4J+4\) as in \Cref{alg:scda-core}. For the analysis, allow any nondecreasing computable replay rule \(F:\mathbb N_{\ge1}\to\mathbb N_{\ge1}\) satisfying
\begin{equation}
  F(x)\ge H+2x\qquad(x\ge1).
  \label{eq:general-wait-condition}
\end{equation}
Keep all control rules unchanged. At every actual output, with final proposed arm \(y\), replace only the replay-length update by
\begin{equation}
  L\gets F(a-\ell),\qquad q\gets a+L,\qquad \ell\gets a.
  \label{eq:general-wait-update}
\end{equation}
The update uses the old \(\ell\), even if the output is incorrect or the final proposal is rejected. The argument \(a-\ell\) is always positive: at the first output, \(\ell=0\) and \(a\ge J-1>0\); later outputs occur at local times strictly greater than the value stored in \(\ell\). The original \protocol{} uses \(F(x)=H+2x\). For a feasible function \(g\), the protocol \(\Pi_g\) uses \(F=F_g\); the required properties of \(F_g\) are verified in Appendix~\ref{sec:tradeoff-upper-proof}.

\paragraph{Execution conventions.}
All executions below follow \Cref{alg:scda-core} from its prescribed initialization, with the replay update in Equation~\eqref{eq:general-wait-update}. Upon arrival, \(a=\ell=\rho=0\), \(m=\SEEK\), \(\nu=\none\), and \(\mathit{seenR}=\false\); mode-specific variables are initialized on entry. Surviving players retain their states across population events. Mode entry preserves \(a,y,\ell,\rho\); a fresh \SEEK{} visit clears \(\mathit{seenR}\), and only actual outputs change \(\ell\).

Population events precede the round-start checks. Before proposing, each player applies its pending mode change and clears \(\nu\), checks replay expiration, and checks sweep qualification, in that order. Round-start states in the proofs are taken after these checks and before proposals. After proposing, each player updates the BUS counter \(\rho\), processes feedback in its current mode, and increments local time \(a\), in that order. Feedback-triggered mode changes take effect in the next round.

\subsection{Correctness and Calibration of Mature Executions}
\label{sec:proof-mature}

For this subsection,  we have \(n\le K\) players, labeled \(1,\ldots,n\) from most to least preferred by BUS. 
Player \(1\) is the root, and \(m^*\) is the market's POSM. For \(1\le i\le n\), define \(\delta_i=2(i-1)\); \Cref{lem:mature} will show that these are the offsets of both \WORK{} entry and output from the root's corresponding rounds. These labels and global round numbers are only for analysis; players need not know their BUS ranks or detect mature entry.

\begin{definition}[Mature entry]
\label{def:mature}
After the round-start updates in round \(c\) and before proposals, suppose the root has just entered \WORK{}, while every other player is in \SEEK{} with \(\rho=0\) and a BUS rejection recorded during its current \SEEK{} visit. Then \(c\) is a \emph{mature entry}, and the resulting \WORK{} modes form a \emph{mature execution}.
\end{definition}

With the staggered entry times below, the \phase{Top} stages share a round in which every player proposes to its first-choice arm. The following lemma proves correctness of the ensuing DA updates, and \Cref{lem:mature} establishes the entry and output times and shows that the matching remains correct as all players enter \REPLAY{}.

\begin{lemma}[Deferred Acceptance with Delayed Proposals]
\label{lem:top-da}
Suppose each player \(i\) enters \WORK{} at \(c+\delta_i\), and no population event occurs in \([c,c+J-1)\). Then
\[
  M_{c+J-2}=m^*.
\]
\end{lemma}

\begin{proof}
Let \(o=c+J-1\) be the root's output round and \(t_{\mathrm{top}}=c+2(n-1)+2K\). At \(t_{\mathrm{top}}\), the last entrant has just reached \phase{Top}, while the root remains in \phase{Top} through \(c+4K\). Since \(n\le K\), all players are in \phase{Top} at \(t_{\mathrm{top}}\), and their application indices are still \(1\). All players remain in \phase{Top} or \phase{DA} on \([t_{\mathrm{top}},o)\).

On consecutive rounds of \([t_{\mathrm{top}},o)\), every accepted player repeats its arm, including across the \phase{Top}--\phase{DA} transition. If displaced, it is replaced by a proposer the arm prefers. Thus an arm's acceptor can only improve after its first acceptance in this interval.

For rejection validity, induct on rounds from \(t_{\mathrm{top}}\), when all application indices are \(1\). Suppose arm \(x\) rejects player \(i\), who processes this rejection in a nonfinal \phase{DA} round, and let \(w\) be the player accepted by \(x\) in that round. Every arm that \(w\) prefers to \(x\) was passed after \(w\) processed a rejection in an earlier nonfinal \phase{DA} round; this set is empty if \(w\) is still in \phase{Top}. By induction, none can be \(w\)'s partner in a stable matching \(m\). If \(m(i)=x\), then \(w\) prefers \(x\) to \(m(w)\), and \(x\) prefers \(w\) to \(i\), so \((w,x)\) blocks \(m\). This contradiction uses only rejections from earlier rounds.

A rejection at a player's last arm would require better actual acceptors at all \(K\) arms, by persistence at its previously rejected arms. These would be \(K\) distinct other players, contradicting \(n\le K\). Hence the cap never prevents an advance, and at most \(n(K-1)\) advances occur. Rejected proposals are repeated during \phase{Top}; otherwise these are player-proposing DA updates.

All players are in \phase{DA} by round \(t_{\mathrm{da}}=c+6K-1\), and
\begin{equation}
  t_{\mathrm{da}}+n(K-1)
  \le c+6K-1+K(K-1)
  =c+J-2<o.
  \label{eq:mature-solve-time}
\end{equation}
Every round in \([t_{\mathrm{da}},t_{\mathrm{da}}+n(K-1)]\) without universal acceptance produces an advance. These \(n(K-1)+1\) rounds therefore include an all-accepted round.

Let \(M\) be the matching in an all-accepted round. Every arm that a player prefers to its partner has rejected that player and retains a better acceptor, so \(M\) is stable. Rejection validity also excludes every such preferred edge from all stable matchings, proving \(M=m^*\). Repeated accepted proposals preserve this matching through round \(o-1\).
\end{proof}

\begin{lemma}[Correctness and Output Offsets of Mature Executions]
\label{lem:mature}
Suppose \(c\) is a mature entry and no population event occurs from \(c\) through \(t_{\mathrm{end}}=c+J+2(n-1)\). Let \(o=c+J-1\) be the root's output round. Player \(i\) starts \WORK{} at \(c+\delta_i\) and outputs at \(o+\delta_i\). The actual matching reaches \(m^*\) by \(c+J-2\) and remains there through \(t_{\mathrm{end}}=o+\delta_n+1\le c+J+2K-2\). Each player outputs \(m^*(i)\) exactly once in \([c,t_{\mathrm{end}}]\) and enters \REPLAY{} with that saved arm at \(o+\delta_i+1\). At \(t_{\mathrm{end}}\), all players have entered \REPLAY{} after their \WORK{} outputs.
\end{lemma}

\begin{proof}
\emph{Entry and output times.}
At mature entry, each remaining player in \SEEK{} has a recorded BUS rejection and \(\rho=0\). While earlier entrants stay in \phase{Park}, the remaining player in \SEEK{} most preferred by BUS receives two consecutive BUS acceptances and enters \WORK{} next round. Induction gives player \(i\)'s \WORK{} entry round as
\begin{equation}
  c+\delta_i=c+2(i-1),\qquad 1\le i\le n.
  \label{eq:mature-starts}
\end{equation}
For \(n\ge2\), the last qualifying pair ends at \(c+2n-3\le c+2K-3\), before the root leaves \phase{Park} at \(c+2K\). Each player in \SEEK{} takes its AA handoff before accumulating the \(J>2\) acceptances needed for a sweep. The case \(n=1\) is immediate.

Since each \WORK{} mode lasts exactly \(J\) rounds, player \(i\)'s output round is
\begin{equation}
  (c+\delta_i)+J-1=o+\delta_i,
  \qquad o=c+J-1.
  \label{eq:mature-outputs}
\end{equation}
By \Cref{lem:top-da}, \(M_{o-1}=m^*\), where \(o-1=c+J-2\).

\emph{Output and replay.}
Every output assigns a replay length at least \(H\), so each new global deadline is at least \(o+\delta_i+H\ge o+H\). Since
\[
  t_{\mathrm{end}}=o+\delta_n+1=o+2(n-1)+1\le o+2K-1<o+H,
\]
no replay expires before all players have entered \REPLAY{}. Inductively, players still in \WORK{} repeat their accepted arms and save them at output, while players already in \REPLAY{} propose to those same arms and remain accepted. Hence the matching stays at \(m^*\) through \(t_{\mathrm{end}}\). Each player outputs exactly once, enters \REPLAY{} at \(o+\delta_i+1\), and remains there through \(t_{\mathrm{end}}\).
\end{proof}

\paragraph{Calibration from consecutive outputs.}
Consider any two mature executions in this fixed market that give consecutive actual outputs for every player. Let \(o_1<o_2\) be the root's output rounds in these executions, and let \(b_i\) be player \(i\)'s arrival round. By \Cref{lem:mature}, its local times \(a_i'\) and \(a_i\) at the first and second outputs satisfy
\[
  a_i'=o_1+\delta_i-b_i,\qquad a_i=o_2+\delta_i-b_i.
\]
Let \(\ell_i\) denote the value of \(\ell\) held by player \(i\) immediately before its second output. Consecutive actual outputs give \(\ell_i=a_i'\), so each player sets the same replay length \(L\), with
\[
  a_i-\ell_i=o_2-o_1,\qquad L=F(o_2-o_1).
\]
Let \(Q=o_2+L\) be the root's global replay deadline and \(Q_i\) player \(i\)'s global replay deadline after its second output. These deadlines retain the output offsets:
\begin{equation}
  Q_i=o_2+\delta_i+L=Q+\delta_i.
  \label{eq:common-wait-deadlines}
\end{equation}
When all players have entered \REPLAY{} after these second outputs, they are calibrated as defined in \Cref{sec:known-analysis}.

\subsection{Sweep Isolation and Mature Entry}
\label{sec:proof-sweep}

The next lemma shows that a sweep accepted at every BUS proposal isolates the other players, and that a successful block followed by a round with no population event establishes a mature entry. Part (i) will also bound recovery time.

\begin{lemma}[Sweep isolation and mature entry]
\label{lem:sweep}
Let player \(z\) enter \SWEEP{} at round \(b\).
\begin{enumerate}[label=(\roman*)]
  \item If \(z\) remains in this visit through round \(d\) and all its BUS proposals in \([b,d]\) are accepted, then throughout this interval, every other active player is in \SEEK{} or \REPLAY{} and schedules no \WORK{} entry.
  \item If this visit completes an all-accepted block at round \(e\) and no population event occurs at \(e+1\), then \(e+1\) is a mature entry with root \(z\).
\end{enumerate}
\end{lemma}

\begin{proof}
\emph{Proof of (i).}
Put \(v=b-J\). Since every action other than an accepted BUS proposal resets \(\rho\), even after saturation, qualification at \(b\) implies that \(z\) is accepted at BUS in
\begin{equation}
  v,v+1,\ldots,b-1.
  \label{eq:sweep-qualification}
\end{equation}
No other player can trigger an AA handoff or complete a successful block in this interval, since both require a final BUS acceptance. Thus no other \WORK{} entry is scheduled.

Any \WORK{} entry pending from \(v-1\) takes effect at \(v\), so all earlier \WORK{} modes finish by \(v+J-1=b-1\). A competing sweep present at \(v\) reaches BUS by \(v+1\) and, if still present, returns to \SEEK{} by \(v+2\). Every other player's counter is reset in each qualification round, excluding new sweeps in \(v+1,\ldots,b\). Since \(J>2\), all other players are in \SEEK{} or \REPLAY{} at \(b\), with zero counters and no pending \WORK{}. Arrivals start in \SEEK{} with \(\rho=0\), and departures simply remove players.

The proposal sequence \(\sigma\) in Equation~\eqref{eq:sweep} leaves BUS only for isolated single rounds, including across failed-block restarts. Non-BUS rejections mark a block as failed without restarting it. Since \(z\) wins every BUS action, every other player's consecutive BUS-acceptance count stays at most one. Thus neither an AA handoff nor a new sweep qualification is possible, and all other players remain in \SEEK{} or \REPLAY{}. This proves (i) for every such interval, regardless of eventual success.

\emph{Proof of (ii).}
On a BUS rejection, the player skips the block-completion test and leaves \SWEEP{} in the following round. Thus a visit completing a successful block at \(e\) has only BUS acceptances, and (i) applies with \(d=e\). No other output occurs: each old replay keeps its saved arm until it ends, and no player in \SEEK{} reenters \REPLAY{}. A player still replaying when its saved arm is visited is rejected. Every arm is visited by \(e-2\), so all old replays have ended by \(e-1\); later arrivals start in \SEEK{}.

The final two proposals are to BUS. In round \(e\), every other present player, including any new arrival, therefore proposes to BUS and is rejected. Each ends the round in \SEEK{} with a rejection recorded during its current visit, \(\rho=0\), and no pending \WORK{}. Since \(z\) wins against all these players, it is the player most preferred by BUS. With no population event at \(c=e+1\), its scheduled \WORK{} entry establishes the mature entry in \Cref{def:mature}.
\end{proof}

\subsection{Proof of Lemma~\ref{lem:known-recovery}}
\label{sec:proof-recovery}

To apply the calibration identity, we must obtain two mature executions whose outputs are consecutive for every player in the final market. For the replay rule \(F\) satisfying Equation~\eqref{eq:general-wait-condition}, define the non-decreasing recovery map
\begin{equation}
 \Psi_F(x)=x+F(x+H)+3H.
 \label{eq:general-recovery-map}
\end{equation}

\begin{lemma}[Recovery to actual consecutive mature outputs]
\label{lem:recovery}
For every dynamic matching instance \(I\) with a nonempty final market, run Algorithm~\ref{alg:scda-core} from its prescribed initialization with the update Equation~\eqref{eq:general-wait-update}. There are two mature executions beginning after \(t_*\). Let \(t_{\mathrm{cal}}\) be the round when all players have entered \REPLAY{} after the second execution's outputs. Then
\begin{equation}
 t_{\mathrm{cal}}\le\Psi_F^{\circ3}(t_*+1).
 \label{eq:recovery-time}
\end{equation}
Each player's two outputs are consecutive in its actual output history, so the second output uses the local time of the first output as its old \(\ell\).
\end{lemma}

\begin{proof}[Proof of \Cref{lem:recovery,lem:known-recovery}]
Let \(n\) be the final population size and \(z\) its root, the player most preferred by BUS. At any output in round \(v\), including an incorrect one, a player entering at round \(b_i\ge1\) has \(0\le\ell\le a=v-b_i\le v\). By monotonicity of \(F\), the value of \(\ell\) before this output gives the global deadline
\begin{equation}
 Q_i=v+F(a-\ell)\le v+F(v).
 \label{eq:replay-deadline-bound}
\end{equation}
Replay expires in its deadline round; acceptance never extends it, and rejection can end it earlier.

\paragraph{First mature execution.}
From round \(t_*+1\), \(z\) is never rejected at BUS. Let \(\beta\) be the first round after \(t_*\) in which \(z\) is in \SWEEP{}. Before \(\beta\), it can produce at most one output from an inherited \WORK{} mode or an old \SEEK{} rejection record, and any such output occurs by \(t_*+J+2\). A fresh \SEEK{} visit clears that record and cannot trigger an AA handoff. By Equation~\eqref{eq:replay-deadline-bound}, every preceding replay ends by \(t_*+J+2+F(t_*+J+2)\). At most \(J\) further qualification rounds therefore give
\begin{equation}
 \beta\le t_*+J+2+F(t_*+J+2)+J
 \le\Psi_F(t_*+1).
 \label{eq:continuing-scan-time}
\end{equation}
For a \SWEEP{} visit already in progress at \(t_*+1\), \Cref{lem:sweep}(i) applies from its original qualified entry, even if population events occurred in the meantime.

Until this \SWEEP{} succeeds, \Cref{lem:sweep}(i) excludes other \WORK{} entries, sweeps, and outputs. All remaining replays therefore expire by \(\beta+F(\beta)\). A residual block and one full block take at most \(4K\) further rounds; the full block succeeds if no earlier one does. Let \(t_{\mathrm{first}}\) be the round when all players have entered \REPLAY{} after their outputs from the resulting mature execution. By \Cref{lem:sweep,lem:mature},
\begin{equation}
 t_{\mathrm{first}}
 \le\beta+F(\beta)+J+6K-2\le\Psi_F(\beta).
 \label{eq:first-recovery-handoff}
\end{equation}
All players are then in correct \REPLAY{}, although their replay lengths can differ.

\paragraph{The next \WORK{} entry is mature.}
Before any new \WORK{} entry, no output replaces the current deadlines, all of which are at most \(t_{\mathrm{first}}+F(t_{\mathrm{first}})\). If no entry has occurred by that round, all current replays have ended. Player \(z\) is then already sweeping or qualifies within \(J\) rounds, and a residual block followed by a full block succeeds within at most \(4K\) more rounds. Thus a new \WORK{} entry occurs by \(t_{\mathrm{first}}+F(t_{\mathrm{first}})+J+4K\); let \(r>t_{\mathrm{first}}\) be the first such round.

Suppose player \(i\) enters \WORK{} through an AA handoff at \(r\). Its current \SEEK{} visit began after \(t_{\mathrm{first}}\) with the rejection flag cleared. The final acceptance run must have length exactly two, or \WORK{} would have started earlier. Hence, in that same visit, \(i\) was rejected at \(r-3\) and accepted at \(r-2,r-1\).

Let \(w\) be the BUS winner at \(r-3\). BUS prefers \(w\) to \(i\), and \(w\) remains present, so it must avoid BUS in both following rounds. A player in \SEEK{} accepted at BUS continues on BUS, including upon sweep qualification; a player accepted while replaying BUS proposes there again, including after expiration; and a player remaining in \SWEEP{} with all its BUS proposals accepted leaves BUS only for isolated rounds. A \WORK{} entry in either intervening round would contradict the choice of \(r\). This two-round absence is impossible, so the first new \WORK{} entry follows a successful sweep and is mature by \Cref{lem:sweep}(ii).

\paragraph{Consecutive outputs and calibration.}
By \Cref{lem:mature}, each player outputs exactly once in each mature execution. No output occurs between \(t_{\mathrm{first}}\) and \(r\), so these are consecutive actual outputs. The last player enters \REPLAY{} after its second output in round \(r+J+2(n-1)\); hence
\begin{equation}
 \begin{aligned}
 t_{\mathrm{cal}}
 &=r+J+2(n-1)\\
 &\le t_{\mathrm{first}}+F(t_{\mathrm{first}})+2J+6K-2\\
 &\le\Psi_F(t_{\mathrm{first}})
 \le\Psi_F^{\circ3}(t_*+1).
 \end{aligned}
 \label{eq:second-recovery-handoff}
\end{equation}
For \(K\ge2\), \(J+1<H\), \(J+6K-2<H\), and \(2J+6K-2<H\); these inequalities and monotonicity justify all three applications of \(\Psi_F\).

Applying Equation~\eqref{eq:common-wait-deadlines} to these two executions gives a common replay length at \(t_{\mathrm{cal}}\), when all players are in correct \REPLAY{}. Taking \(F=F_g\) proves \Cref{lem:known-recovery}.
\end{proof}

\subsection{Proof of Lemma~\ref{lem:known-maintenance}}
\label{sec:proof-maintenance}

After calibration, each maintenance window starts at the root's replay deadline and ends when all players have returned to \REPLAY{} after their next \WORK{} outputs. We show that these windows have bounded length and preserve calibration.

\begin{lemma}[Bounded maintenance under a general replay rule]
\label{lem:maintenance}
Suppose two mature executions in a fixed nonempty market with POSM \(m^*\) give consecutive actual outputs for every player, and the population remains fixed thereafter. After all players have entered \REPLAY{} following the second execution, outside maintenance windows the realized matching equals \(m^*\) and all players have the same replay length.

Each window lasts at most \(U=2K^2+14K\) rounds and contains a mature execution giving every player's next actual output. At the end of each window, all players are in \REPLAY{} with a common replay length.

Let \(L_0\) be the common replay length after all players have entered \REPLAY{} following the second execution, and for \(j\ge1\), let \(L_j\) be the common replay length between the \(j\)-th and \((j+1)\)-st maintenance windows. Then
\[
  L_0\ge H,\qquad L_{j+1}\ge F(L_j)\qquad(j\ge0).
\]
\end{lemma}

\begin{proof}[Proof of \Cref{lem:maintenance,lem:known-maintenance}]
Let \(n\) be the number of players, label them from most to least preferred by BUS, and put \(\delta_i=2(i-1)\) for \(1\le i\le n\) as in Appendix~\ref{sec:proof-mature}. Index the second given mature execution by \(0\); its common replay length satisfies \(L_0\ge H\). We construct the subsequent mature executions inductively. For execution \(j\ge0\), let \(o_j\) be the root's output round at which it sets \(L_j\), \(t_j=o_j+2n-1\) the round when all players have entered \REPLAY{} after its outputs, and \(Q_j=o_j+L_j\) the root's global replay deadline.

At round \(t_j\), the deadlines are \(Q_j+\delta_i\) by Equation~\eqref{eq:common-wait-deadlines}, and \(L_j\ge H>2K\) gives \(Q_j>t_j\). Throughout \([t_j,Q_j)\), all players remain in correct \REPLAY{} and keep the common replay length \(L_j\); every other current replay ends by \(Q_j+2K-2\), possibly earlier after rejection. Once the root sweeps, \Cref{lem:sweep}(i) excludes other \WORK{} entries, sweeps, and outputs. We distinguish the two possible cases for the root's saved arm and write \(c\) for its next \WORK{} entry round.

\emph{If \(m^*(1)\ne\BUS\)}, the root enters a fresh \SEEK{} visit at \(Q_j\) with \(\rho=0\) and its rejection record cleared. Its \(J\) BUS acceptances in \(Q_j,\ldots,Q_j+J-1\) cannot trigger its own AA handoff and prevent every other AA handoff. Any other player replaying BUS is rejected at \(Q_j\), before its own deadline, resetting its counter; players replaying non-BUS arms already have zero counters. Thus no competing sweep or \WORK{} mode begins. Since \(J>2K-2\), all others are in \SEEK{} when the root enters \SWEEP{} at \(Q_j+J\), and the first block succeeds:
\begin{equation}
  c=Q_j+J+2K.
  \label{eq:maintenance-nonbus-entry}
\end{equation}

\emph{If \(m^*(1)=\BUS\)}, all \(L_j\ge H>J\) proposals in \(o_j,\ldots,Q_j-1\), including the output round \(o_j\), are accepted at BUS. The counter is saturated, so expiration followed by qualification starts \SWEEP{} at \(Q_j\). All other replays end by \(Q_j+2K-2\), so the second block, starting at \(Q_j+2K\), succeeds unless the first already did. Hence
\begin{equation}
  c\le Q_j+4K\le Q_j+J+2K.
  \label{eq:maintenance-bus-entry}
\end{equation}

In both cases, no output occurs between \(t_j\) and \(c\), and \(c\le Q_j+J+2K\). By \Cref{lem:sweep}(ii) and \Cref{lem:mature}, the next execution is mature and gives each player's next actual output with the same offsets. The root's next output is at \(o_{j+1}=c+J-1\). Let \(C_j\) be the number of rounds from \(Q_j\) to this output. Then
\begin{equation}
  C_j=o_{j+1}-Q_j,\qquad 0<C_j\le2J+2K-1.
  \label{eq:maintenance-overhead}
\end{equation}
Calibration therefore applies again with \(o_{j+1}-o_j=L_j+C_j\):
\begin{equation}
  L_{j+1}=F(L_j+C_j)\ge H+2(L_j+C_j)>2L_j.
  \label{eq:wait-doubling}
\end{equation}

The last player enters \REPLAY{} one round after its output, so \(t_{j+1}=o_{j+1}+2(n-1)+1\le o_{j+1}+2K-1\), giving
\begin{equation}
  t_{j+1}-Q_j\le C_j+2K-1\le2J+4K-2=2K^2+14K=U.
  \label{eq:maintenance-window-budget}
\end{equation}
The \((j+1)\)-st maintenance window is \([Q_j,t_{j+1})\). Thus it lasts at most \(U\) rounds and costs at most \(U\) mismatch rounds. Since \(L_{j+1}\ge H>2K\), the next deadline \(Q_{j+1}\) follows \(t_{j+1}\). Throughout \([t_{j+1},Q_{j+1})\), all players remain in correct \REPLAY{} with the common replay length \(L_{j+1}\), completing the induction. Monotonicity gives \(L_{j+1}=F(L_j+C_j)\ge F(L_j)\). Applied with \(F=F_g\) to the pair supplied by \Cref{lem:recovery}, these bounds prove \Cref{lem:known-maintenance}.
\end{proof}

\paragraph{Finite-prefix accounting.}
Write \(D_F(T;I)\) for the mismatch cost under replay rule \(F\), and let
\[
 N_F(T)=\#\{m\ge0:F^{\circ m}(H)<T\}.
\]
Start with the pair supplied by \Cref{lem:recovery}, so \(t_0=t_{\mathrm{cal}}\), and use the notation introduced in the proof of \Cref{lem:maintenance}. Since \(L_0\ge H\) and \(L_{j+1}\ge F(L_j)\), monotonicity gives \(L_j\ge F^{\circ j}(H)\). The \((j+1)\)-st maintenance window is \([Q_j,t_{j+1})\) and meets the first \(T\) rounds only if \(Q_j\le T\). Since \(Q_j=o_j+L_j\) and \(o_j\ge1\), it satisfies \(F^{\circ j}(H)\le L_j<T\). At most \(N_F(T)\) windows contribute, each costing at most \(U\), including an unfinished last window. Charging at most one per round before calibration and using \Cref{lem:recovery} gives
\begin{equation}
 D_F(T;I)\le
 \min\{T,\Psi_F^{\circ3}(t_*+1)+U N_F(T)\}.
 \label{eq:general-wait-accounting}
\end{equation}
This also covers prefixes ending before calibration. If the final market is empty, \(D_F(T;I)\le\min\{T,t_*-1\}\), so the same bound holds; at \(T=0\), both sides are zero.

\subsection{Proof of Lemma~\ref{lem:known-scale-rule}}
\label{sec:tradeoff-upper-proof}

\begin{proof}
Fix a feasible function \(g\), as defined in \Cref{sec:known-analysis}, and put
\[
 M_g=\max\left\{1,\sup_{t\ge1}\frac{t}{g(t)}\right\}<\infty.
\]
The asymptotic linear lower bound controls the ratios for all sufficiently large \(t\), and positivity controls the remaining finite set. The constant \(M_g\) is used only in the analysis. For the sequence~\((s_j)\) and function \(F_g\) in Equation~\eqref{eq:budget-waiting-rule}, write
\[
 A_g(x)=\max\bigl(\{g(s_j)-s_j:j\ge0,\ 64s_j\le x\}\cup\{0\}\bigr),
 \qquad F_g(x)=H+2x+A_g(x).
\]
The maximum is non-negative and non-decreasing in \(x\), even though the individual differences need not increase with \(j\). Hence \(F_g\) is non-decreasing and satisfies Equation~\eqref{eq:general-wait-condition}.

Since \(s_{j+1}\ge(1+1/M_g)s_j\), evaluation visits \(O(1+\log(x+1))\) eligible values of \(s_j\), with a constant depending only on \(g\). It calls \(g(s_j)\) only while \(64s_j\le x\), and maintains the maximum of the differences. The assumed polynomial-time computability of \(g\), together with integer additions and comparisons, therefore gives a polynomial-time evaluation of \(F_g\). The protocol \(\Pi_g\) uses Equation~\eqref{eq:general-wait-update} with this function, fixed by \(K,g\) before the instance and horizon; it does not need \(M_g\) as an input.

\emph{Recovery.}
The recovery map is
\[
 \Psi_{F_g}(x)=3x+6H+A_g(x+H).
\]
We show
\begin{equation}
 \Psi_{F_g}^{\circ3}(t+1)\le C_g(t),\qquad
 C_g(t)=
 \begin{cases}
  105H,&1\le t<H,\\
  118(t+g(t)),&t\ge H.
 \end{cases}
 \label{eq:budget-recovery-envelope}
\end{equation}
For an integer \(t\ge H\), choose \(j\ge0\) such that \(s_j\le t<s_{j+1}\). Then
\[
 t+1\le s_{j+1}=s_j+g(s_j)\le t+g(t),\qquad H\le s_{j+1}.
\]
For any \(z\ge1\) with \(z+H<64s_{j+1}\), only indices \(k\le j\) can be activated, so \(A_g(z+H)\le\max_{k\le j}s_{k+1}=s_{j+1}\). Writing \(z_0=t+1\) and \(z_{i+1}=\Psi_{F_g}(z_i)\) gives
\[
 \begin{array}{lll}
 z_0+H\le2s_{j+1},&z_1\le10s_{j+1},\\
 z_1+H\le11s_{j+1},&z_2\le37s_{j+1},\\
 z_2+H\le38s_{j+1},&z_3\le118s_{j+1}\le118(t+g(t)).
 \end{array}
\]
Every input stays below \(64s_{j+1}\). For \(1\le t<H\), \(z_0\le H\), and the three inputs to \(A_g\) are at most \(2H,10H,34H\), below the first threshold \(64H\). Thus all three additional terms vanish and
\[
 z_3=27(t+1)+78H\le105H.
\]
This proves Equation~\eqref{eq:budget-recovery-envelope} without any bound on \(g(t+1)/g(t)\). The function \(C_g\) is non-decreasing, including at \(H\).

\emph{Iteration count.}
Since \(F_g(x)\ge2x\), six iterations from \(s_m\) reach at least \(64s_m\), and
\[
 F_g(64s_m)\ge H+128s_m+g(s_m)-s_m
             \ge s_m+g(s_m)=s_{m+1}.
\]
Monotonicity therefore gives, for every \(m\ge0\),
\[
 F_g^{\circ(7m)}(H)\ge s_m,\qquad
 N_{F_g}(T)\le7N_g(T).
\]
Put \(k=N_g(T)=\#\{m\ge0:s_m<T\}\). Then \(F_g^{\circ(7k)}(H)\ge s_k\ge T\), proving the count bound, including when \(T\) equals an iterate.

Since \(t\le M_g g(t)\) and \(H\le16K^2\), Equation~\eqref{eq:budget-recovery-envelope} gives the first bound in \Cref{lem:known-scale-rule} for every \(t\ge1\), with a constant depending only on \(g\). The iteration comparison gives the second bound with constant \(7\) for every \(T\ge1\); when \(T\le H\), both counts are zero. In particular, both bounds hold for \(1\le t\le T\), as stated in the lemma. When \(g(t)=t\), every difference in \(A_g\) vanishes, so \(F_g(x)=H+2x\) for all \(x\ge1\).
\end{proof}

Combining Equation~\eqref{eq:general-wait-accounting} with the recovery and iteration bounds gives, for every \(T\ge0\),
\begin{equation}
 D_{\Pi_g}(T;I)
 \le\min\{T,118(t_*+g(t_*))+105H+7U N_g(T)\}.
 \label{eq:budget-finite-dynamic}
\end{equation}
For a static instance, \(t_*=1<H\), so the sharper recovery estimate in Equation~\eqref{eq:budget-recovery-envelope} gives
\begin{equation}
 \sup_{I:\,t_*=1}D_{\Pi_g}(T;I)\le105H+7U N_g(T).
 \label{eq:budget-finite-static}
\end{equation}
Here \(H\le16K^2\) and \(U\le9K^2\), so the static bound has constants uniform in \(K,g,I,T\). For \(T>H\), we have \(N_g(T)\ge1\), so the startup term is absorbed by \(O(K^2N_g(T))\), as used in the joint proof of \Cref{thm:known,thm:general-budget-upper} in \Cref{sec:known-analysis}.

\subsection{Proof of Corollary~\ref{cor:flexible-sublinear}}
\label{sec:known-recovery-guarantee}

\begin{proof}
Fix a feasible function \(g\) and \(\eta\in(0,1)\), and first consider a nonempty final market. Starting at \(t_0=t_{\mathrm{cal}}\), use \(o_j,L_j,Q_j,C_j,t_j\) as defined in the proof of \Cref{lem:maintenance}. Equations~\eqref{eq:maintenance-overhead} and~\eqref{eq:wait-doubling} give
\[
 o_{j+1}=o_j+L_j+C_j,\qquad
 L_{j+1}=F_g(L_j+C_j)\ge H+2L_j,\qquad C_j>0.
\]
Thus
\[
 L_j\ge(2^{j+1}-1)H,\qquad
 Q_{j+1}-Q_j=C_j+L_{j+1}\ge(2^{j+2}-1)H.
\]
Every maintenance window \([Q_j,t_{j+1})\) costs at most \(U\), and \(Q_0>t_{\mathrm{cal}}\). Put
\[
 m_\eta=\min\{m\in\mathbb Z_{\ge0}:(2^{m+2}-1)H\ge U/\eta\},\qquad
 Q_{K,\eta}=(m_\eta+1)U.
\]
Charge the first \(m_\eta\) windows separately. Later starts are at least \(U/\eta\) apart, so at most \(1+\eta h/U\) occur in any \(h\) consecutive rounds. In particular,
\begin{equation}
 D_{\Pi_g}(t_{\mathrm{cal}}+h;I)-D_{\Pi_g}(t_{\mathrm{cal}};I)
 \le Q_{K,\eta}+\eta h \qquad(h\ge0).
 \label{eq:budget-tail-discount}
\end{equation}
This difference counts rounds \(t_{\mathrm{cal}}+1,\ldots,t_{\mathrm{cal}}+h\). Because \(U/H\le1\), \(Q_{K,\eta}=O(K^2[1+\log(1/\eta)])\), independently of \(g\).

\emph{The post-event window.}
The first \(t_{\mathrm{cal}}-t_*\) rounds of the event-excluding window cost at most \(t_{\mathrm{cal}}-t_*\le C_g(t_*)\). Splitting a longer window at \(t_{\mathrm{cal}}\) and using Equation~\eqref{eq:budget-tail-discount} gives, for every \(h\ge1\),
\begin{equation}
 L_{\Pi_g}(h;I)
 \le \min\{h,C_g(t_*)+Q_{K,\eta}+\eta h\}.
 \label{eq:budget-finite-recovery}
\end{equation}
Empty final markets have zero post-event cost.

\emph{Recovery coefficients.}
For \(0<\varepsilon<1\), take \(\eta=\varepsilon/2\) and
\[
 c_\varepsilon=
 \frac{2(118(1+M_g)+105H+Q_{K,\varepsilon/2})}{\varepsilon}>0.
\]
Since \(g(t_*)\ge1\) and \(C_g(t_*)\le118(1+M_g)g(t_*)+105H\), the condition \(h\ge c_\varepsilon g(t_*)\) implies \(L_{\Pi_g}(h;I)\le\varepsilon h\). Thus the same \(\Pi_g\) has convergence speed \(g\); its rule has no accuracy parameter.
\end{proof}

%% file: appendices/B_tradeoffs.tex
\section{Lower Bounds and Optimality}
\label{app:tradeoff-proofs}
This appendix proves \Cref{thm:recovery-maintenance} and the iteration-count results in \Cref{sec:tradeoff-insights}, and gives the computational details for the exponential function. We use the known-preference setting, in which each arm accepts its most-preferred proposer in each round. The lower bounds allow private state, private randomness, adaptive proposals, and accepted utilities.

\subsection{Proof of Theorem~\ref{thm:recovery-maintenance}}
\label{sec:tradeoff-lower-proof}

\begin{proof}[Proof of \Cref{thm:recovery-maintenance}]
Fix \(K,g,\Pi\) as in the theorem. Recurrence \(s_0=H\), \(s_{j+1}=s_j+g(s_j)\), and count \(N_g(T)=\#\{j\ge0:s_j<T\}\) are well-defined for every nondecreasing positive integer value function \(g\).

\paragraph{Static bound.}
In the static matching instance \(I_0\), players \(p_1,p_2\) enter in round \(1\) and remain forever, both ranking \(0\succ1\succ\cdots\succ K-1\). Arm \(0\) prefers \(p_1\) to \(p_2\); every other arm prefers \(p_2\) to \(p_1\). 
Choose deterministic utilities consistent with these rankings.
The unique stable matching is \(\{(p_1,0),(p_2,1)\}\). For each \(\tau\ge2\), let dynamic instance \(I_\tau\) remove \(p_1\) at the start of round \(\tau\); thereafter the POSM assigns \(p_2\) to \(0\). 
The construction of \(I_0\) and \(I_\tau\) does not depend on \(\Pi\), \(g\), or \(T\).

Write
\[
 m(T)=\mathbb E D_\Pi(T;I_0),\qquad
 L_\Pi(h;I_\tau)=
 \mathbb E[D_\Pi(\tau+h;I_\tau)-D_\Pi(\tau;I_\tau)].
\]

Couple the executions of \(\Pi\) on \(I_0\) and \(I_\tau\) using the same private randomness for each player in both executions. They are identical through round \(\tau-1\). The unannounced departure does not change \(p_2\)’s feedback from any arm other than \(0\), since each such arm prefers \(p_2\) to \(p_1\). Thus, by induction, \(p_2\)’s proposals coincide until its first proposal to arm \(0\) at or after \(\tau\). Consequently, for every \(h\ge1\), the probability of no such proposal in rounds \(\tau,\ldots,\tau+h\) is the same in both executions.

In the static execution, every proposal by \(p_2\) to \(0\) causes mismatch. The expected number of such proposals in \(\{\tau,\ldots,\tau+h\}\) is at most \(m(\tau+h)-m(\tau-1)\). With probability at least \(\max\{0,1-m(\tau+h)+m(\tau-1)\}\), there is no such proposal. On that event all \(h\) rounds \(\tau+1,\ldots,\tau+h\) are mismatched in the departure execution. Therefore
\begin{equation}
 L_\Pi(h;I_\tau)\ge
 h\max\{0,1-m(\tau+h)+m(\tau-1)\},
 \qquad \tau\ge2,\ h\ge1.
 \label{eq:tradeoff-window-coupling}
\end{equation}
Including \(\tau\) in the static charge accounts for detection in the event round, which the recovery cost itself excludes.

In particular, Equation~\eqref{eq:tradeoff-window-coupling} implies
\begin{equation}
 L_\Pi(h;I_\tau)\le h/2
 \quad\Longrightarrow\quad
 m(\tau+h)-m(\tau-1)\ge1/2.
 \label{eq:tradeoff-single-window-charge}
\end{equation}
Let \(c_{1/2}\) be the constant in \Cref{def:sublinear-mismatch}
for \(\varepsilon=1/2\), and set
\(q=\max\{1,\lceil c_{1/2}\rceil\}\). Define
\begin{equation}
     x_0=2,\qquad h_j=qg(x_j),\qquad x_{j+1}=x_j+h_j+1.
     \label{eq:xj}
\end{equation}
The recovery window of \(I_{x_j}\) is \(\{x_j+1,\ldots,x_{j+1}-1\}\), of length \(h_j\). Since \(h_j\ge c_{1/2}g(x_j)\), the recovery condition bounds its expected cost by \(h_j/2\). Equation~\eqref{eq:tradeoff-single-window-charge} therefore forces at least \(1/2\) expected static mismatch in the augmented interval \(\{x_j,\ldots,x_{j+1}-1\}\). These intervals are disjoint. If
\[
 n(T)=\#\{j:x_{j+1}\le T+1\},
\]
then \(m(T)\ge n(T)/2\), since all these intervals lie among the first \(T\) rounds.

Let \(\Phi(x)=x+g(x)\), and note that \(x_0=2\le H=s_0\). Since \(g\) is nondecreasing and positive,
\[
 \Phi^{\circ(q+1)}(x)\ge x+(q+1)g(x)
 \ge x+qg(x)+1.
\]
Thus \(x_j\le s_{(q+1)j}\). Since \(x_{n(T)+1}>T+1\), also \(s_{(q+1)(n(T)+1)}>T\), giving \(N_g(T)\le(q+1)(n(T)+1)\). We obtain the following bound
\begin{equation}
 \mathbb E D_\Pi(T;I_0)\ge
 \frac12\max\left\{0,\frac{N_g(T)}{q+1}-1\right\}
 \qquad(T\ge0).
 \label{eq:tradeoff-finite-lower}
\end{equation}
Every iterate is finite, and the sequence is unbounded, so \(N_g(T)\to\infty\). For fixed \(K\), the coefficient \(q\) depends only on the chosen protocol and function \(g\). Thus Equation~\eqref{eq:tradeoff-finite-lower} proves the static bound on the same instance \(I_0\) for every protocol.
\paragraph{Dynamic bound.}
Fix \(t_*\ge2\). We construct a distribution over dynamic matching instances with last population-event time \(t_*\). Player \(b\) enters in round \(1\) and remains forever. We also introduce distinct potential players \(c_1,\ldots,c_{t_*}\). All players rank \(0\succ1\succ\cdots\succ K-1\). Arm \(0\) ranks every \(c_r\) above \(b\), while every other arm ranks \(b\) above every \(c_r\). Complete the arm rankings arbitrarily and choose consistent deterministic utilities. When \(b\) is alone, its POSM assignment is \(0\); with one additional player \(c_r\), the unique stable matching is \(\{(c_r,0),(b,1)\}\).

Before execution, sample independent bits \(\xi_1,\ldots,\xi_{t_*-1}\), each equal to \(0\) or \(1\) with probability \(1/2\), independently of \(\Pi\)’s private randomness. For each \(r<t_*\), if \(\xi_r=1\), player \(c_r\) is active only in round \(r\); otherwise, it never enters. Player \(c_{t_*}\) enters in round \(t_*\) and remains forever. At most two players are active in any round, and no player reenters. Each player is initialized according to \(\Pi\) upon entry; population events do not reinitialize \(b\).

For each \(r<t_*\), player \(b\) has no observation of \(\xi_r\) before proposing. Conditional on its local history, \(\xi_r\) remains fair and independent of its proposal. Let \(p_0,p_1\) be its conditional probabilities of proposing to arms \(0,1\). Its POSM arm is \(0\) when \(\xi_r=0\) and \(1\) when \(\xi_r=1\), so the conditional probability of making the required proposal is
\[
\tfrac12p_0+\tfrac12p_1\le\tfrac12.
\]
Thus the probability of mismatch is at least \(1/2\) in each round.

Summing over \(r<t_*\), the prefix cost averaged over both the bits and \(\Pi\)’s randomness is at least \((t_*-1)/2\). At least one fixed bit string has expected cost under \(\Pi\) at least this average. Let \(I'\) be the corresponding instance and write \(m(T)=\mathbb E D_\Pi(T;I')\). Then
\begin{equation}
 m(T)\ge m(t_*-1)\ge\frac{t_*-1}{2}\ge\frac{t_*}{4}
 \qquad(T\ge t_*).
 \label{eq:tradeoff-population-prefix}
\end{equation}
The choice of \(I'\) may depend on \(\Pi\) and \(t_*\), but not on \(T\) or the realized private randomness.

For each \(\tau\ge t_*+1\), obtain a comparison instance \(I'_\tau\) from \(I'\) by removing player \(c_{t_*}\) at the start of round \(\tau\), leaving the utilities and all other population events unchanged. Its last population event is at \(\tau\). Couple the executions as in the static case. Since every arm other than \(0\) accepts \(b\) with the same utility in both executions, \(b\)’s proposals coincide until its first proposal to arm \(0\) at or after \(\tau\).

Every such proposal causes mismatch in \(I'\); if none occurs in rounds \(\tau,\ldots,\tau+h\), all \(h\) rounds \(\tau+1,\ldots,\tau+h\) are mismatched in \(I'_\tau\). Including round \(\tau\) in the cost on \(I'\), the same argument as in Equation~\eqref{eq:tradeoff-window-coupling} gives
\begin{equation}
 L_\Pi(h;I'_\tau)\ge
 h\max\{0,1-m(\tau+h)+m(\tau-1)\}
 \qquad(\tau\ge t_*+1,\ h\ge1).
 \label{eq:tradeoff-dynamic-coupling}
\end{equation}

Let \(j_\star=\min\{j\ge0:s_j\ge t_*+1\}\). 
Repeat the construction in Equation~\eqref{eq:xj} with the same \(q\), replacing \(x_0=2\) by \(x_0=s_{j_\star}\), and apply the recovery condition to each comparison instance \(I'_{x_j}\). 
As in the static proof, each augmented interval incurs expected mismatch cost at least \(1/2\) on \(I'\). Hence at most \(2m(T)\) such intervals are completed by round \(T\), and the same iteration comparison, with initial offset \(j_\star\), gives
\[
N_g(T)\le j_\star+(q+1)(2m(T)+1).
\]
Since \(s_j\ge j+2\), we have \(j_\star\le t_*-1\). Combining this with the prefix bounds \(t_*\le4m(T)\) and \(1\le2m(T)\) yields
\begin{equation}
 \mathbb E D_\Pi(T;I')\ge
 \frac{t_*+N_g(T)}{12(q+1)}
 \qquad(T\ge t_*\ge2).
 \label{eq:tradeoff-cumulative-finite}
\end{equation}
The coefficient is independent of \(t_*\) and \(T\), and the selected instance works for all \(T\ge t_*\), proving the dynamic bound.
\end{proof}

No protocol can have convergence speed \(g\) for bounded \(g\). Since \(N_g(T)=\Omega(T)\), the static lower bound gives \(\mathbb E D_\Pi(T;I_0)=\Omega(T)\), whereas \Cref{def:sublinear-mismatch} requires this expectation to be \(o(T)\).


\subsection{Proofs for Section~\ref{sec:tradeoff-insights}}
\label{sec:tradeoff-growth-proof}

Fix \(K\ge2\). We first prove the general growth relations in Equation~\eqref{eq:budget-growth-regimes}. We then verify the feasibility of the functions in \Cref{tab:representative-budgets} and derive their iteration counts.

\paragraph{General growth relations.}
Fix a feasible function \(g\).
Since \(g(t)=\Omega(t)\) and \(g(t)>0\) for every \(t\ge1\), there is a constant \(\delta_g>0\) such that \(g(t)\ge\delta_g t\) for all \(t\ge1\).
Consequently, \(N_g(T)\to\infty\) and
\(N_g(T)=O(\log T)\) as \(T\to\infty\).

\begin{proof}[Proof of Equation~\eqref{eq:budget-growth-regimes}]
Put \(y_j=\log s_j\) and \(z_j=\log y_j\). The recurrence \(s_{j+1}=s_j+g(s_j)\) gives
\begin{equation}
 y_{j+1}-y_j=\log\!\left(1+\frac{g(s_j)}{s_j}\right).
 \label{eq:budget-log-increment}
\end{equation}
For \(T>H\), let \(n=N_g(T)\).
By the definition of \(N_g(T)\), we have \(s_{n-1}<T\le s_n\).
Taking logarithms gives
\[
 y_{n-1}<\log T\le y_n,
 \qquad
 z_{n-1}<\log\log T\le z_n.
\]

Suppose \(\log g(t)/\log t\to\alpha\in[1,\infty]\).
Since
\[
\max\{s_j,g(s_j)\} \le s_{j+1} \le 2\max\{s_j,g(s_j)\},
\]
we have \(y_{j+1}/y_j\to\alpha\).
Thus \(z_{j+1}-z_j\to\log\alpha\), and Cesaro averaging gives $z_j / j\longrightarrow\log\alpha$,
where \(\log\infty=\infty\).
We now consider the three cases.

\emph{Case \(\alpha=1\).}
The upper horizon bound gives
\[
 \frac{\log\log T}{n}
 \le \frac{z_n}{n}
 \longrightarrow 0.
\]
Hence \(N_g(T)=\omega(\log\log T)\).

\emph{Case \(1<\alpha<\infty\).}
Both \(z_{n-1}/n\) and \(z_n/n\) tend to \(\log\alpha\).
The horizon bounds therefore give
\[
 N_g(T)\sim\frac{\log\log T}{\log\alpha}.
\]
In particular, \(N_g(T)=\Theta(\log\log T)\).

\emph{Case \(\alpha=\infty\).}
The lower horizon bound gives
\[
 \frac{\log\log T}{n}
 > \frac{z_{n-1}}{n}
 \longrightarrow\infty.
\]
Hence \(N_g(T)=o(\log\log T)\).
\end{proof}


\paragraph{Representative functions.}
The expressions in \Cref{tab:representative-budgets} omit rounding to integers for readability. 
We use the precise integer-valued definitions below, verify their feasibility, and derive their iteration counts.

\begin{proof}[Proof of \Cref{tab:representative-budgets}]
The parameters \(\alpha>1\) and \(\delta>0\) are fixed rational numbers.
For a positive integer input \(t\), let \(b=1+\lfloor\log t\rfloor\) be its bit length.
As above, write \(y_j=\log s_j\) and \(z_j=\log y_j\). 

\textit{Linear function.}
The function \(g(t)=t\) is nondecreasing, positive integer-valued, and polynomial-time computable, so it is feasible. The recurrence gives \(s_j=H2^j\), and hence $N_g(T)=\Theta(\log T)$.

\textit{Logarithmically enlarged functions.}
We use \(g(t)=t(1+\lfloor\log t\rfloor)=tb\). This function is nondecreasing and satisfies \(g(t)\ge t\). It is polynomial-time computable in \(b\), since \(b\) is obtained directly from the input representation and \(tb\) is computed by integer multiplication. Thus \(g\) is feasible.

The logarithmic recurrence gives 
\[y_{j+1}-y_j=\log(2+\lfloor y_j\rfloor)=\Theta(\log y_j)\]
for sufficiently large \(j\). Consider intervals \([2^k,2^{k+1})\) in the \(y_j\)-coordinate. If \(j\) is the first index with \(y_j\ge2^k\), then, for sufficiently large \(k\), we have 
\[0\le y_j-2^k\le\log(2+2^k)=O(k).\]
Applying this bound at both endpoints gives a traversed distance of \(2^k+O(k)\). Since the increments inside the interval are \(\Theta(k)\), traversal takes \(\Theta(2^k/k)\) steps.

When \(\log T\in[2^k,2^{k+1})\), the preceding complete interval \([2^{k-1},2^k)\) provides \(\Omega(2^k/k)\) steps. Summing the completed intervals and bounding the last partial interval by a full one gives \(O(2^k/k)\) steps. Therefore, \(N_g(T)=\Theta(\log T/\log\log T)\).

\textit{Polynomial functions.}
For fixed \(\alpha=p/q>1\), with positive integers \(p,q\),
\[
 \lceil t^{p/q}\rceil
 =\min\{m\in\mathbb N_{\ge1}:m^q\ge t^p\}.
\]
Integer binary search computes this value, including equality cases, in polynomial time in \(b\): the searched integers have \(O(b)\) bits, and comparisons use exact integer powers. This function is also nondecreasing and satisfies \(g(t)=\Omega(t)\).

For \(g(t)=\lceil t^\alpha\rceil\), \(\alpha>1\), the general classification gives \(N_g(T)\sim\log\log T/\log\alpha\).

\textit{The functions \(g_\delta\).}
Fix positive integers \(p,q\), put \(\delta=p/q\), and define
\begin{equation}
 g_\delta(t)=t\,2^{\lfloor(\lfloor\log t\rfloor)^\delta\rfloor},
 \qquad t\ge1.
 \label{eq:computable-delta-scale}
\end{equation}
Let \(b\) be the input bit length and put \(n=b-1=\lfloor\log t\rfloor\). The exponent is the integer
\[
 k=\max\{m\in\mathbb Z_{\ge0}:m^q\le n^p\}
   =\lfloor n^\delta\rfloor.
\]
It can be computed exactly by integer binary search on \([0,n^p+1)\), including equality cases. For fixed \(p,q\), the endpoints and powers used in the comparisons have \(O(\log(b+1))\) bits, and the search uses \(O(\log(b+1))\) comparisons. The output is obtained by appending \(k\) zero bits to the input, and its length satisfies
\[
 b+k\le b+(b-1)^\delta=O(b+b^\delta).
\]
Thus reading the input, computing the exponent, and writing the output all take polynomial time in \(b\). When \(t=1\), we have \(n=0\), and the search returns \(k=0\), giving \(g_\delta(1)=1\). As \(t\) increases, both \(n\) and \(k\) are nondecreasing, including at bit-length boundaries. Hence \(g_\delta\) is nondecreasing and satisfies \(g_\delta(t)\ge t\).

For \(g=g_\delta\) in Equation~\eqref{eq:computable-delta-scale}, put \(k_j=\lfloor(\lfloor y_j\rfloor)^\delta\rfloor\). For real \(y\ge2\), the mean value theorem and \(\lfloor y\rfloor\ge y/2\) give
\[
 0\le y^\delta-(\lfloor y\rfloor)^\delta
 \le C y^{\delta-1}.
\]
Rounding the power down changes it by less than one. Also, \(0<\log(1+2^{-k})\le1\) for \(k\ge0\). Therefore, the actual integer orbit satisfies
\begin{equation}
 y_{j+1}-y_j
 =\log\!\left(1+2^{k_j}\right)
 =y_j^\delta+O(y_j^{\delta-1}+1)
 \sim y_j^\delta.
 \label{eq:double-floor-increment}
\end{equation}
Here \(g_\delta(t)\ge t\) implies \(y_j\ge\log H+j\), so the relative error \(O(y_j^{-1}+y_j^{-\delta})\) tends to zero.

\emph{Case \(0<\delta<1\).}
The increment estimate gives \(y_{j+1}/y_j\to1\). By the mean value theorem, for some \(\xi_j\in[y_j,y_{j+1}]\),
\[
 y_{j+1}^{1-\delta}-y_j^{1-\delta}
 =(1-\delta)\xi_j^{-\delta}(y_{j+1}-y_j)
 \longrightarrow1-\delta.
\]
Summing these increments gives \(y_j^{1-\delta}\sim(1-\delta)j\). For \(n=N_g(T)\), the horizon bounds \(y_{n-1}<\log T\le y_n\) therefore yield \(N_g(T)=\Theta((\log T)^{1-\delta})\).

\emph{Case \(\delta>1\).}
The increment estimate gives \(y_{j+1}/y_j^\delta\to1\). Consequently, \(z_{j+1}=\delta z_j+\varepsilon_j\), where \(\varepsilon_j=\log(y_{j+1}/y_j^\delta)\to0\). Choose \(j_0\) so that \(|\varepsilon_j|\le1\) for every \(j\ge j_0\) and \(z_{j_0}>1/(\delta-1)\). Expanding the recurrence and bounding the geometric sum of errors gives, for every \(m\ge0\),
\[
 \delta^m\!\left(z_{j_0}-\frac1{\delta-1}\right)
 \le z_{j_0+m}
 \le
 \delta^m\!\left(z_{j_0}+\frac1{\delta-1}\right).
\]
The lower coefficient is positive, so \(z_j=\Theta(\delta^j)\) and \(\log z_j=j\log\delta+O(1)\). For \(n=N_g(T)\), taking one more logarithm in the horizon bounds gives \(\log z_{n-1}<\log\log\log T\le\log z_n\). It follows that
\[N_g(T)=\log\log\log T/\log\delta+O(1),\]
and hence \(N_g(T)=\Theta(\log\log\log T)\).

At \(\delta=1\), we have \(t^2/2<g_1(t)\le t^2\), so the general growth relation with \(\alpha=2\) applies.

The five functions in \Cref{tab:representative-budgets} are feasible. For fixed \(K\) and \(g\), their optimal static maintenance orders follow from Corollary~\ref{cor:flexible-sublinear} and \Cref{thm:general-budget-upper,thm:recovery-maintenance}. 
The constants and thresholds in these iteration-count estimates may depend on \(K\) and the fixed rational shape parameters.
\end{proof}

\begin{lemma}[Comparison of iteration counts]
\label{lem:recovery-scale-comparison}
Let \(\widetilde g:\mathbb N_{\ge1}\to\mathbb N_{\ge1}\) be nondecreasing. If \(g(t)\le\widetilde g(t)\le c\,g(t)\) for every \(t\ge1\), then, with \(d=\max\{1,\lceil c\rceil\}\),
\[
 N_{\widetilde g}(T)\le N_g(T)\le dN_{\widetilde g}(T)
 \qquad(T\ge0).
\]
In particular, \(g(t)=\Theta(\widetilde g(t))\) implies \(N_g(T)=\Theta(N_{\widetilde g}(T))\).
\end{lemma}

\begin{proof}
Put 
\[
 Q_g(x)=x+g(x),\qquad
 d=\max\{1,\lceil c\rceil\}.
\]
Monotonicity gives \(Q_g^{\circ d}(x)\ge x+d g(x)\ge x+\widetilde g(x)\), and hence
\[
 Q_g(x)\le Q_{\widetilde g}(x)\le Q_g^{\circ d}(x).
\]
Iterating from the common initial value \(H\) gives the claimed inequalities. If two functions are within constant factors only for sufficiently large \(t\), positivity lets us enlarge the comparison constant to cover the finitely many initial values. Applying the preceding inequalities to their pointwise minimum and maximum proves the final assertion.
\end{proof}

\subsection{The Exponential Function}
\label{sec:tradeoff-exponential}

\paragraph{Iteration count.}
For \(g(t)=2^t\), let \(E(x)=2^x\). Since \(E(x)\le x+2^x\le E^{\circ2}(x)\) for \(x\ge2\), monotone iteration from \(s_0=H\) gives \(E^{\circ j}(H)\le s_j\le E^{\circ(2j)}(H)\). With \(\log^*T=\min\{m\in\mathbb Z_{\ge0}:\log^{\circ m}T\le1\}\), the horizon sandwich gives \((\log^*T-\log^*H)/2\le N_g(T)\le\log^*T-\log^*H+1\) for \(T>H\). Hence \(N_g(T)=\Theta(\log^*T)\).

\paragraph{Computation in local time.}
The binary representation of \(2^t\) has \(t+1\) bits, so \(g(t)=2^t\) is not polynomial-time computable given \(t\) in binary. 
Consider the following variant of $\Pi_g$.
It doesn't compute $L$ or $q$ explicitly. 
Rather, it saves two consecutive output times from \WORK{}, $a$ and $\ell$, instead of only $\ell$.
For every local clock $\hat{a}$ after $a$, it checks whether $\hat{a}-a < L$ (that is, whether \REPLAY{} hasn't finished yet) by checking whether there exists $j\geq 0$ such that $64 s_j\leq x$ and 
$\hat{b}\triangleq \hat{a}-a-(H+2x) < 2^{s_j}-s_j$, with $x=a-\ell$.
To check the latter, it suffices to consider $s_j\leq 2\log \hat{b}$, as $2^{2\log\hat{b}}- 2\log\hat{b} = \hat{b}^2 - 2\log\hat{b}>\hat{b}$ for $\hat{b}$ large enough.
For such $s_j$'s, $2^{s_j}$ is polynomial-time computable given $\hat{b}$ in binary.
Thus the player can determine whether \REPLAY{} has finished or not in polynomial time given the player's local clock $\hat{a}$ in binary.



%% file: appendices/C_algorithm2.tex
\section{The SC-PC Protocol and Its Analysis}
\label{app:learning-proofs}
Appendix~\ref{sec:rcpc:rules} gives the complete \textsc{SC-PC} protocol, starting with the main protocol for the player side, Algorithm~\ref{alg:rcpc-player}, then the protocol for the arm side, \Cref{alg:rcpc-arm}. The subroutines used by Algorithm~\ref{alg:rcpc-player} are then introduced in Algorithms~\ref{alg:rcpc-state-update},~\ref{alg:rcpc-collection},~\ref{alg:rcpc-work-action}, and~\ref{alg:rcpc-work-feedback}.
\subsection{The \textsc{SC-PC} Protocol}
\label{sec:rcpc:rules}
\label{sec:rcpc:algorithms}
\label{sec:learning-algorithm}
\label{sec:learning-maintenance}
\input{sections/learning/player-algorithm}

\subsection{Collection and Recovery}
\label{sec:rcpc:control}
\label{sec:rcpc:preliminaries}

Fix a nonempty population of $n\le K$ active players for each collection or handoff interval below. We use the $\BUS$, $\PARK$, \SEEK{}, \WORK{}, and \REPLAY{} notation of Section~\ref{sec:known}. A player collects while in \SEEK{} or \phase{Root}. A \emph{root} is a player in \phase{Root}. The \emph{highest} member of a set comes first among its members in the private handle order at $\BUS$. All event times below are global rounds unless identified as local clocks. The arguments apply to every history reachable from the prescribed initialization, including histories with statistical errors.

\subsubsection{Scale Envelope and Recovery Constants}
Cold recovery starts from the local states inherited at a given round. We first bound how long a player can remain in an inherited \REPLAY{} visit.

\begin{lemma}[Scale envelope under the prescribed initialization]
\label{lem:rcpc:envelope}
Let $a_0<a_1<\cdots$ be one player's local clock readings in the final rounds of its successive \WORK{} invocations, and let $P_m$ be its scale after the update at $a_m$. Then
\begin{equation}
 H_L\le P_m\le H_L+\left\lfloor\frac{a_m-a_0}{2}\right\rfloor.
\label{eq:rcpc:envelope}
\end{equation}
If the player finishes \WORK{} in global round $O$, the replay deadline it sets in that round corresponds to a global round at most $3O/2+H_L$. Later rejections can only shorten the visit, and the player cannot enter fresh \SEEK{} before local time $\ell+H_L$.
\end{lemma}
\begin{proof}
Initially $P_0=H_L$. Set $X_m=P_m-H_L$ and $\delta_m=a_m-a_{m-1}$. Equation~\eqref{eq:rcpc:output} gives
\[
 X_m=\max\left\{X_{m-1},\left\lfloor\frac{3X_{m-1}+2\delta_m}{4}\right\rfloor\right\},
 \qquad
 0\le X_m-X_{m-1}
 \le\left\lfloor\frac{\delta_m}2\right\rfloor.
\]
Summing proves Equation~\eqref{eq:rcpc:envelope} for all outputs, including short ones. Differences between local times equal differences between global rounds. Thus an output in round $O$, with first output in round $O_0\ge0$, sets a deadline of at most
\[
 O+H_L+\left\lfloor\frac{O-O_0}2\right\rfloor\le\frac32O+H_L.
\]
Replay feedback occurs only at $a<q$, hence $a+1\le q$. Since $q\ge\ell+H_L$ at entry, the update Equation~\eqref{eq:rcpc:rejection} preserves this lower bound and never increases $q$.
\end{proof}

\paragraph{Constants for collection and recovery.}
For the collection and recovery bounds, define
\begin{equation}\begin{aligned}
 L_{s,L}&=2K(K-1)(G_L+1),\\
 C_0&=2\tau_L+2E_L+3J_L+2\chi_L+G_L+R_{c,L}+L_{s,L}+K+32\\
    &=39K^3-24K^2+156K-26,\\
 V_B&=\tau_L+KG_L+2=5K^3-2K^2+28K-7,\\
 C_R&=\frac52C_0+H_L+\kappa_L+16=\frac52C_0+8K+12.
\end{aligned}\label{eq:rcpc:constants}
\end{equation}
Here $L_{s,L}$ bounds the scan duration, $C_0$ bounds the time to output while the player retains highest priority in the collection set, $V_B$ bounds recall of players replaying $\BUS$, and $C_R$ is the additive term in the cold recovery bound. Each is $O(K^3)$.

\subsubsection{Collection and Mature Handoff}

\paragraph{Mature episodes and collection sets.}
Suppose a player enters \WORK{} from \phase{Root} in round $c_0$. Here $c_0$ indexes a global round, distinct from the local clearing register $c$. The entry is \emph{mature} if, after the round-start transitions and before proposals, every other active player is in armed \SEEK{}, $\BUS$ uses control service, every other arm has started and was empty in the preceding two rounds, and no earlier \WORK{} or \REPLAY{} visit remains. The \emph{mature episode} runs from $c_0$ through the first round $S$ in which all players are in \REPLAY{}. Lemma~\ref{lem:rcpc:handoff} shows that each player executes \WORK{} exactly once in this interval. Maturity is a condition used in the analysis.

For the analysis starting at round $v$, the \emph{collection set} initially contains every active player except those replaying an arm other than $\BUS$. Thus initial members include players in \WORK{} or replaying $\BUS$, as well as players collecting in \SEEK{} or \phase{Root}. Each excluded player joins on leaving its current \REPLAY{} visit. A member stays in the set across subsequent mode changes, so the set contains every proposer to $\BUS$. A \emph{higher-priority join} occurs when a previously excluded player joins with higher priority at $\BUS$ than every current member.

\paragraph{Control service and the work handoff.}
\begin{lemma}[Control service and the highest member of the collection set]
\label{lem:rcpc:control-service}
In a fixed population under control service, an arm accepts a player proposing to it every round within $KG_L$ rounds, and one proposing in every round of a fixed parity within $2KG_L$ rounds. A residual business lease adds at most $\tau_L$ rounds if no two consecutive empty rounds renew it.

Suppose $\BUS$ uses control service throughout an interval, $z$ remains the highest-priority member of a collection set containing every proposer to $\BUS$, and $z$ proposes to $\BUS$ at least once in each pair of consecutive rounds in the interval. Then $\BUS$ cannot reject $K$ successive proposals from $z$, counting only the rounds in which $z$ proposes to $\BUS$. Nor can $\BUS$ accept the same lower-priority player in three consecutive global rounds of the interval.
\end{lemma}

\begin{proof}
Until a player's first acceptance at the arm, its timestamp there stays fixed. Any player accepted ahead of it on a fair round receives a newer timestamp and cannot precede it on another fair round. Since $n\le K$, at most $K$ fair proposals suffice. For a player proposing every round, these occur every $G_L$ rounds. For a player proposing in every round of a fixed parity, every second fair round has that parity because $G_L$ is odd, so these proposals occur every $2G_L$ rounds. A lease that is never renewed expires within $\tau_L$ rounds.

On an ordinary control round $\BUS$ accepts $z$ whenever $z$ proposes. Consecutive proposals by $z$ are at most two rounds apart, whereas successive fair rounds are $G_L>2$ rounds apart. Hence two successive proposals by $z$ cannot both be rejected. If a lower-priority player is accepted in round $t$, consider $z$'s first proposal in $t+1$ or $t+2$. On an ordinary round $z$ wins. On a fair round $z$'s last acceptance predates $t$, so the arm cannot choose the lower-priority player ahead of $z$. The same argument applies on the first proposal of a higher-priority player joining from non-$\BUS$ replay if the lower player's acceptance occurred during collection.
\end{proof}

\begin{lemma}[Mature handoff and service intervals]
\label{lem:rcpc:handoff}
Let $c_0$ be a mature entry in a fixed population of $n$ players. For each player $i$, write $c_i$ for its global \WORK{} entry round and $O_i=c_i+J_L-1$ for its final \WORK{} round, when it saves the arm it will replay. The initiating root enters at $c_0=\min_i c_i$, and $c_i\in[c_0,c_0+\kappa_L]$ for all $i$. The first common replay round is $S=\max_i O_i+1$. In $[c_0,S]$ each player enters \WORK{} and outputs exactly once. Every arm uses business service from the first nonparking \WORK{} round $c_0+d_L$ through $S+\chi_L$.

The common interval
\begin{equation}
 I_0=[c_0+\kappa_L+h_{D,L},\,c_0+f_L)\cap\mathbb Z
\label{eq:rcpc:common-window}
\end{equation}
contains $W_L$ rounds. Throughout this interval every player's \WORK{} index is in its deletion window and certification is still allowed. Throughout the rejection run for any deletion, including one outside $I_0$, every player either proposes to its held arm or makes an exploratory proposal satisfying the cooldown.
\end{lemma}

\begin{proof}
\noindent\emph{Bound the entry span.} The initiating root's final commitment acceptance leaves every armed follower with no positive $\BUS$ acceptance suffix. No entrant leaves $\PARK$ before $c_0+d_L=c_0+\kappa_L+2$. Until then, three consecutive ordinary acceptances send the highest-priority remaining follower into \WORK{} in the next round. A fair round that rejects this follower interrupts a run of at most two acceptances and costs at most three rounds. A sole remaining follower is accepted even on a fair round.

Between interruptions, $G_L-1=2K+2$ ordinary rounds suffice for at least $\lfloor(2K+2)/3\rfloor$ followers to qualify for \WORK{} if that many remain. Two such gaps suffice for $2\lfloor(2K+2)/3\rfloor\geq K-2$ followers, leaving at most one and ruling out a third interruption. For $K=2$, the sole follower needs at most three rounds. For $3\leq K\leq5$, one gap already suffices for $K-2$ followers, so at most one interruption gives $3(K-1)+3\leq4K-3$ rounds. For $K\geq6$, the bound is $3(K-1)+6\leq4K-3=\kappa_L$.

\noindent\emph{Maintain business service and identify the common window.} Write $\delta=\max_i c_i-c_0\leq\kappa_L$, so $S=c_0+\delta+J_L$. Since $d_L=\kappa_L+2$, all players park in rounds $c_0+\kappa_L$ and $c_0+\kappa_L+1$, renewing the lease at the empty $\BUS$. Every other arm was empty in the two rounds before $c_0$. Even this earlier renewal covers all rounds through $S+\chi_L$, since $J_L+\kappa_L+\chi_L+1=J_L+2\chi_L<\tau_L$. For $t\in I_0$ and every $j$, the local \WORK{} index satisfies $h_{D,L}\le t-c_j<f_L$. The length of $I_0$ is $f_L-\kappa_L-h_{D,L}=W_L$.

\noindent\emph{Check deletion and certification endpoints.} A deletion by player $i$ in round $t$ requires $K$ rejections starting no earlier than $c_i+h_{D,L}$. At any round $u$ of this run, player $j$ has index $u-c_j\ge h_{D,L}-\kappa_L=\chi_L+2>d_L$, since $c_j-c_i\le\kappa_L$. The end of the run satisfies
\[
 t\leq c_i+f_L-1
 \leq c_0+\kappa_L+f_L-1
 =c_0+J_L-2<\min_j O_j.
\]
Thus all players have finished parking and none has reached output or replay. Every exploratory proposal, including a filler, satisfies the cooldown.

The latest certification is by $c_0+\kappa_L+f_L=c_0+J_L-1=\min_i O_i$, before that round's feedback. Also, $\min_i O_i+H_L\geq\max_i O_i+2=S+1$. Each player therefore stays in \REPLAY{} from round $O_i+1$ through round $S$, which rules out a second \WORK{} entry in $[c_0,S]$.
\end{proof}

\paragraph{From collection to mature entry.}
Suppose player $i$ enters fresh \SEEK{} in global round $w$ and retains highest priority in the collection set. Its local clearing deadline $c$ corresponds to global round $b_i+c$. If another player $j$ enters \WORK{} in round $c_j\leq w+2$, it outputs at $c_j+J_L-1$. Its minimum replay deadline is $c_j+J_L+\chi_L$, at least two rounds before the collector's initial clearing deadline $w+J_L+\chi_L+4$. A later entry $c_j\geq w+3$ before \phase{Root} nomination requires three $\BUS$ acceptances in rounds $c_j-3,c_j-2,c_j-1$. The collector proposes to $\BUS$ in each, so the third rejection moves its clearing deadline to at least $c_j+J_L+\chi_L+2$. After nomination its $\BUS$ proposals continue at least every other round. Lemma~\ref{lem:rcpc:control-service} then blocks new qualifying runs by lower-priority players.

Thus, by the start of the scan, all earlier \WORK{} invocations have ended and their minimum replay deadlines have passed. A player with a longer deadline may still be replaying its saved arm. Of the collector's last two $\BUS$ proposals before the scan, at least one wins ordinary control service and rejects every player still replaying $\BUS$ after its minimum deadline. If the root skips clearing, its nomination acceptances suffice. The two-round margin allows the resulting fresh \SEEK{} entry to take effect. At each other arm, a scan acceptance rejects every player still replaying that arm, sending it to fresh \SEEK{} in the next round.

Let $s$ be the last non-$\BUS$ scan acceptance. All recalled players have entered fresh \SEEK{} by $s+1$, so the latest guard occupies $s+1,\ldots,s+E_L$. Commitment starts at $s+1$ and lasts at least $E_L+K$ rounds. Its final $K$ accepted proposals to $\BUS$ give each recalled seeker $K$ post-guard rejections. Even an inherited root scanning away from $\BUS$ at the start of the $R_{c,L}=2K+1$ acceptance suffix receives its $K$th rejected $\BUS$ proposal by suffix round $2K$. In round $2K$ it either makes that proposal or is already in armed \SEEK{}. In round $2K+1$ every such root is in armed \SEEK{}. The other seekers propose to $\BUS$ in both rounds, and all are armed by the end of the suffix. Every arm other than $\BUS$ has therefore started and is empty in both rounds. Since $G_L-1=2K+2\geq R_{c,L}$, there are enough consecutive ordinary rounds for this suffix.

Hence the highest active player establishes maturity when it completes collection after a fresh entry. We will also use this argument for the first \WORK{} entry following a mature episode.

\begin{lemma}[An output-round bound under highest priority]
\label{lem:rcpc:one-output}
Fix a population and a starting round $v$ in a history reachable from the prescribed initialization. Let $z$ have highest priority among all active players or among the members of the collection set at $v$. If $z$ is in \REPLAY{} at $v$, let $O<v$ be the global round in which it finished the \WORK{} invocation preceding this visit. Otherwise, either $z$ finishes \WORK{} in a global round $O\in[v,v+C_0]$, or a higher-priority player joins the collection set during that interval. If $z$ has highest priority among all active players, no higher-priority player can join.
\end{lemma}

\begin{proof}
The \REPLAY{} case is immediate, and an inherited \WORK{} invocation has at most $J_L$ rounds left. Suppose $z$ starts in \SEEK{} or \phase{Root}. Follow it until a higher-priority join or its own output. Until it enters \WORK{}, $z$ proposes to $\BUS$ at least every other round. The table bounds elapsed time while $z$ stays highest. If it enters \WORK{} earlier, only the remaining rounds of that invocation are needed.

\begin{center}
\renewcommand{\arraystretch}{1.1}
\begin{tabular}{@{}p{0.52\linewidth}l@{}}
\toprule
Stage & Allowance in rounds\\
\midrule
Expire the $\BUS$ lease and establish a root & $\tau_L+E_L+6$\\
Clear inherited work and minimum replay grace & $2(J_L+\chi_L+4)$\\
Expire occupied-arm leases and finish the scan & $\tau_L+L_{s,L}$\\
Commit and complete work & $E_L+K+G_L+R_{c,L}+J_L$\\
\midrule
Total & $C_0-18$\\
\bottomrule
\end{tabular}
\end{center}

\noindent\emph{Establish control service.} The inherited lease at $\BUS$ expires within $\tau_L$ rounds. Set $t_c=v+\tau_L$ and $t_1=t_c+E_L+6$. The highest seeker finishes its guard within $E_L$ rounds and receives three consecutive acceptances within six more. An inherited root proposes to $\BUS$ by $t_c+1$. If its inherited rejection suffix causes demotion, it enters armed \SEEK{} by $t_c+2$ and \WORK{} within six further rounds, with $8\le E_L+6$. Otherwise an ordinary acceptance clears the suffix and Lemma~\ref{lem:rcpc:control-service} prevents demotion. Thus by $t_1$ either $z$ has entered \WORK{} or it is a root that stays in \phase{Root} until it enters \WORK{} or a higher-priority player joins.

\noindent\emph{Clear old work and finish scanning.} The root's frozen clearing deadline is at most $J_L+\chi_L+4$ rounds away. While $z$ collects under control service, Lemma~\ref{lem:rcpc:control-service} prevents $\BUS$ from accepting any given lower-priority player in three consecutive global rounds from $t_c$ onward. A lower player's nomination to \phase{Root} or entry into \WORK{} requires three such acceptances immediately beforehand, so the last possible new entry is $t_c+2$. Any lower player that enters \WORK{} by $t_c+2$ consequently finishes that invocation by $t_c+J_L+1$, with minimum replay deadline at most $t_c+J_L+\chi_L+2$. A longer replay can continue beyond that round until a rejection or expiry. The second allowance of $J_L+\chi_L+4$ rounds covers the remaining \WORK{} rounds and minimum grace. After $t_c+2$, each other player can continue at most one inherited \phase{Root} visit, giving at most $(K-1)^2$ further non-$\BUS$ scan acceptances in total.

Allow another $\tau_L$ rounds for the leases at arms still receiving proposals from old replayers to expire. This allowance runs concurrently across arms. Put $t_2=t_1+2(J_L+\chi_L+4)+\tau_L$. If $z$ is still collecting at $t_2$, it is scanning or committing. No other player is in \WORK{}, and any arm still occupied by an old replayer uses control service. At such a target $z$ proposes every other round, so its scan succeeds within $2KG_L$ rounds while the replayer stays. At any target without an old replayer, each rejection of $z$ requires a scan acceptance by an inherited root. Counting intervening $\BUS$ proposals and $z$'s own scan acceptances gives a remaining scan time of at most
\[
 2KG_L(K-1)+2\bigl((K-1)^2+(K-1)\bigr)
 =L_{s,L}.
\]
This includes each target's preceding $\BUS$ proposal, even at the first target. The root updates its target after the acceptance and enters commitment in the round after the last scan acceptance.

\noindent\emph{Commit and output.} The minimum commitment duration and a full suffix of ordinary acceptances take at most $E_L+K+G_L+R_{c,L}$ rounds. The root then outputs within $J_L$ more rounds. The allowances sum to
\[
 2\tau_L+2E_L+3J_L+2\chi_L+G_L+R_{c,L}+L_{s,L}+K+14
 =C_0-18.
\]
This includes next-round transitions and the output round itself, proving the bound while $z$ stays highest.
\end{proof}

\subsubsection{Cold and Warm Recovery}

\paragraph{Cold recovery from inherited states.}
After the population stabilizes, cold recovery starts from whatever states the active players have reached, even after incorrect comparisons.

\begin{lemma}[Cold recovery, simultaneously for every starting time]
\label{lem:rcpc:cold}
On every execution from the prescribed initialization with nonempty final population, fix any integer $v\geq t_*+1$. There is a mature episode in which every player resets its remaining-arm set strictly after $v$. The first common replay round $T(v)$ of the earliest such episode satisfies
\begin{equation}
 T(v)\leq \frac32 v+C_R.
\label{eq:rcpc:cold}
\end{equation}
This bound holds simultaneously for all such $v$, regardless of earlier statistical errors.
\end{lemma}

\begin{proof}
\noindent\emph{Obtain a fresh entry.} Let $z$ have highest priority at $\BUS$ in the final population. If it enters fresh \SEEK{} at $v$, take $w=v$. In all other cases use the global round $O$ from Lemma~\ref{lem:rcpc:one-output}. If $z$ is already in \REPLAY{} at $v$, then $O<v$ is its preceding final \WORK{} round. Otherwise $z$ finishes \WORK{} in round $O\in[v,v+C_0]$. In both cases let $w$ be the round when the replay visit following $O$ expires. By Lemma~\ref{lem:rcpc:envelope}, in all cases $w\ge v$ and
\[
 w\leq\frac32(v+C_0)+H_L.
\]
The entry at $w$ clears the inherited qualification.

\noindent\emph{Complete a mature collection.} During its $E_L$-round guard, $z$ proposes to $\BUS$ continuously, so the old lease expires before qualification. Under control service $z$ cannot receive the $K$ consecutive rejections needed to become armed. Its next \WORK{} entry therefore follows its own \phase{Root} visit. The clearing, recall, and commitment argument above gives armed followers, two empty rounds at every other arm, and no remaining old \WORK{} or \REPLAY{} visit. This entry is mature.

Lemma~\ref{lem:rcpc:one-output} from $w$, followed by Lemma~\ref{lem:rcpc:handoff}, bounds common replay by
\[
 w+C_0+\kappa_L+1
 \leq\frac32v+\frac52 C_0+H_L+\kappa_L+1
 \leq\frac32v+C_R.
\]
Each player resets its remaining-arm set on entering \WORK{} strictly after $w\ge v$. The argument holds on each realized path for every integer $v$, including times selected from the full history.
\end{proof}

\paragraph{Warm recovery between successive invocations of \WORK{}.}
Warm recovery starts at the first common replay round of a mature episode. It bounds the time from the first fresh entry to the next common replay and identifies which local output time each player uses at its next scale update.

\begin{lemma}[Warm collection between successive invocations of \WORK{}]
\label{lem:rcpc:warm}
Suppose a mature episode reaches common replay in round $S$ and the population stays fixed. Let $q_0$ be the first global round after $S$ in which any player enters fresh \SEEK{}. The first \WORK{} entry after $q_0$ initiates a mature episode. Its first common replay round $S'$, when all players are replaying their outputs from that episode, satisfies
\begin{equation}
 S'-q_0\leq V_B+KC_0+\kappa_L+1.
\label{eq:rcpc:warm-time}
\end{equation}
Each player completes exactly one \WORK{} invocation between the two common replay rounds. Its scale update in the final round of that invocation uses the local time $\ell$ saved in its final \WORK{} round of the episode ending at $S$. This holds regardless of statistical errors or replay correctness.
\end{lemma}

\begin{proof}
\noindent\emph{Recall players replaying $\BUS$.} Lemma~\ref{lem:rcpc:handoff} gives $q_0\geq S+1$. The first fresh player proposes to $\BUS$ every round until accepted, by $q_0+\tau_L+KG_L$ under the residual lease and fair control service. In that round $\BUS$ rejects every player still replaying it. For each player $i$, the minimum replay deadline from its preceding \WORK{} invocation is $O_i+H_L\le S+\chi_L$ in global rounds. By Equation~\eqref{eq:rcpc:rejection}, rejection before this deadline sets it as the new deadline, and rejection at or after it triggers fresh entry next round. Thus every player replaying $\BUS$ at $S$ has entered fresh \SEEK{} by $q_0+V_B$.

\noindent\emph{Establish maturity of the first new \WORK{} entry.} Start the collection set at $q_0$. Until the first \WORK{} entry it contains all fresh players and all players replaying $\BUS$. Those still outside propose to the non-$\BUS$ arms they saved for replay. The first fresh player proposes to $\BUS$ at least every other round, preventing lease renewal, and even the earliest fresh guard lasts until control service begins. Membership persists, so the current highest member has been highest since joining. It proposes to $\BUS$ at least every other round, including any rounds spent replaying $\BUS$. After its fresh guard it cannot become armed. By Lemma~\ref{lem:rcpc:control-service}, no lower armed seeker can be accepted in three consecutive global rounds while it is highest. If a higher-priority player joins in either of the two rounds after the seeker's first acceptance, it proposes to $\BUS$ immediately. Even on a fair round the seeker loses, since the joiner's last $\BUS$ acceptance was no later than $S$ and the seeker's run began after $S$. Thus the first new \WORK{} entry must come from \phase{Root}.

That root's guard ends after every minimum replay deadline $O_i+H_L\le S+\chi_L$, since $q_0>S$ and $E_L>\chi_L$. Its nomination acceptances at $\BUS$ and its subsequent scan acceptances recall all remaining replayers by the round after the last scan acceptance. By then every player has joined the collection set. A higher-priority collector would interrupt the final commitment suffix, so the completing root is highest. The minimum commitment duration gives the last recalled seeker a full guard and $K$ subsequent rejections. The final $2K+1$ acceptances demote any remaining roots and leave every arm other than $\BUS$ empty for two rounds. This proves maturity.

\noindent\emph{Bound the time to the next common replay.} After the first $V_B$ rounds, split the remaining collection period at higher-priority joins. Each raises the highest member's priority, so there are at most $K$ intervals. By Lemma~\ref{lem:rcpc:one-output}, each ends in a higher-priority join or an output within $C_0$ rounds. The last allowance includes the root's \WORK{} invocation. Lemma~\ref{lem:rcpc:handoff} adds at most $\kappa_L+1$ rounds to common replay. If \WORK{} starts during the initial allowance, its remaining duration and the handoff also fit this bound. This proves Equation~\eqref{eq:rcpc:warm-time}.

No player completes \WORK{} before the first new \WORK{} entry. Its mature handoff gives each player exactly one \WORK{} invocation through $S'$. Only the final round of \WORK{} updates $\ell$, so the scale update in this invocation uses the time saved in the episode ending at $S$.
\end{proof}

\subsection{Stable-Edge Protection, Absorption, and Sample Progress}
\label{sec:rcpc:stable}

Fix a mature episode during which the same $n$ players are present. Statements about persistence after the episode assume that the population stays fixed until the next entry into fresh \SEEK{}. We first show how coverage protects stable edges, then bound the samples collected before a failed episode ends.

\subsubsection{Comparison Coverage and Good Episodes}
\label{sec:rcpc:coverage}
At a player or arm with local clock $a$, the comparison of two options $x,z$ with positive counts is \emph{covered} if
\begin{equation}
 \left|(\widehat\theta_x-\widehat\theta_z)-(\theta_x-\theta_z)\right|
          \le\beta_a(x,z).
\label{eq:rcpc:covered}
\end{equation}
A comparison with a zero count has infinite radius and cannot exclude either option. If every positive-count pair in a nonempty set $Q$ is covered, its true best option belongs to $C_a(Q)$. Thus if $C_a(Q)=\{x\}$, then $x$ is the true best option in $Q$.

Let $S$ be the first round in which all players are in \REPLAY{}. Call the episode \emph{good} if, from the first remaining-set reset through the end of round $S$, every positive-count pair is covered at the start and end of each round. This includes every pair of arms for each player and every pair of present players for each arm, whether or not the protocol compares the pair. In particular, coverage holds when a later deletion witness first certifies an arm.

In SC-PC, a player \emph{holds} an arm from certification in the current \WORK{} invocation until deletion, or through the ensuing \REPLAY{} if it keeps the arm. Several players may hold the same arm, although the arm accepts only one proposer per round. Call the episode \emph{successful} if, at $S$, every player holds an arm and their saved replay arms $y_i$ are pairwise distinct. Otherwise it \emph{fails}. A good episode may fail when the confidence sets still contain several candidates.

\subsubsection{Stable-Edge Protection and POSM Absorption}

\paragraph{The POSM benchmark.}
By Section~\ref{sec:market}, the POSM assigns each player its most preferred arm among those it can receive in a stable matching.

\begin{lemma}[A held witness for every deletion]
\label{lem:rcpc:witness}
If player $i$ deletes arm $k$ in a mature good episode, then in at least one of the $K$ consecutive rejection rounds that triggered deletion, another player $j$ held $k$ and proposed to it, and $k$ preferred $j$ to $i$. The arm may accept a different proposer in that round.
\end{lemma}
\begin{proof}
During these $K$ rounds, $k$ accepts one of at most $K-1$ other players in each round. Some player $w$ is accepted twice, at arm ages $r<s$ with $s-r\le K-1$. Lemma~\ref{lem:rcpc:handoff} excludes parking and replay, and the cooldown excludes an exploratory proposal by $w$ at $s$. Thus $w$ holds $k$ at $s$. If $k$ prefers $w$ to $i$, then $w$ is the required witness.

Otherwise let $z$ be the player that $k$ most prefers among its proposers at $s$. If $z\ne i$ is a holder, it is the required witness. In the remaining cases, either $z=i$ has been rejected throughout the window, or $z$ is exploring and has no acceptance at $k$ in the preceding $K-1$ rounds. In either case, the last-acceptance ages at $k$ just before service at $s$ satisfy
\[
 \operatorname{last}[z]<r\le\operatorname{last}[w].
\]
By Lemma~\ref{lem:rcpc:handoff}, $k$ uses business service throughout the rejection run. Coverage keeps $z$ eligible among the current proposers, so the arm's oldest-timestamp rule cannot select $w$, a contradiction.
\end{proof}

\begin{lemma}[Stable-edge protection and absorption]
\label{lem:rcpc:stable}
A mature good episode deletes no edge of any stable matching. Each player $i$ that holds an arm weakly prefers it to its assignment $m_t^*(i)$ in the fixed market. Call the first round in which all players hold pairwise distinct arms after choosing their proposals and before arm service \emph{absorption}. The players realize the POSM in that round and in every subsequent round before the first entry into fresh \SEEK{}, provided the population stays fixed. Such an episode with $n$ players has at most $(n-1)^2$ deletions.
\end{lemma}
\begin{proof}
\noindent\emph{Stable edges.} Suppose $(i,k)$ is the first deleted edge of a stable matching, choosing any one in case of a tie. Lemma~\ref{lem:rcpc:witness} gives a holder $j$ whom $k$ prefers to $i$. When $j$ certified $k$, its partner in that matching was still in its remaining set. This holds even for a certificate and a deletion in the same round, since certificates use the round-start sets and deletions follow feedback. By coverage, $k$ was $j$'s best remaining arm. Hence $(j,k)$ blocks the matching, a contradiction. In particular, each player's POSM arm is still available whenever it certifies an arm.

\noindent\emph{Persistent holders and absorption.} For any arm that has been held, consider the player it most prefers among all its holders up to the current round. This player still holds the arm, since deleting it would have required an even better holder at the time of deletion. Thus every arm ever held still has a holder, including across output and replay. Suppose now that every player holds a distinct arm. If player $i$ prefers arm $k$ to its held arm, it must have deleted $k$ before certifying its current arm. The deletion witness held $k$ and was preferred to $i$, so $k$'s current holder is also preferred to $i$. This excludes every blocking pair, including a deviation to an unoccupied arm. The resulting stable assignment is weakly preferred by every player to its POSM assignment and hence equals the POSM by strict preferences. Each occupied arm now receives just one proposal and accepts it. The holders keep proposing to those arms through output and replay until a player next enters fresh \SEEK{}.

\noindent\emph{Deletion count.} The first time $n$ distinct arms have been held, their persistent holders account for all $n$ players, giving absorption. Before then, deletions concern at most $n-1$ ever-held arms. Each such arm's best holder over the episode never deletes it, and each of the other players deletes it at most once. Thus there are at most $(n-1)^2$ deletions.
\end{proof}

\subsubsection{Samples Collected by Failed Episodes}
\label{sec:rcpc:progress}

We use the minimum preference gap $\Delta_{\min}$ of the final market from Equation~\eqref{eq:learning-gap} to bound the samples needed to distinguish two options. For an integer horizon $T\ge2$, define the analysis threshold
\begin{equation}
                 Q_T=1+\left\lfloor40\Delta_{\min}^{-2}\log T\right\rfloor.
\label{eq:rcpc:threshold}
\end{equation}
If $a\le T$ and both counts are at least $Q_T$, then $a+2\le T^2$ and $\log(a+2)\le2\log T$, so
\[
 \beta_a(x,z)\le\sqrt{\frac{10\log T}{Q_T}}<\frac{\Delta_{\min}}{2}.
\]
Under coverage, a gap of at least $\Delta_{\min}$ then excludes the inferior option whenever the superior option is present. Neither $Q_T$ nor the gap is a protocol input.

Fix a mature good episode of the final population with common replay by $T$. All relevant ages are at most $T$. An acceptance of player $i$ by arm $k$ is \emph{insufficient} if their common count $N_i(k)$ is below $Q_T$ just before that acceptance. Each such acceptance counts once, even if several occur in one round or the same pair contributes in several rounds.

Before the certification cutoff, a player that creates a probe target needs more samples on that edge. We show that each block of $K$ rounds with no recent insufficient acceptance must contain a deletion or reach the POSM. The bound on deletions then gives a sample quota for each failed episode.

\begin{lemma}[Fixed probe targets remain insufficiently sampled in the common window]
\label{lem:rcpc:target}
Every exploratory target maintained in $I_0$ has player count below $Q_T$, even if it has left the current confidence set.
\end{lemma}
\begin{proof}
Such a target was created before the player's certification cutoff. Stable-edge protection keeps the remaining set nonempty, and coverage keeps its best arm among the candidates. By Equation~\eqref{eq:learning-gap}, the player-side gap between that best arm and any inferior candidate is at least $\Delta_{\min}$. If all candidates had counts at least $Q_T$, coverage and Equation~\eqref{eq:rcpc:threshold} would let that best arm exclude every inferior candidate. The resulting singleton would be certified before target creation. Thus the least-sampled candidate selected as target has count below $Q_T$.

While the target is retained, its count cannot change. An accepted probe or a new certificate clears it, and a filler cannot choose the target while it is on cooldown. The conclusion therefore holds even after the target leaves the confidence set. It need not hold for targets created after the certification cutoff.
\end{proof}

\begin{lemma}[A cooled block and the amortized inequality]
\label{lem:rcpc:amortized}
Consider $K$ consecutive rounds in $I_0$ with no insufficient acceptance in the block or the preceding $K-1$ rounds. If the block contains no deletion, absorption at the POSM occurs by its last round. Consequently, if absorption has not occurred by the last round of a prefix $\mathcal J$ of $I_0$, then
\begin{equation}
 |\mathcal J|\le Kd(\mathcal J)+(2K-1)b(\mathcal J)+2K-2,
\label{eq:rcpc:amortized}
\end{equation}
where $d(\mathcal J)$ counts deletions and $b(\mathcal J)$ counts insufficient acceptances in $\mathcal J$, counting each accepted edge separately.
\end{lemma}
\begin{proof}
\noindent\emph{A block with neither kind of event.} At its first action, each player holds an arm, possibly certifying it then, or explores. If player $i$ explores with target $k$, then $N_i(k)<Q_T$ by Lemma~\ref{lem:rcpc:target}. Since this count never decreases, any acceptance of $i$ at $k$ in the preceding $K-1$ rounds would have been insufficient. The cooldown therefore allows $i$ to probe $k$ in the first round of the block. Every subsequent probe acceptance would be insufficient as well. Without acceptance or deletion, the player's statistics and remaining set stay fixed. Increasing age only widens the confidence radii, so a nonsingleton cannot become a singleton. The player therefore probes for $K$ rounds and deletes, a contradiction. All players must hold from their first action.

If two fixed holders share an arm, let $z$ be its most-preferred proposer. By Lemma~\ref{lem:rcpc:handoff}, the arm uses business service throughout $I_0$, so coverage keeps $z$ eligible in every round of the block. If $z$'s edge is insufficient, its timestamp cannot change without an insufficient acceptance. Each other proposer can be served ahead of it at most once, so the arm accepts $z$ within $K$ rounds, a contradiction. If its edge is sufficient, the arm excludes every sufficient inferior holder because its mean utility for $z$ exceeds its mean utility for that holder by at least $\Delta_{\min}$. The block rules out insufficient acceptances, so the arm accepts only $z$ throughout it. Any inferior holder then deletes after $K$ rejections. Thus all held arms are distinct, and Lemma~\ref{lem:rcpc:stable} gives absorption.

\noindent\emph{Charge the rounds before absorption.} Mark the first $K-1$ rounds of $\mathcal J$ and each insufficient-acceptance round together with the next $K-1$ rounds, stopping at the end of $\mathcal J$. This marks at most $Kb(\mathcal J)+K-1$ rounds and leaves at most $b(\mathcal J)+1$ unmarked intervals. Divide each unmarked interval into disjoint $K$-round blocks and a remainder. Each full block contains a deletion, since otherwise absorption would persist to the end of $\mathcal J$. An interval containing $d$ deletions therefore has at most $Kd+K-1$ rounds. Summing gives
\[
 |\mathcal J|\le Kb(\mathcal J)+K-1+Kd(\mathcal J)+(b(\mathcal J)+1)(K-1),
\]
which is Equation~\eqref{eq:rcpc:amortized}. Marking the initial $K-1$ rounds also covers cooldowns caused by acceptances before $\mathcal J$.
\end{proof}

\begin{lemma}[Sample density and the first fresh entry]
\label{lem:rcpc:density}
For the nonempty final market, fix $T\ge2$ and the threshold $Q_T$ from Equation~\eqref{eq:rcpc:threshold}. Every failed mature good episode with common replay by $T$ contains at least $b_K=2K^2$ insufficient acceptances in its common work window $I_0$. There are at most $\lfloor Q_T/2\rfloor$ such episodes.

For any mature good episode, suppose the population stays fixed through the first entry into fresh \SEEK{} after its first common replay round $S$. If the episode succeeds, the players replay the POSM until this entry. If it fails, the entry occurs by $S+\chi_L$.
\end{lemma}
\begin{proof}
\noindent\emph{Sample quota and failure count.} Absorption during $I_0$ persists through output and common replay, so a failed episode cannot be absorbed there. If it had at most $b_K-1$ insufficient acceptances in $I_0$, Lemma~\ref{lem:rcpc:stable} and Equation~\eqref{eq:rcpc:amortized} would give
\[
 W_L\le K(K-1)^2+(2K-1)(b_K-1)+2K-2=W_L-1,
\]
a contradiction. Let $N_i(k;t)$ count the acceptances of player $i$ by arm $k$ through round $t$. The potential
\[
 \Phi_T(t)=\sum_{i\in\mathcal P_{t_*}}\sum_{k\in\mathcal K}
        \max\{0,Q_T-N_i(k;t)\}\le nKQ_T\le K^2Q_T
\]
never increases and drops by one at each insufficient acceptance. Distinct mature episodes have disjoint common windows, so at most $\lfloor K^2Q_T/b_K\rfloor=\lfloor Q_T/2\rfloor$ can fail, regardless of intervening successes or additional samples.

\noindent\emph{First fresh entry.} By Lemma~\ref{lem:rcpc:handoff}, all deletions occur before the earliest output round $\min_i O_i$. Once players have chosen their proposals in that round, all certifications are also complete. In a successful episode each player $i$ therefore already holds the arm it will save as $y_i$. These arms are pairwise distinct, so every player's final \WORK{} proposal is accepted and every player takes the long replay branch. By Lemma~\ref{lem:rcpc:stable}, these players replay the POSM until the next entry into fresh \SEEK{}.

In a failed episode, if some player $i$ takes the short replay branch, its deadline is $O_i+H_L\le S+\chi_L$. Otherwise every player takes the long branch, so two players must have saved the same arm. Both propose to it in round $S$, and at least one replay proposal is rejected. For a rejected player $i$, the shortened deadline satisfies
\[
 \max\{S+1,O_i+H_L\}\le S+\chi_L.
\]
Thus a failed episode has a fresh entry by $S+\chi_L$, while minimum grace places every such entry after $S$.
\end{proof}

\subsection{Counting Episodes and Mismatch Cost}
\label{sec:rcpc:maintenance}

A mature episode runs from its first \WORK{} entry $c_0$ through its first common replay round $S$. To charge mismatch after a good episode, we use the interval $[S,S')$, where $S'$ is the first common replay round of the next mature episode. It includes replay from the current episode and collection and work for the next one. Only the episode ending at $S$ must be good for this interval bound. The sample quota bounds how often good episodes fail, and growth of the replay scales bounds how often they succeed.

\paragraph{Mismatch cost between common replay rounds.}
For the cumulative mismatch cost of Section~\ref{sec:objective}, take $D(0;I)=0$. On an interval of global rounds $[s,t)$ with $1\le s\le t$, the cost is
\[
 D([s,t);I):=\sum_{r=s}^{t-1}\indicator\{M_r\ne m_r^*\}
             =D(t-1;I)-D(s-1;I).
\]
Define the per-interval cost bound
\begin{equation}\begin{aligned}
 C_F&=KC_0+V_B+\kappa_L+\chi_L+16\\
    &=39K^4-19K^3+154K^2+10K+4.
\end{aligned}\label{eq:rcpc:interval-constant}
\end{equation}
The four terms allow for collection, recall of BUS replayers, the spread of output rounds, and the delay before fresh entry after failure. The additive constant covers endpoints. The bound $C_F=O(K^4)$ counts mismatched rounds. A long successful replay can make the interval much longer while contributing zero mismatch.

\begin{lemma}[At most $C_F$ mismatched rounds in a warm interval]
\label{lem:rcpc:interval-cost}
Suppose a mature good episode reaches common replay at $S$ and the population remains fixed through the first common replay round $S'$ of the next mature episode. Then
\[
 D([S,S');I)\le C_F.
\]
For every integer horizon $T\ge S$, the same bound holds on $[S,\min\{S',T+1\})$, including when $T$ falls inside the next collection or work invocation.
\end{lemma}
\begin{proof}
Let $q_0$ be the first fresh entry after $S$. By Lemma~\ref{lem:rcpc:density}, $[S,q_0)$ has zero cost after success and length at most $\chi_L$ after failure. Lemma~\ref{lem:rcpc:warm} then bounds the duration of the remaining interval for every outcome of the next episode. Hence
\[
 D([S,S');I)\le \chi_L+(S'-q_0)
 \le \chi_L+V_B+KC_0+\kappa_L+1\le C_F.
\]
Truncation only reduces the mismatch count. Goodness is needed for the episode that reached $S$. After an uncovered comparison, the players might replay an incorrect assignment to distinct arms and receive no rejection. Cold recovery accounts for these rounds.
\end{proof}

\paragraph{Scale growth and the number of successes.}
\begin{lemma}[Minimum-scale growth after a successful episode]
\label{lem:rcpc:growth}
Consider consecutive mature episodes $s,s+1$ in a fixed population. Suppose $s$ is good and successful and $s+1$ reaches common replay. Let $O_{i,s}$ be player $i$'s global output round in episode $s$, and let $P_{i,s}$ be its individual scale immediately after that output. With $O_s$ the earliest output round and $p_s$ the smallest of these scales, write
\[
 O_{i,s}=O_s+\delta_{i,s},\qquad O_s=\min_iO_{i,s},\qquad
 0\le\delta_{i,s}\le\kappa_L,\qquad p_s=\min_iP_{i,s}.
\]
Even if the successor fails,
\begin{equation}
 p_{s+1}\ge
 \left\lfloor\frac{5p_s+H_L+2(J_L-\kappa_L-1)}4\right\rfloor
 >\frac54p_s.
\label{eq:rcpc:growth}
\end{equation}
Each $P_{i,s}$ is nondecreasing across that player's outputs, so the minimum scale is nondecreasing as well. Along any consecutive chain of mature good episodes, the number $N_S(T)$ of successes with common replay by $T$ satisfies
\begin{equation}
 N_S(T)\le1+\left\lfloor\log_{5/4}\left(1+\frac{T}{2H_L}\right)\right\rfloor.
\label{eq:rcpc:success-count}
\end{equation}
\end{lemma}
\begin{proof}
\noindent\emph{Growth.} After success, the POSM has one proposer per occupied arm until the first fresh entry, even if business leases expire. Every replay proposal is accepted. Thus the first global round $q_s$ in which any player enters fresh \SEEK{} after episode $s$ is
\[
 q_s=\min_i(O_{i,s}+P_{i,s})\ge O_s+p_s.
\]
Each player's next \WORK{} invocation starts at or after $q_s$ and lasts $J_L$ rounds. By Lemma~\ref{lem:rcpc:warm}, player $i$ produces no intervening output between rounds $O_{i,s}$ and $O_{i,s+1}$. Its local clock advances by one per round, so the elapsed local time used in its scale update is
\[
 \Delta_{i,s+1}:=O_{i,s+1}-O_{i,s}\ge p_s+J_L-\kappa_L-1.
\]
Substituting $P_{i,s}\ge p_s$ and this interval into Equation~\eqref{eq:rcpc:output} proves the first inequality in Equation~\eqref{eq:rcpc:growth}. Since $H_L+2(J_L-\kappa_L-1)>4$ and $\lfloor x\rfloor>x-1$, the second inequality is strict. Every actual output performs the update, so neither success nor coverage is required of the successor.

\noindent\emph{Count successes.} Outputs in a counted episode precede $T$, so Equation~\eqref{eq:rcpc:envelope} gives $H_L\le p_s\le H_L+T/2$. If $N_S(T)\ge1$, each of its first $N_S(T)-1$ successes has a completed immediate successor in the counted chain. Growth at those transitions and monotonicity through failures yield
\[
 (5/4)^{N_S(T)-1}H_L\le H_L+T/2.
\]
This proves Equation~\eqref{eq:rcpc:success-count} whenever there is a success. The case of zero successes is immediate. The additive one allows the last success an unfinished successor.
\end{proof}

\paragraph{Combining the episode counts.}
For $T\ge2$, combine Lemma~\ref{lem:rcpc:density} with Equation~\eqref{eq:rcpc:success-count}. In a consecutive mature good chain of the final market,
\begin{equation}
 N_{\mathrm{fail}}(T)+N_S(T)
 \le\left\lfloor\frac{Q_T}{2}\right\rfloor+1
 +\left\lfloor\log_{5/4}\left(1+\frac{T}{2H_L}\right)\right\rfloor,
\label{eq:rcpc:episode-count}
\end{equation}
where $N_{\mathrm{fail}}(T)$ counts failures with common replay by $T$. Thus the logarithmic maintenance term is paid once over the entire chain.

\subsection{Concentration and Recovery from Statistical Errors}
\label{sec:rcpc:probability}

The preceding bounds apply to a chain of mature good episodes. We now bound the time until such a chain begins, using concentration and two cold recoveries. A maximal inequality at prescribed sample counts handles both adaptive sampling and random arm startup.

\subsubsection{Adaptive Concentration}

Call each player or arm an \emph{owner} of its own records. A stream for player $i$ consists of its rewards when matched to a given arm $k$, with mean $\mu_{i,k}$. The stream for arm $k$ and player $i$ consists of the arm's rewards on the same edge, with mean $\pi_{k,i}$. Each owner receives at most one sample per global round.

Let $\mathcal F_{t-1}$ be the history of all actions and observations through round $t-1$. Let $\mathcal H_t$ add the proposals and acceptance decisions in round $t$, and let $\mathcal F_t$ also include the utility observations in that round. By the assumption in Section~\ref{sec:learning-model}, an accepted observation from stream $x$ lies in $[0,1]$ and has mean $\theta_x$ conditional on $\mathcal H_t$. We analyze one owner's observation at a time, before conditioning on any other current observation. A union bound then covers all owners without requiring independence.

\begin{lemma}[Time-uniform concentration for prescribed sample counts]
\label{lem:rcpc:concentration}
Fix an owner, two distinct streams, and deterministic positive integers $n,r$. Weight the first $n$ accepted samples of the first stream by $1/n$, the first $r$ of the second by $-1/r$, and all other global rounds by zero. Let $X_t$ be the weighted centered sum through round $t$ and $V_t$ the sum of squared weights. For every $b>0$,
\begin{equation}
 \Pr\left\{\sup_t|X_t|>b\right\}
 \le2\exp\left\{-\frac{2b^2}{1/n+1/r}\right\}.
\label{eq:rcpc:fixed-count}
\end{equation}
For owner $o$ and a fixed family of at most $K$ streams, let $\mathrm{Bad}_{o,a}$ be the event that $o$ reaches local age $a$ and some positive-count pair in the family is uncovered at the start or end of that round, whether or not $o$ compares the pair. Both snapshots use the radius at age $a$ and the observations available at that snapshot. Then
\begin{equation}
 \Pr(\mathrm{Bad}_{o,a})
 \le\frac{K(K-1)}2\frac{a(a+1)}{(a+2)^5}.
\label{eq:rcpc:local-bad}
\end{equation}
Both conclusions cover random arm startup times.
\end{lemma}

\begin{proof}
\noindent\emph{A maximal bound at prescribed counts.} Let $w_t$ be the round-$t$ weight and, when nonzero, let $R_t,\theta_t$ be the selected reward and mean. The weight is fixed by $\mathcal H_t$ before the reward, and truncation gives
\[
 V_t\le v_{n,r}:=1/n+1/r.
\]
Conditional Hoeffding's lemma gives
\[
 \mathbb E\left[e^{\lambda w_t(R_t-\theta_t)}\mid\mathcal H_t\right]
 \le e^{\lambda^2w_t^2/8}.
\]
Zero weights give zero increments. Conditioning successively on $\mathcal H_t$ and $\mathcal F_{t-1}$ makes
\[
 \mathcal E_t(\lambda)=\exp\{\lambda X_t-\lambda^2V_t/8\}
\]
a nonnegative supermartingale with respect to $(\mathcal F_t)_{t\ge0}$, with $\mathcal E_0(\lambda)=1$.

Ville's inequality bounds the chance that this process ever exceeds $\exp\{\lambda b-\lambda^2v_{n,r}/8\}$ by its reciprocal. Choosing $\lambda=4b/v_{n,r}$ and repeating for $-X_t$ gives Equation~\eqref{eq:rcpc:fixed-count}, using the ownerwise bounded-increment construction of \cite{HRMS20}. This argument uses the conditional reward assumption, not independence between accepted samples.

\noindent\emph{Union over counts and comparisons.} At counts $n,r$, the truncated sum is the empirical difference error. Thus coverage failure lies in a maximal event from Equation~\eqref{eq:rcpc:fixed-count}, without conditioning on realized counts or assuming conditional IID samples. With
\[
 b=\sqrt{\frac{5\log(a+2)}2v_{n,r}},
\]
its probability is at most
\[
 2\exp\{-5\log(a+2)\}
 =2(a+2)^{-5/\ln2}\le2(a+2)^{-5}.
\]
Each age-$a$ snapshot contains at most $a+1$ observations, allowing $a(a+1)/2$ positive count pairs $n+r\le a+1$ and at most $\binom K2$ unordered stream pairs. Their union proves Equation~\eqref{eq:rcpc:local-bad}. The same maximal events cover both snapshots without another factor of two. At age zero, two counts cannot both be positive. A comparison involving a zero count has infinite radius.

\noindent\emph{Random startup times.} The process uses all global rounds, with zero weights before startup and whenever no sample arrives. The supremum already includes the random round at each fixed arm age, requiring neither a union over startup times nor conditioning on them.
\end{proof}

\subsubsection{Recovery after the Last Uncovered Comparison}

To bound the expected sum of the local ages at which coverage last fails, set
\begin{equation}
 C_Y=\frac{K^2(K-1)}4.
\label{eq:rcpc:last-bad-constant}
\end{equation}

\begin{lemma}[The last uncovered comparison and two cold recoveries]
\label{lem:rcpc:last-bad}
Suppose the final population is nonempty. Consider all pairs of arms for each final player and all pairs of final players for each arm. Almost surely there is a mature episode whose remaining-set resets and \WORK{} rounds all occur after the last uncovered comparison in these families, if any. The episode reaches common replay in round $R^*$, with
\begin{equation}
 \mathbb E R^*
 \le\frac94(t_*+1)+\frac52C_R+\frac32(1+C_Y).
\label{eq:rcpc:last-bad-cost}
\end{equation}
All subsequent comparisons in these families are covered. The cost before $R^*$ includes all mismatch during recovery from earlier errors.
\end{lemma}

\begin{proof}
The first recovery bounds every arm's startup time, so we can locate an uncovered comparison in global time from its local age. The second gives each player a fresh \WORK{} invocation after the last such comparison. Both recoveries retain the accumulated samples.

The oblivious schedule fixes $\mathcal P_{t_*}$. For each player in $\mathcal P_{t_*}$, use the streams for its $K$ arms. For each arm, use the streams for players in $\mathcal P_{t_*}$. These families involve at most $2K$ owners. Departed players' records persist, but those players make no proposals after $t_*$.

\noindent\emph{Bound the local age at the last coverage failure.} Define $Y_o$ on each execution by
\[
 Y_o:=\sup\bigl(\{0\}\cup\{a\ge1:\mathrm{Bad}_{o,a}\text{ occurs}\}\bigr)
 \le\sum_{a\ge1}a\mathbf1\{\mathrm{Bad}_{o,a}\}.
\]
Tonelli's theorem and Equation~\eqref{eq:rcpc:local-bad} give
\[
 \mathbb E\sum_oY_o
 \le K^2(K-1)\sum_{a\ge1}\frac{a^2(a+1)}{(a+2)^5}
 <\frac{K^2(K-1)}4=C_Y.
\]
Indeed, $a^2(a+1)/(a+2)^5\le a/(a+2)^3$, and $x/(x+2)^3$ is nonincreasing for $x\ge1$, so
\[
 \sum_{a\ge1}\frac a{(a+2)^3}
 \le\sum_{a=1}^{4}\frac a{(a+2)^3}
 +\int_4^\infty\frac{x}{(x+2)^3}\,dx
 =\frac{8989}{36000}<\frac14.
\]
Finite expectation makes every $Y_o$ finite almost surely, without independence between owners or coverage failures.

\noindent\emph{First recovery.} Apply Lemma~\ref{lem:rcpc:cold} at $v=t_*+1$. Its root scan has started every arm by $\frac32(t_*+1)+C_R$, regardless of statistical errors. All final players were born by $t_*$. An owner's local clock $a$ corresponds to the global round of its birth or startup plus $a$. Let $Z$ be the maximum of $t_*+1$ and the last global round with an uncovered comparison in these families, taking $Z=t_*+1$ if there is none. Thus
\[
 Z\le\frac32(t_*+1)+C_R+\sum_oY_o,
 \qquad
 \mathbb EZ\le\frac32(t_*+1)+C_R+C_Y.
\]
\noindent\emph{Second recovery.} At the end of round $Z$, earlier errors may still affect the players' work states. Lemma~\ref{lem:rcpc:cold} holds on each execution for every integer starting time. Applied at $v=Z+1$, it gives a mature episode with all remaining sets reset after $Z$ and common replay at
\[
 R^*\le\frac32(Z+1)+C_R.
\]
Choose $R^*$ to be the earliest such common replay round. Each candidate event is measurable in $Z$ and the execution, so this countable minimum is measurable. Set $R^*=0$ on the null event of no candidate. On the probability-one event above, all comparisons in the specified families are covered throughout this episode and thereafter. Taking expectations proves Equation~\eqref{eq:rcpc:last-bad-cost}.

The pathwise bound applies at $Z+1$ even though players cannot identify the last coverage failure from their observations. Both recoveries retain statistics, replay scales, and local output times. Charging $[1,R^*)$ covers erroneous stable-edge deletions, inherited armed or root states, and long replays in which players save distinct arms but realize the wrong matching.
\end{proof}

\subsection{Proof of Theorem~\ref{thm:learning}}
\label{sec:rcpc:theorem}

We combine the episode bounds from Section~\ref{sec:rcpc:maintenance} with the expected recovery time from Section~\ref{sec:rcpc:probability}, including any final incomplete collection or work interval.

\begin{proof}
\noindent\emph{Charge the finite prefix.} First suppose the final population is nonempty and $T\ge2$. On the probability-one event of Lemma~\ref{lem:rcpc:last-bad}, charge every round before $R^*$. This covers the prefix if $T<R^*$. Otherwise set $S_0=R^*$ and follow the consecutive mature episodes from warm collection. All these episodes are good because their remaining sets reset after $Z$ and every later comparison is covered.

Let $S_s$ be the first common replay round of episode $s$ in this chain. Count every $s$ with $S_s\le T$, including an episode whose latest output round is $T-1$ and whose common replay starts at $T$. For each counted episode, Lemma~\ref{lem:rcpc:interval-cost} bounds the cost from $S_s$ up to the next common replay or the horizon, on the interval
\[
 [S_s,\min\{S_{s+1},T+1\}).
\]
These partition $[R^*,T+1)$, including unfinished collection or \WORK{}. Equation~\eqref{eq:rcpc:episode-count} bounds their number, yielding
\begin{equation}
 D(T;I)\le R^*+C_F\left[
 \left\lfloor\frac{Q_T}{2}\right\rfloor+1
 +\left\lfloor\log_{5/4}\left(1+\frac{T}{2H_L}\right)\right\rfloor
 \right].
\label{eq:rcpc:pathwise-prefix}
\end{equation}
Each counted failed episode obtains its sample quota in a completed common work window before $T$, so $Q_T$ is applied only at local clocks at most $T$. The scale argument likewise uses only successor outputs recorded before the horizon.

\noindent\emph{Take expectations.} Equations~\eqref{eq:rcpc:pathwise-prefix} and~\eqref{eq:rcpc:last-bad-cost}, together with $D(T;I)\le T$, give
\begin{equation}
 \begin{split}
 \mathbb ED(T;I)\le\min\Bigg\{T,\;&
 \frac94(t_*+1)+\frac52C_R+\frac32(1+C_Y)\\
 &+C_F\left[
 \left\lfloor\frac{Q_T}{2}\right\rfloor+1
 +\left\lfloor\log_{5/4}\left(1+\frac{T}{2H_L}\right)\right\rfloor
 \right]\Bigg\}.
 \end{split}
\label{eq:rcpc:unfolded}
\end{equation}
\noindent\emph{Simplify with uniform constants.} For $K\ge2$, the table gives bounds obtained from Equation~\eqref{eq:rcpc:constants}, \eqref{eq:rcpc:interval-constant}, and \eqref{eq:rcpc:last-bad-constant}. They hold uniformly over instances and horizons.

\begin{center}
\renewcommand{\arraystretch}{1.15}
\begin{tabular}{@{}lll@{}}
\toprule
Quantity & Elementary upper bound & Bound used below\\
\midrule
$C_0$ & $39K^3+156K$ & $78K^3$\\
$C_Y$ & $K^2(K-1)/4$ & $K^3/4$\\
$C_R$ & $(5/2)(78K^3)+8K+12$ & $199K^3$\\
$C_F$ & $39K^4+154K^2+10K+4$ & $80K^4$\\
$(5/2)C_R+(3/2)(1+C_Y)$ & $(5/2)(199K^3)+(3/2)(1+K^3/4)$ & $500K^3$\\
\bottomrule
\end{tabular}
\end{center}
Put $x=\log T\ge1$ and $\delta=\Delta_{\min}^{-2}\ge1$. Since $Q_T=1+\lfloor40\delta x\rfloor$, we have $\lfloor Q_T/2\rfloor\le1/2+20\delta x$. Also $H_L\ge1$, $T\ge2$, and $\log(5/4)>1/5$ give $1+T/(2H_L)\le T$ and bound the bracket in Equation~\eqref{eq:rcpc:unfolded} by
\[
 \frac32+(20\delta+5)x\le28\delta x.
\]
Since $K\ge2$ and $\delta x\ge1$, the recovery constant satisfies $500K^3\le500K^4\delta x$. Consequently,
\[
 500K^3+80K^4(28\delta x)
 \le(500+80\cdot28)K^4\delta x
 =2740K^4\delta x.
\]
Finally, $2740\le4000$ and $(9/4)(t_*+1)\le(9/2)t_*$ give
\begin{equation}
 \mathbb ED(T;I)
 \le\min\left\{T,\frac92t_*+
 4000K^4\Delta_{\min}^{-2}\log T\right\}.
\label{eq:rcpc:uniform-log}
\end{equation}
\noindent\emph{Endpoint horizons.} At $T=1$, the same bound follows from $\mathbb ED(1;I)\le1\le t_*$. Theorem~\ref{thm:learning} therefore holds for every integer $T\ge1$, with one constant for all admissible parameters, instances, and reward processes, and no extra $\log K$ factor.

\noindent\emph{Empty final population.} Both matchings are empty from round $t_*$ onward, so
\[
 D(T;I)\le\min\{T,t_*-1\},
\]
without a gap parameter. A permanently empty market has zero cost.
\end{proof}

%% file: sections/learning/player-algorithm.tex
On arrival in round \(b_i\), player \(i\) sets its local clock to zero and starts in \SEEK{}, proposing to BUS (arm~\(0\)). Algorithm~\ref{alg:rcpc-player} gives its actions in each round while it is active. We write \(a\) for the player's local clock reading \(a_i=t-b_i\), \(m\) for its current mode, and \(\nu\) for a change of mode or stage scheduled for the next round. The player initializes its replay scale \(P\) to the minimum waiting length \(H_L\) and leaves \(\ell\), its local clock reading at the last \WORK{} output, unset (\(\bot\)).

At the start of each round the player applies any scheduled change, then checks whether its replay deadline \(q\) has expired. In \REPLAY{} it proposes to its saved arm \(y\). In \phase{Root}, \(s\) denotes the stage. During a scan, \(j\) is the current non-BUS arm and \(\mathrm{busNext}\) indicates whether the next proposal goes to BUS. The player makes one proposal and observes whether it was accepted or rejected. On acceptance both sides record their own realized utility as in Section~\ref{sec:information}. The player then processes the outcome for its current mode and advances its local clock. All subroutines use the player's local state.
\begin{figure}[htbp]
    \centering
    \includegraphics[width=0.75\linewidth]{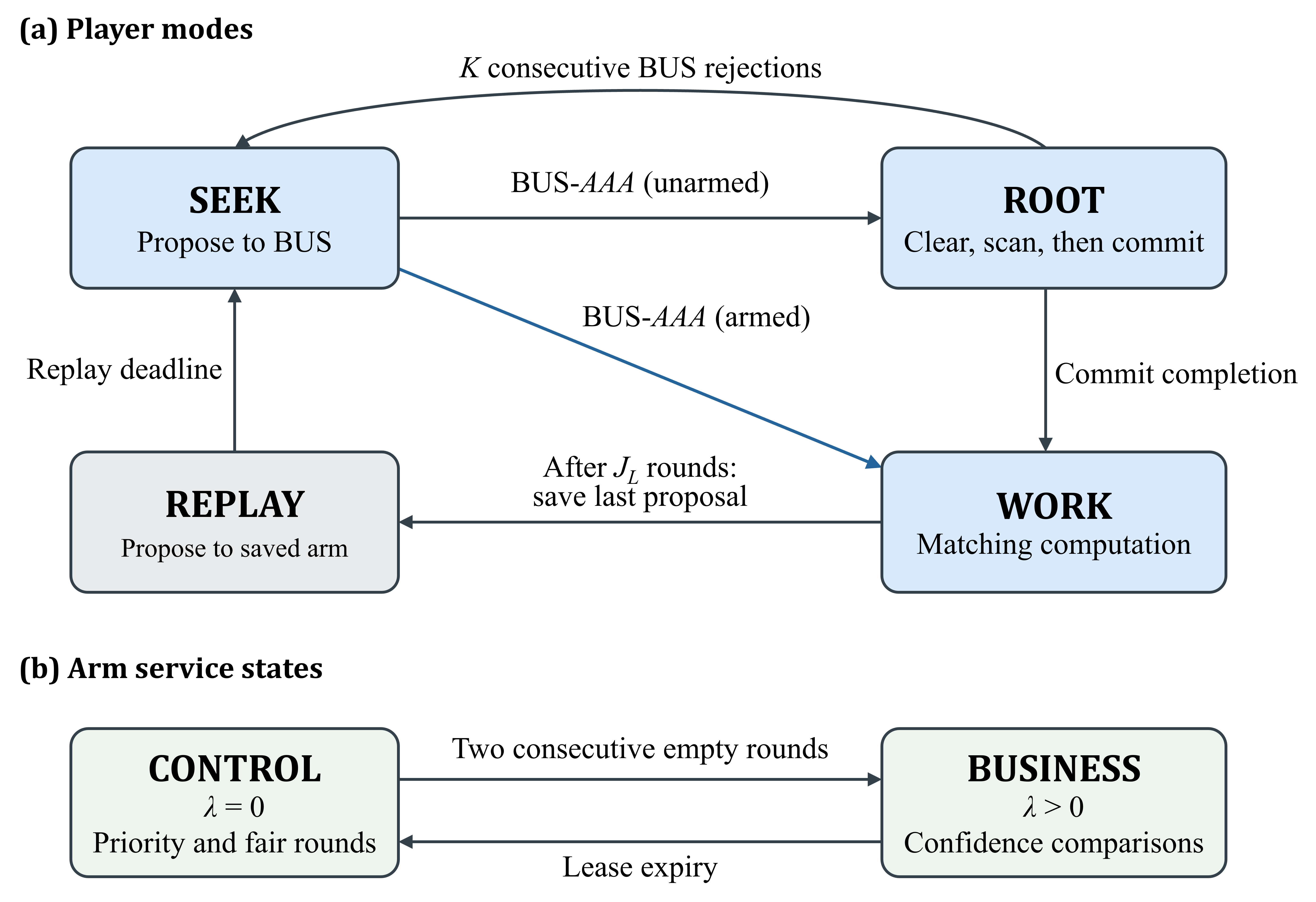}
    \caption{State transitions in SC-PC: (a) player modes and (b) arm service states. Player mode transitions triggered by feedback take effect in the next round; replay expiry is processed before the current proposal.}
\label{fig:scpc-modes}
\end{figure}
\begin{algorithm}[H]
\caption{\textsc{SC-PC} for player \(i\)}
\label{alg:rcpc-player}\hypertarget{scpc:player}{}
\raggedright
\algrenewcommand{\algorithmicindent}{0.75em}
\begin{algorithmic}[1]
\State \hypertarget{scpc:initialization}{}For every \(v\in\mathcal K\), set \(N_v=U_v=0\)\label{line:rcpc-player-init-start}
\Statex \hspace{\algorithmicindent}and \(\operatorname{lastA}[v]=-\infty\).
\State \(a\gets0;\quad P\gets H_L;\quad\ell\gets\bot;\quad\nu\gets\none\)
\State \Call{StateUpdate}{$\SEEK_{\mathrm{fresh}}$}\label{line:rcpc-player-init-end}
\While{player \(i\) is active}
  \If{\(\nu\ne\none\)}\hypertarget{scpc:scheduler}{}
    \State \Call{StateUpdate}{$\nu$}; \(\nu\gets\none\)
  \EndIf
  \If{\(m=\REPLAY\) and \(a\ge q\)}\label{line:rcpc-player-expiry-start}
    \State \Call{StateUpdate}{$\SEEK_{\mathrm{fresh}}$}\label{line:rcpc-player-expiry-end}
  \EndIf
  \If{\(m=\WORK\)}
    \State \((v,\mathit{type})\gets\)\Call{WorkAction}{}
  \ElsIf{\(m=\REPLAY\)}
    \State \(v\gets y\)
  \ElsIf{\(m=\phase{Root}\), \(s=\phase{Scan}\), and \(\neg\mathrm{busNext}\)}
    \State \(v\gets j\)
  \Else
    \State \(v\gets\BUS\)
  \EndIf
  \State Propose to \(v\) once and observe \(o\in\{\A,\R\}\).
  \If{\(o=\A\)}\label{line:rcpc-player-records-start}
    \State Observe player \(i\)'s realized utility \(r\) and set \(N_v\gets N_v+1\)
    \State \(U_v\gets U_v+r;\quad\operatorname{lastA}[v]\gets a\)\label{line:rcpc-player-records-end}
  \EndIf
  \If{\(m\in\{\SEEK,\phase{Root}\}\)}
    \State \Call{CollectionFeedback}{$v,o$}
  \ElsIf{\(m=\WORK\)}
    \State \Call{WorkFeedback}{$v,\mathit{type},o$}
  \ElsIf{\(o=\R\)}\hypertarget{scpc:replay}{}\label{line:rcpc-player-replay-start}
    \State \(q\gets\max\{a+1,\ell+H_L\}\) \Comment{\(m=\REPLAY\).}\label{line:rcpc-player-replay-end}
  \EndIf
  \State \(a\gets a+1\)
\EndWhile
\end{algorithmic}
\end{algorithm}

\paragraph{Parameters and utility observations.}
As in \protocol{}, \(K\) is the number of arms, \(\mathcal K=\{0,\ldots,K-1\}\), and arm~\(1\) serves as PARK. The parameters of \textsc{SC-PC} are
\begin{equation}\begin{aligned}
 &\chi_L=4K-2, \quad \kappa_L=\chi_L-1, \quad H_L=d_L=\chi_L+1, \quad b_K=2K^2,\\
 &W_L=K(K-1)^2+(2K-1)b_K=5K^3-4K^2+K,\\
 &h_{D,L}=2\chi_L+1, \quad f_L=W_L+3\chi_L, \quad J_L=W_L+4\chi_L,\\
 &G_L=2K+3, \quad R_{c,L}=2K+1, \quad \tau_L=J_L+2\chi_L+3,\\
 &E_L=\tau_L+1, \quad M_L=E_L+K .
\end{aligned}\label{eq:rcpc:parameters}
\end{equation}

Here \(J_L\) is the number of rounds in each \WORK{} invocation. The timing margin \(\chi_L\) allows players to enter \WORK{} at different rounds. Under the entry conditions in Appendix~\ref{sec:rcpc:control}, their entry rounds differ by at most \(\kappa_L\), and their deletion windows overlap for at least \(W_L\) rounds. The proof uses \(b_K\) to count the samples collected during this overlap.

A player keeps a sample count \(N_x\) and a sum \(U_x\) of its observed utilities for each arm \(x\). An arm keeps the corresponding records for each proposer it recognizes. When arm \(k\) accepts player \(i\), both record the acceptance on their own clocks, the player in \(\operatorname{lastA}[k]\) and the arm in \(\operatorname{last}[i]\). Counts and sums start at zero and acceptance times at \(-\infty\). Each acceptance gives both sides a sample, including when the observed utility is zero. Rejected proposals yield no utility observations. These records persist across mode changes. We write \(N_i(k)\) for the number of times arm \(k\) has accepted player \(i\).

For a player \(i\) comparing arms, the mean utility of option \(k\) is \(\theta_k=\mu_{i,k}\). For an arm \(k\) comparing proposers, it is \(\theta_i=\pi_{k,i}\). Using its own local clock \(a\) and observations through the previous round, each side computes
\begin{equation}
 \widehat\theta_x=\begin{cases}U_x/N_x,&N_x>0,\\0,&N_x=0,\end{cases}
 \qquad B(a)=5\log(a+2),\qquad
 \beta_a(x,z)=\sqrt{\frac{B(a)}2\left(\frac1{N_x}+\frac1{N_z}\right)}.
\label{eq:rcpc:radius}
\end{equation}

We use base-two logarithms throughout this appendix and write \(\ln\) for the natural logarithm. If either count is zero, set \(\beta_a(x,z)=+\infty\). For a nonempty set \(Q\), the options that the pairwise confidence comparisons have not ruled out as best are
\begin{equation}
 C_a(Q)=\{x\in Q:\widehat\theta_z-\widehat\theta_x\le\beta_a(z,x)
                       \text{ for every }z\in Q\}.
\label{eq:rcpc:candidates}
\end{equation}

Every empirical maximizer belongs to \(C_a(Q)\). A player uses this set to compare its remaining arms, and an arm in \phase{Business} uses it to compare its current proposers.

\paragraph{Arm service.}
When an arm first receives a proposal, it starts its local clock at \(a=0\) and enters \phase{Control}. Its clock then advances in every round, including rounds with no proposals. In Algorithm~\ref{alg:rcpc-arm}, \(R\) denotes the current proposers, \(e\) counts consecutive empty rounds, and \(\lambda\) is the time remaining in \phase{Business}. The arm uses the value of \(\lambda\) at the start of the round. For \(R\ne\emptyset\), the eligible proposers are
\begin{equation}
 Q=\begin{cases}
       C_a(R),&\lambda>0,\\
       R,&\lambda=0\text{ and }a\equiv0\pmod{G_L},\\
       \{\min_{\prec}R\},&\lambda=0\text{ and }a\not\equiv0\pmod{G_L},
    \end{cases}
\label{eq:rcpc:arm-service}
\end{equation}

Here \(\prec\) is the arm's fixed private order of proposers, which may differ from its utility ranking. In \phase{Control} (\(\lambda=0\)), rounds with \(a\equiv0\pmod{G_L}\) are \emph{fair} and the remaining rounds are \emph{ordinary}. Among eligible proposers, those the arm has never accepted come first because their last acceptance times are \(-\infty\). Two consecutive empty rounds start or renew \(\tau_L\) rounds of \phase{Business} service, and each further empty round renews it.

\begin{algorithm}[H]
\caption{\textsc{SC-PC} at an arm}
\label{alg:rcpc-arm}\hypertarget{scpc:arm}{}
\raggedright
\algrenewcommand{\algorithmicindent}{0.75em}
\begin{algorithmic}[1]
\State Upon receiving its first proposal, set \(a=e=\lambda=0\).
\State In this and every subsequent round, observe the set \(R\) of current proposers. For each previously unseen \(i\in R\), initialize \(N_i=U_i=0\) and \(\operatorname{last}[i]=-\infty\).
\State The round starts in \phase{Control} if \(\lambda=0\) and in \phase{Business} if \(\lambda>0\).
\If{\(R\ne\emptyset\)}\label{line:rcpc-arm-select-start}
  \State In \phase{Control}, set \(Q\gets R\) if \(a\equiv0\pmod{G_L}\), and \(Q\gets\{\min_{\prec}R\}\) otherwise.
  \State In \phase{Business}, set \(Q\gets C_a(R)\).
  \State Accept the proposer \(i\in Q\) minimizing \(\operatorname{last}[i]\), breaking ties by \(\prec\), and reject the other proposers.\label{line:rcpc-arm-select-end}
  \State Observe the arm's utility \(r\) and set \(N_i\gets N_i+1\), \(U_i\gets U_i+r\), and \(\operatorname{last}[i]\gets a\).\label{line:rcpc-arm-records}
\Else
  \State Accept nobody.
\EndIf
\State \hypertarget{scpc:lease}{}Set \(e\gets e+1\) if \(R=\emptyset\), and \(e\gets0\) otherwise.\label{line:rcpc-arm-lease-start}
\State If \(e\ge2\), set \(\lambda\gets\tau_L\). Otherwise set \(\lambda\gets\max\{\lambda-1,0\}\).\label{line:rcpc-arm-lease-end}
\State Set \(a\gets a+1\) and repeat, including during empty rounds.
\end{algorithmic}
\end{algorithm}


\begin{algorithm}[H]
\caption{\textsc{StateUpdate}\((u)\)}
\label{alg:rcpc-state-update}\hypertarget{scpc:entry}{}
\raggedright
\algrenewcommand{\algorithmicindent}{0.75em}
\begin{algorithmic}[1]
\If{\(u\in\{\SEEK_{\mathrm{fresh}},\SEEK_{\mathrm{armed}}\}\)}\label{line:rcpc-entry-seek-start}
  \State \(m\gets\SEEK;\quad x\gets0\)
  \State \(\mathrm{armed}\gets(u=\SEEK_{\mathrm{armed}})\)
  \State \(c\gets a+J_L+\chi_L+4\)
  \State \textbf{if} \(\mathrm{armed}\) \textbf{then} \(b\gets a\); \textbf{else} \(b\gets a+E_L\)\label{line:rcpc-entry-seek-end}
\ElsIf{\(u=\phase{Root}\)}
  \State \(m\gets\phase{Root};\quad x\gets0\)\Comment{Retain \(c\).}
  \If{\(a<c\)}
    \State \(s\gets\phase{Clear}\)
  \Else
    \State \(s\gets\phase{Scan};\quad j\gets1\)
    \State \(\mathrm{busNext}\gets\true\)
  \EndIf
\ElsIf{\(u=\phase{Scan}\)}
  \State \(s\gets\phase{Scan};\quad j\gets1\)
  \State \(\mathrm{busNext}\gets\true\)
\ElsIf{\(u=\phase{Commit}\)}
  \State \(s\gets\phase{Commit};\quad v_C\gets a\)\Comment{Local age at \phase{Commit} entry.}
  \State \(x\gets\min\{x,0\}\)
\ElsIf{\(u=\WORK\)}
  \State \(m\gets\WORK;\quad h\gets0;\quad S_i\gets\mathcal K\)
  \State \(\mathrm{held}\gets\bot;\quad\mathrm{target}\gets\bot;\quad z\gets0\)
\Else\Comment{\(u=\REPLAY\).}
  \State \(m\gets\REPLAY\)
\EndIf
\end{algorithmic}
\end{algorithm}

\paragraph{\SEEK{} and \phase{Root}.}
On arrival or replay expiry, the player makes a \emph{fresh entry} into \SEEK{}. It clears its qualification flag and proposes to BUS throughout a guard of \(E_L\) rounds. In Algorithm~\ref{alg:rcpc-state-update}, \(b\) is the local age at which BUS outcomes begin to count toward qualification. On fresh entry this is the first round after the guard. The local deadline \(c\) is the earliest age at which the player can begin scanning if it enters \phase{Root}. After \(K\) consecutive rejected BUS proposals in \phase{Root}, the player returns to \SEEK{} with the flag set (\emph{armed}) and \(b=a\), so qualification begins immediately.

In \SEEK{} and \phase{Root}, the player uses a signed counter \(x\) for successive BUS acceptances (positive) or rejections (negative). Proposals to other arms during a scan leave \(x\) unchanged. In \SEEK{}, the player proposes to BUS every round, and only BUS outcomes from local age \(b\) onward count toward qualification. A run of \(K\) such rejections arms the player for the rest of the visit. Three consecutive such acceptances send an armed player to \WORK{} and an unarmed player to \phase{Root} in the next round. Throughout \SEEK{}, including the guard, the player sets \(c=a+J_L+\chi_L+3\) whenever \(x\le-3\), giving earlier \WORK{} invocations time to finish.

A player entering \phase{Root} clears \(x\). If \(a<c\), it proposes to BUS until age \(c\), when the scan starts. Otherwise it starts scanning immediately. The scan visits arms \(1,\ldots,K-1\) in order, alternating between BUS and the current arm until that arm accepts. In \phase{Commit}, the player proposes only to BUS. On entry it clears a positive \(x\) so that only acceptances during \phase{Commit} count toward \WORK{}. It retains a negative \(x\) because \(K\) successive BUS rejections take precedence in every \phase{Root} stage, even when the run crosses a stage boundary. The player enters \WORK{} in the next round once it has spent at least \(M_L=E_L+K\) rounds in \phase{Commit} and its last \(R_{c,L}\) proposals have all been accepted. This minimum duration leaves time for the last player recalled by the scan to finish its guard and receive \(K\) qualifying BUS rejections. Algorithm~\ref{alg:rcpc-collection} specifies these tests, and Figure~\ref{fig:scpc-modes} shows the resulting mode transitions.

\begin{algorithm}[H]
\caption{\textsc{CollectionFeedback}\((v,o)\)}
\label{alg:rcpc-collection}
\raggedright
\algrenewcommand{\algorithmicindent}{0.75em}
\begin{algorithmic}[1]
\If{\(v=\BUS\)}\label{line:rcpc-collection-qualify-start}
  \State \textbf{if} \(o=\A\) \textbf{then} \(x\gets\max\{x,0\}+1\)
  \State \textbf{else} \(x\gets\min\{x,0\}-1\)
\EndIf
\If{\(m=\SEEK\)}\hypertarget{scpc:seek}{}
  \State \textbf{if} \(x\le-3\) \textbf{then} \(c\gets a+J_L+\chi_L+3\)
  \If{\(a\ge b\), \(x<0\), and \(\min\{-x,a-b+1\}\ge K\)}
    \State \(\mathrm{armed}\gets\true\)
  \EndIf
  \If{\(a\ge b+2\) and \(x\ge3\)}
    \State \textbf{if} \(\mathrm{armed}\) \textbf{then} \(\nu\gets\WORK\)
    \State \textbf{else} \(\nu\gets\phase{Root}\)\label{line:rcpc-collection-qualify-end}
  \EndIf
  \State \Return
\EndIf
\If{\(v=\BUS\) and \(x=-K\)}\hypertarget{scpc:root}{}\label{line:rcpc-collection-root-start}
  \State \(\nu\gets\SEEK_{\mathrm{armed}}\); \Return
\EndIf
\If{\(s=\phase{Clear}\)}
  \State \textbf{if} \(a+1\ge c\) \textbf{then} \(\nu\gets\phase{Scan}\)
\ElsIf{\(s=\phase{Scan}\)}
  \If{\(v=\BUS\)}
    \State \(\mathrm{busNext}\gets\false\)
  \Else
    \State \(\mathrm{busNext}\gets\true\)
    \If{\(o=\A\)}
      \State \textbf{if} \(j=K-1\) \textbf{then} \(\nu\gets\phase{Commit}\)
      \State \textbf{else} \(j\gets j+1\)
    \EndIf
  \EndIf
\ElsIf{\(a+1\ge v_C+M_L\) and \(x\ge R_{c,L}\)}
  \State \(\nu\gets\WORK\)\Comment{\(s=\phase{Commit}\).}\label{line:rcpc-collection-root-end}
\EndIf
\end{algorithmic}
\end{algorithm}


\paragraph{Proposals in \WORK{}.}
On entry, player \(i\) resets its remaining-arm set to \(S_i=\mathcal K\), clears its held arm and target, and sets its rejection counter \(z=0\). Algorithm~\ref{alg:rcpc-work-action} chooses the proposal after \(h\) completed \WORK{} rounds, starting at \(h=0\).

After \(d_L\) rounds at PARK, the player proposes to its held arm whenever it has one. If it holds none, \(h\le f_L\), \(S_i\ne\emptyset\), and \(C_a(S_i)=\{v\}\), it certifies \(v\) and records it as held before proposing, even if \(v\) then rejects it. A player still holding no arm explores. When it needs a target, it chooses a least-sampled arm in \(C_a(S_i)\), breaking ties by arm index and using \(C_a(\mathcal K)\) if \(S_i\) is empty. It keeps the target until an accepted probe, deletion, or certification, even if the confidence set changes.

A player \emph{probes} its target if that arm has not accepted it in the preceding \(K-1\) rounds, in any mode. Otherwise it proposes to the smallest-index \emph{filler} meeting the same condition among all \(K\) arms, including deleted arms. At most \(K-1\) arms could have accepted the player in those rounds, so a filler exists.

\begin{algorithm}[H]
\caption{\textsc{WorkAction}\(()\)}
\label{alg:rcpc-work-action}\hypertarget{scpc:work-action}{}
\raggedright
\algrenewcommand{\algorithmicindent}{0.75em}
\begin{algorithmic}[1]
\If{\(h<d_L\)}\label{line:rcpc-work-priority-start}
  \State \(z\gets0\); \Return \((\PARK,\mathrm{park})\)
\EndIf
\If{\(\mathrm{held}\ne\bot\)}
  \State \Return \((\mathrm{held},\mathrm{hold})\)
\EndIf
\If{\(S_i\ne\emptyset\), \(h\le f_L\), and \(|C_a(S_i)|=1\)}
  \State \(v\gets\) the unique element of \(C_a(S_i)\)
  \State \(\mathrm{held}\gets v;\quad\mathrm{target}\gets\bot;\quad z\gets0\)
  \State \Return \((v,\mathrm{hold})\)\label{line:rcpc-work-priority-end}
\EndIf
\If{\(\mathrm{target}=\bot\)}
  \State \textbf{if} \(S_i\ne\emptyset\) \textbf{then} \(Q\gets C_a(S_i)\)
  \State \textbf{else} \(Q\gets C_a(\mathcal K)\)
  \State \(\mathrm{target}\gets\arg\min_{v\in Q}(N_v,v)\)
  \State \(z\gets0\)
\EndIf
\If{\(a-\operatorname{lastA}[\mathrm{target}]\ge K\)}\label{line:rcpc-work-cooldown-start}
  \State \Return \((\mathrm{target},\mathrm{probe})\)
\EndIf
\State \(z\gets0\)
\State \(v\gets\min\{w\in\mathcal K:a-\operatorname{lastA}[w]\ge K\}\)
\State \Return \((v,\mathrm{filler})\)\label{line:rcpc-work-cooldown-end}
\end{algorithmic}
\end{algorithm}

\paragraph{\WORK{} feedback and \REPLAY{}.}
During the deletion window \(h_{D,L}\le h<f_L\), each rejection at a held arm or on a probe increments \(z\). Choosing a new target or certifying an arm resets \(z\) before the proposal, so a rejected probe or holding proposal to that arm starts the new count at one. A filler proposal, an acceptance, or a round outside the window resets \(z\). When \(z=K\), the player removes the proposed arm from \(S_i\) and clears its held arm, target, and counter. Certification remains possible at \(h=f_L\), after the deletion window has closed.

Player \(i\) outputs in its final \WORK{} round (\(h=J_L-1\)). Its output arm \(y\) is the arm \(v\) it proposed to in that round, even if the proposal was rejected. Its output age is the local clock reading \(a\) in that round, which corresponds to global round \(O_i=b_i+a\) in the analysis. Algorithm~\ref{alg:rcpc-work-feedback} uses the previous output age \(\ell\) and scale \(P\) to choose the waiting length \(L\), then saves \(a\) as the new \(\ell\).
\begin{equation}\begin{aligned}
 \widetilde L&=\begin{cases}H_L,&\ell=\bot,\\H_L+2(a-\ell),&\ell\ne\bot,\end{cases}
 &P^+&=\max\left\{P,\left\lfloor\frac{3P+\widetilde L}{4}\right\rfloor\right\},\\
 L&=\begin{cases}P^+,&\mathsf{held}\ne\bot\text{ and the final proposal is accepted},\\
                  H_L,&\text{otherwise},\end{cases}\\[-2pt]
 (P,\ell,y,q)&\leftarrow(P^+,a,v,a+L).
\end{aligned}\label{eq:rcpc:output}
\end{equation}

The held-and-accepted case takes the \emph{long} branch (\(L=P^+\)), even if \(P^+=H_L\). All other cases take the \emph{short} branch (\(L=H_L\)). In round \(O_i+1\) the player enters \REPLAY{} and proposes to \(y\). Acceptance preserves \(q\), while rejection sets
\begin{equation}
                     q\leftarrow\max\{a+1,\ell+H_L\}.
\label{eq:rcpc:rejection}
\end{equation}

At the first round with \(a\ge q\), the player enters fresh \SEEK{} before choosing its arm (Algorithm~\ref{alg:rcpc-player}). If the deadline is never shortened, the player replays \(y\) at local ages \(\ell+1,\ldots,\ell+L-1\) and proposes to BUS at age \(\ell+L\).

\begin{algorithm}[H]
\caption{\textsc{WorkFeedback}\((v,\mathit{type},o)\)}
\label{alg:rcpc-work-feedback}\hypertarget{scpc:work-feedback}{}
\raggedright
\algrenewcommand{\algorithmicindent}{0.75em}
\begin{algorithmic}[1]
\If{\(\mathit{type}=\mathrm{probe}\) and \(o=\A\)}
  \State \(\mathrm{target}\gets\bot;\quad z\gets0\)
\EndIf
\If{\(h_{D,L}\le h<f_L\), \(\mathit{type}\in\{\mathrm{hold},\mathrm{probe}\}\), and \(o=\R\)}\label{line:rcpc-work-delete-start}
  \State \(z\gets z+1\)
\Else
  \State \(z\gets0\)
\EndIf
\If{\(z=K\)}
  \State \(S_i\gets S_i\setminus\{v\}\)
  \State \(\mathrm{held}\gets\bot;\quad\mathrm{target}\gets\bot;\quad z\gets0\)\label{line:rcpc-work-delete-end}
\EndIf
\If{\(h<J_L-1\)}
  \State \(h\gets h+1\); \Return
\EndIf
\State \hypertarget{scpc:output}{}Using the previous \(P,\ell\), compute \(\widetilde L\) and \(P^+\) from Equation~\eqref{eq:rcpc:output}.\label{line:rcpc-output-start}
\State \textbf{if} \(\mathrm{held}\ne\bot\) and \(o=\A\) \textbf{then} \(L\gets P^+\)
\State \textbf{else} \(L\gets H_L\)
\State \((P,\ell,y,q)\gets(P^+,a,v,a+L)\)
\State \(\nu\gets\REPLAY\)\label{line:rcpc-output-end}
\end{algorithmic}
\end{algorithm}

%% file: appendices/D_Claim1.tex
\section{Proof of Claim~\ref{cor:learning-static-lower}}
\label{app:learning-lower}
We adapt the single-edge change-of-measure argument of ~\cite{SBA21}.
In each alternative instance, we change one arm-side mean so that a player--arm pair enters the POSM. Each match of that pair in the original instance contributes one mismatch round.

We write $\log$ for the base-two logarithm. Both relative entropy $\operatorname{KL}$ and its Bernoulli specialization $\operatorname{kl}$ use natural logarithms. Fix a protocol $\Pi$, including its proposal and acceptance rules, independently of the unknown means. For this lower bound, whenever player $i$ and arm $k$ are matched, they independently observe Bernoulli utilities with means $\mu_{i,k}$ and $\pi_{k,i}$, respectively. The draws are independent across pairs and rounds and independent of the protocol's randomization. Rejected proposals yield no utility observations.

\begin{lemma}[Information required by a one-edge alternative]
\label{lem:learning-lower:one-edge}
Let \(I\) and \(J\) be static Bernoulli instances with strict preferences and the same population. They differ only in the mean of one endpoint of edge \(e\), which is \(p\) in \(I\) and \(q\) in \(J\), with \(p,q\in(0,1)\). Suppose \(e\) is absent from the POSM of \(I\) and present in the POSM of \(J\). For each \(\alpha>0\), assume
\[
 \mathbb E_I[D_\Pi(T;I)]=o(T^\alpha)
 \qquad\text{and}\qquad
 \mathbb E_J[D_\Pi(T;J)]=o(T^\alpha).
\]
If \(N_e(T)\) counts the rounds through \(T\) in which \(e\) is accepted, then
\[
 \liminf_{T\to\infty}\frac{\mathbb E_I[N_e(T)]}{\ln T}
 \ge\frac1{\operatorname{kl}(p,q)}.
\]
\end{lemma}

\begin{proof}
Use the same initial local information and proposer identities in both instances. Let \(\mathbb P_I^T\) and \(\mathbb P_J^T\) be the laws of the full transcript through round \(T\), including the protocol's randomization, all proposals and acceptance decisions, and both rewards on every accepted edge. Given the randomization and the past transcript, the players' proposals and the arms' responses to them determine the current matching before any new rewards are drawn. These rules are the same in both instances. Conditional on that matching and the past, each accepted edge \((i,k)\) contributes two independent Bernoulli observations with player mean \(\mu_{i,k}\) and arm mean \(\pi_{k,i}\). Let \(N_{i,k}(T)\) count acceptances of \((i,k)\) through \(T\). The chain rule for relative entropy gives
\begin{equation}
 \operatorname{KL}(\mathbb P_I^T,\mathbb P_J^T)
 =\sum_{i,k}\mathbb E_I[N_{i,k}(T)]
 \left\{\operatorname{kl}(\mu_{i,k}^{I},\mu_{i,k}^{J})
             +\operatorname{kl}(\pi_{k,i}^{I},\pi_{k,i}^{J})\right\}.
\label{eq:learning-lower:kl-transcript}
\end{equation}
Only one mean changes, so this identity reduces to
\[
 \operatorname{KL}(\mathbb P_I^T,\mathbb P_J^T)
 =\mathbb E_I[N_e(T)]\operatorname{kl}(p,q).
\]
We apply the change-of-measure argument of \cite[Lemma 1]{KCG16} to this transcript.

Since the POSM contains \(e\) only in \(J\), the pathwise bounds are
\[
              N_e(T)\le D_\Pi(T;I)\ \text{in }I,\qquad
              T-N_e(T)\le D_\Pi(T;J)\ \text{in }J.
\]
Fix \(\alpha\in(0,1)\) and let \(A_T=\{N_e(T)<T/2\}\). Markov's inequality and the two expected-cost bounds give
\[
 \begin{aligned}
 \mathbb P_I^T(A_T^c)&\le 2\mathbb E_I[D_\Pi(T;I)]/T=o(T^{\alpha-1}),\\
 \mathbb P_J^T(A_T)&\le 2\mathbb E_J[D_\Pi(T;J)]/T=o(T^{\alpha-1}).
 \end{aligned}
\]
Applying data processing to the indicator of \(A_T\) and bounding its binary entropy by \(\ln 2\) gives
\[
 \operatorname{KL}(\mathbb P_I^T,\mathbb P_J^T)
 \ge \mathbb P_I^T(A_T)\ln\frac1{\mathbb P_J^T(A_T)}-\ln 2.
\]
After division by \(\ln T\), the right-hand side has lower limit at least \(1-\alpha\). Letting \(\alpha\downarrow0\) and using \eqref{eq:learning-lower:kl-transcript} gives
\begin{equation}
 \liminf_{T\to\infty}\frac{\mathbb E_I[N_e(T)]}{\ln T}
                 \ge\frac1{\operatorname{kl}(p,q)}.
\label{eq:learning-lower:one-edge}
\end{equation}
\end{proof}

\begin{proof}[Proof of Claim~\ref{cor:learning-static-lower}]
Fix \(K\ge2\) and \(0<\Delta\le\min\{1/8,1/[4(K-1)]\}\), as in Claim~\ref{cor:learning-static-lower}.

\textit{Original instance and unique stable matching.}
In the original instance \(I\), players \(p_0,\ldots,p_{K-1}\) enter in round \(1\) and stay forever. The arms are \(0,\ldots,K-1\). Choose player means
\[
 \mu_{p_0,j}=\frac12+(K-1-j)\Delta,
 \qquad
 \mu_{p_r,j}=\begin{cases}
       3/4,&j=r-1,\\
       (j+1)\Delta,&j\ne r-1,
      \end{cases}\qquad 1\le r<K.
\]
On each arm $j<K-1$, set
\[
 \pi_{j,p_{j+1}}=\frac12,\qquad
 \pi_{j,p_0}=\frac12-\Delta,\qquad
 \pi_{j,p_r}=r\Delta\quad(r\notin\{0,j+1\}).
\]
On the last arm, set $\pi_{K-1,p_0}=3/4$ and $\pi_{K-1,p_r}=r\Delta$ for $r\ge1$. All means lie in $(0,1)$ and all local rankings are strict. In each pair $(p_r,r-1)$, $r\ge1$, the endpoints are mutual first choices, so the pair occurs in every stable matching. The remaining player $p_0$ receives arm $K-1$, giving a unique stable matching.

\textit{The minimum preference gap.}
Consecutive arms in $p_0$'s ranking differ by exactly $\Delta$ in mean utility.
For every other player, the means below its first choice are distinct multiples of $\Delta$ no larger than $K\Delta\le 1/4+\Delta\le3/8$, whereas its first choice has mean $3/4$.
At each arm $j<K-1$ the two highest means are $1/2$ and $1/2-\Delta$. The lower means are distinct multiples of $\Delta$ bounded by $(K-1)\Delta\le1/4$, so each is separated from $1/2-\Delta$ by at least $1/4-\Delta\ge\Delta$.
At the last arm the same bound separates every lower mean from $3/4$, and distinct lower means differ by at least $\Delta$.
Thus every gap in \eqref{eq:learning-gap} is at least $\Delta$, and $\Delta_{\min}=\Delta$.

\textit{A one-edge alternative.}
For each \(0\le j<K-1\) and \(0<\varepsilon<1/8\), form \(I_{j,\varepsilon}\) by changing only \(\pi_{j,p_0}\) to \(1/2+\varepsilon\). The \(K-2\) pairs \((p_r,r-1)\) with \(1\le r<K\) and \(r\ne j+1\) are still mutual first choices. Once these pairs are fixed, \(p_0\) prefers \(j\) to \(K-1\) and arm \(j\) prefers \(p_0\) to \(p_{j+1}\). Thus every stable matching contains \((p_0,j)\), leaving arm \(K-1\) for \(p_{j+1}\). Again the stable matching is unique. The only pairs that change are
\[
 \{(p_0,K-1),(p_{j+1},j)\}
 \quad\longrightarrow\quad
 \{(p_0,j),(p_{j+1},K-1)\}.
\]
If \(0<\varepsilon<\Delta\), the alternative has \(\Delta_{\min}=\varepsilon\). Arm \(j\) ranks \(p_0\) above \(p_{j+1}\) by \(\varepsilon\), and every other comparison has gap at least \(\Delta\). Fix any uniformly efficient protocol \(\Pi\). Its expected-cost guarantee applies to every fixed static Bernoulli instance with strict preferences, so it applies to \(I\) and each \(I_{j,\varepsilon}\) even when their minimum gaps differ.

Write \(N_j(T)=N_{p_0,j}(T)\). Applying \Cref{lem:learning-lower:one-edge} to \(I\) and \(I_{j,\varepsilon}\) gives
\[
 \liminf_{T\to\infty}\frac{\mathbb E_I[N_j(T)]}{\ln T}
       \ge\frac1{\operatorname{kl}(1/2-\Delta,1/2+\varepsilon)}.
\]
The expectation on the left is under the same original instance \(I\) for every \(\varepsilon\). Having taken \(T\to\infty\) for fixed \(\varepsilon\), we can now let \(\varepsilon\downarrow0\) on the right by continuity of \(\operatorname{kl}\).

\textit{From samples to mismatch rounds.}
Each acceptance of one of these edges in the original instance incurs a mismatch. All these edges share player $p_0$, so at most one can be accepted in any round. Hence
\[
 \sum_{j=0}^{K-2}N_j(T)\le D_\Pi(T;I).
\]
For fixed $K$, the finite sum gives
\[
 \liminf_{T\to\infty}\frac{\mathbb E_I[D_\Pi(T;I)]}{\ln T}
 \ge\sum_{j=0}^{K-2}\liminf_{T\to\infty}
       \frac{\mathbb E_I[N_j(T)]}{\ln T}
 \ge\frac{K-1}{\operatorname{kl}(1/2-\Delta,1/2)}.
\]

\textit{The KL constant and logarithm conversion.}
Let \(f(p)=\operatorname{kl}(p,1/2)\). Then \(f(1/2)=f'(1/2)=0\) and \(f''(p)=1/[p(1-p)]\le64/15\) on \([3/8,1/2]\). Since \(\Delta\le1/8\), Taylor's theorem gives
\[
 \operatorname{kl}(1/2-\Delta,1/2)
 \le\frac{32}{15}\Delta^2
 <4(\ln2)\Delta^2.
\]
Since \(\ln T=(\ln 2)\log T\), the bound on the original instance becomes
\[
 \liminf_{T\to\infty}\frac{\mathbb E_I[D_\Pi(T;I)]}{\log T}
 \ge\frac{(\ln2)(K-1)}{\operatorname{kl}(1/2-\Delta,1/2)}
 \ge\frac{K-1}{4\Delta^2}.
\]
The same bound holds for a uniformly efficient central controller that knows the player means, chooses a feasible matching each round, and observes both utilities on every matched pair. The revealed player means coincide in \(I\) and \(I_{j,\varepsilon}\). Given the past and the controller's randomization, the matching is chosen before the current utilities are observed, so the same chain rule gives \eqref{eq:learning-lower:kl-transcript}.
\end{proof}

\paragraph{Comparison with binary stability.}
Binary stable regret counts rounds in which the realized matching is unstable~\cite{basu2025super}. In the original instance and each alternative, the POSM is the only stable matching, so binary stable regret and the mismatch cost count exactly the same rounds. In a market with several stable matchings, a round spent at a different stable matching contributes one to the mismatch cost and zero to binary stable regret.

%% file: sections/ai-methodology.tex
\section{AI Methodology}
\label{sec:ai-methodology}

We used ChatGPT and Codex for manuscript preparation and consistency checks throughout the paper, as well as for technical discussions and mathematical checks of the algorithms and analyses. The authors take full responsibility for the correctness, originality, and integrity of the paper, including all proofs and references.